\documentclass[aps,prb,reprint]{revtex4-1}

\usepackage{amsmath}
\usepackage{amssymb}
\usepackage{graphicx}
\usepackage{epstopdf}
\usepackage{dcolumn}
\usepackage{cases}
\usepackage{mathtools}
\usepackage{dcolumn}
\usepackage{color}
\usepackage{bbold}
\usepackage{bm}

\DeclarePairedDelimiter\bra{\langle}{\rvert}
\DeclarePairedDelimiter\ket{\lvert}{\rangle}

\newcommand\zero{\mathbb{0}}
\newcommand\one{\mathbb{1}}

\renewcommand\vec[1]{\bm{#1}}

\newcommand\RED[1]{#1}
\newcommand\BLUE[1]{#1}

\begin{document}
 
\title{\RED{Electrical manipulation of semiconductor 
spin qubits within the $g$-matrix formalism}}
 
\author{Benjamin Venitucci} 
\author{L\'eo Bourdet}
\author{Daniel Pouzada}
\author{Yann-Michel Niquet}
\email{yniquet@cea.fr}
\affiliation{Univ. Grenoble Alpes, CEA, INAC-MEM-L\_Sim, F-38000 Grenoble, France}

\begin{abstract}
We discuss the modeling of the electrical manipulation of spin qubits in the linear-response regime where the Rabi frequency is proportional to the magnetic field and to the radio-frequency electric field excitation. We show that the Rabi frequency can be obtained from a generalized $g$-tensor magnetic resonance formula featuring a $g$-matrix and its derivative $g^\prime$ with respect to the electric field (or gate voltage) as inputs. These matrices can be easily calculated from the wave functions of the qubit at zero magnetic field. The $g$-matrix formalism therefore provides the complete dependence of the Larmor and Rabi frequencies on the orientation of the magnetic field at very low computational cost. It also provides a compact model for the control of the qubit, and a simple framework for the analysis of the effects of symmetries on the anisotropy of the Larmor and Rabi frequencies. \RED{The $g$-matrix formalism applies to a wide variety of electron and hole qubits, and we focus on a hole qubit in a silicon-on-insulator nanowire as an illustration. We show that the Rabi frequency of this qubit shows a complex dependence on the orientation of the magnetic field, and on the gate voltages that control the symmetry of the hole wave functions. We point out that the qubit may be advantageously switched between two bias points, one where it can be manipulated efficiently, and one where it is largely decoupled from the gate field but presumably longer lived. We also discuss the role of residual strains in such devices in relation to recent experiments.}
\end{abstract}

\maketitle

\section{Introduction}
\label{sectionIntro}

Single spins in semiconductors have attracted a lot of interest in the prospect of storing and manipulating quantum information.\cite{Kane98,Loss98} Spin quantum bits (qubits) have actually been demonstrated in different semiconductor materials (GaAs,\cite{Petta05,Koppens06} Si\cite{Pla12}...) in the last two decades.\cite{Hanson07} They are usually based on spins confined to an impurity, or to a lithography or electrically defined quantum dot. Silicon\cite{Zwanenburg13} is, in particular, a very attractive host for spin qubits as it can be isotopically purified in order to get rid of the nuclear spins that may interact with the electron spins and speed up decoherence.\cite{Tyryshkin12} Single qubits and two qubit gates with high fidelity have already been reported in silicon,\cite{Veldhorst14,Veldhorst15b,Takeda16,Yoneda18,Watson18} as well as hole spin qubits.\cite{Maurand16,Crippa18}

Spins can be manipulated by a radio-frequency (RF) magnetic field resonant with the Zeeman splitting between the up and down states (Electron Spin Resonance, ESR).\cite{Koppens05,Pla12,Veldhorst14} However, in order to address a specific qubit, it may be more desirable to manipulate the spin with the RF electric field from a local gate (Electric Dipole Spin Resonance, EDSR\cite{Rashba91,Rashba08}). For that purpose, the real space motion of the carrier in this RF electric field must be coupled to its spin. Such a spin-orbit coupling (SOC) is actually intrinsic to semiconductor materials;\cite{Winkler03} this relativistic effect can be described semi-classically as the action of the magnetic field created by the nuclei moving in the frame of the carrier onto its spin. Intrinsic SOC is stronger in heavier semiconductors, and its effects are larger in the valence than in the conduction band. Yet the electrical manipulation of electron spins based on intrinsic SOC has been reported in many materials,\cite{Kato03,Nowack07,NadjPerge10,vandenBerg13} including silicon where SOC is particularly weak in the conduction band.\cite{Corna18}

The electrical manipulation of spins based on intrinsic SOC has usually been described by two mechanisms. The first mechanism, named $g$-tensor magnetic resonance ($g$-TMR),\cite{Kato03} exploits the anisotropy and spatial dependence of the gyromagnetic factors:\cite{Pingenot11,Schroer11,Takahashi13,Ares13,Voisin16} the RF electric field modulates the confinement potential and wave function of the qubit, hence the $g$-factors and the spin precession (Larmor) vector, which results in a rotation of the spin at resonance. In the second mechanism, the RF electric field shakes the wave function of the qubit as a whole in a quasi-harmonic potential; spin-orbit interactions build up during this motion and drive the spin rotation.\cite{Rashba03,Golovach06,Flindt06,Nowack07,NadjPerge10,vandenBerg13} The shape of the wave function and the $g$-factors do not change under the RF excitation, so that the contribution from this mechanism can not be anticipated from the measurement of the Zeeman splitting as a function of gate voltage.\cite{Crippa18} These two mechanisms are, however, non-exclusive manifestations of SOC and do, therefore, co-exist in general. In the linear response regime (where both Larmor and Rabi frequencies are proportional to the magnetic field $B$), they can actually be described in a unified picture by a generalized $g$-matrix formalism.\cite{Crippa18} This $g$-matrix and its derivative with respect to gate voltage $g^\prime$ can be characterized experimentally by Zeeman splitting and Rabi frequency measurements, in order to get a comprehensive picture of the operation of the qubit.

Here we show that the $g$-matrix formalism is also a very efficient framework for the modeling of spin qubits. It provides the complete map of Larmor and Rabi frequencies as a function of the orientation of the magnetic field with only the wave functions of the qubit at zero field as input. It is also well suited to the analysis of the effects of symmetries on the response of the qubit. We discuss, in particular, the impact of mirror planes on the shape of $g$, $g^\prime$, and on the anisotropy of the Rabi frequency map. We also show how the $g$-matrix formalism can be supplemented by the analysis of an equivalent perturbation series that provides additional insights into the physics of the device. The present formalism \RED{can be used to interpret experimental data and model electrical manipulation on a wide variety of electron and hole qubits, and we choose a hole qubit on silicon-on-insulator (SOI) as an illustration.\cite{Maurand16,Crippa18} The Rabi frequency of this qubit displays complex dependences on the orientation of the magnetic field and gate voltages, which can be understood from the symmetry of the wave functions. The calculated $g$-factor anisotropy is, however, different from recent experimental data. We suggest that non-intentional strains might explain this discrepancy, which highlights their importance and opens the way for strain engineering in qubit devices.}

\RED{We introduce the perturbation series for the Rabi frequency and the equivalent $g$-matrix formalism in section \ref{sectionModels}, then discuss the effects of symmetries on the orientational dependence of the Rabi frequency in section \ref{sectionSymmetries}; We finally detail the application to a hole spin qubit on SOI in section \ref{sectionSOI}.}

\section{Models}
\label{sectionModels}

\begin{figure}
\includegraphics[width=0.66\columnwidth]{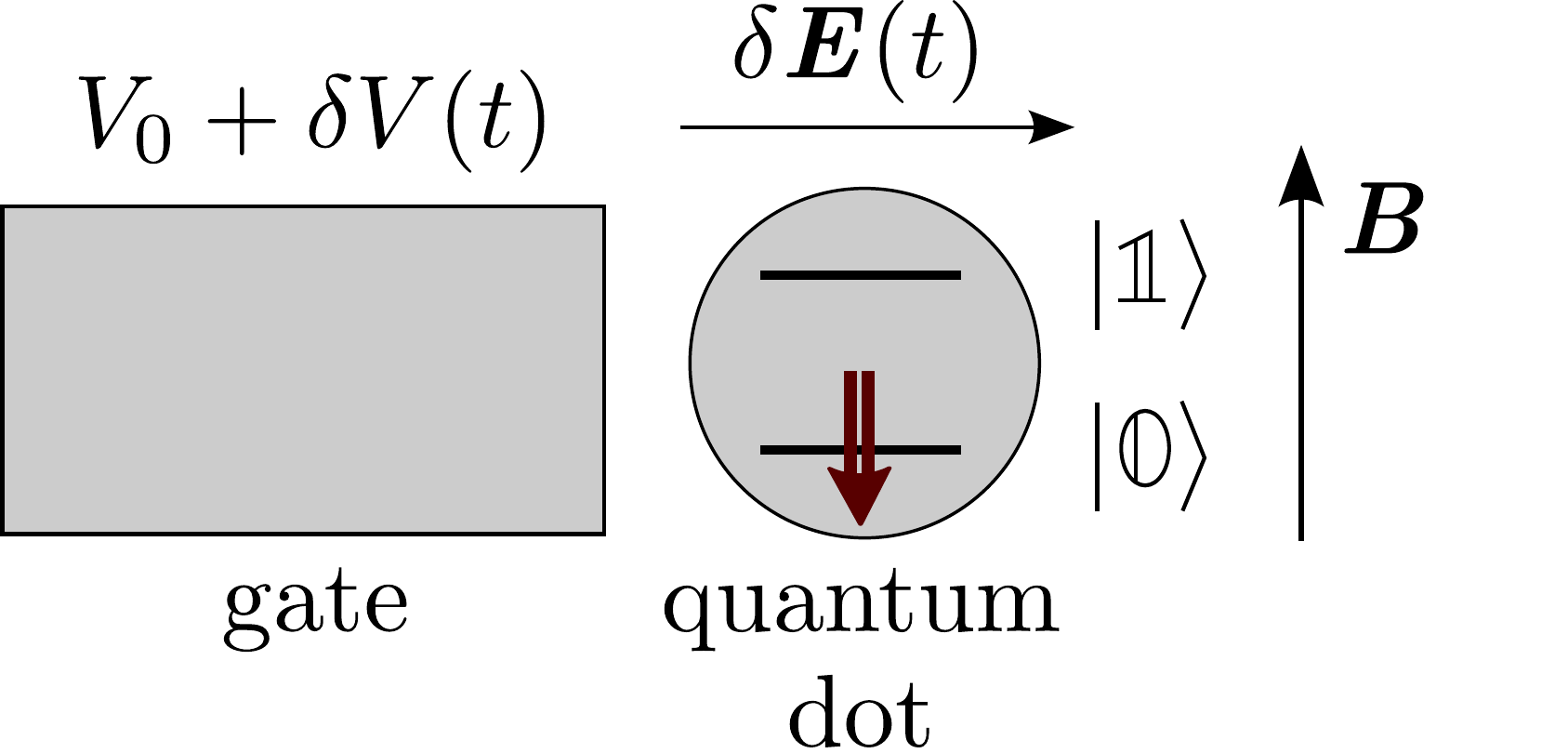}
\caption{\RED{Model of a qubit. A radio-frequency modulation $\delta V(t)$ on a gate drives rotations between the states $\ket{\zero}$ and $\ket{\one}$ of a quantum dot in a static magnetic field $\vec{B}$. $\delta\vec{E}(t)$ is the induced electric field modulation.}}
\label{figModel}
\end{figure}

We consider a quantum dot in a homogeneous, static magnetic field $\vec{B}$ \RED{(see Fig. \ref{figModel})}. The carriers in this dot are controlled by a gate at potential $V$. The system can, therefore, be characterized by a Hamiltonian $H(V,\vec{B})$. For practical purposes, we expand $H(V,\vec{B})$ in powers of $\vec{B}$:
\begin{align}
H(V,\vec{B})&=H_0(V) \nonumber \\
&-B_xM_{1,x}-B_yM_{1,y}-B_zM_{1,z}+{\cal O}(B^2) \nonumber \\
&=H_0(V)-\vec{B}\cdot\vec{M}_1+{\cal O}(B^2)\,,
\label{eqH}
\end{align}
where $M_{1,\alpha}=-\partial H/\partial B_\alpha|_{\vec{B}=\vec{0}}$. We discard all higher order terms as we focus on linear response throughout the paper. We assume that $\vec{M}_1=(M_{1,x}, M_{1,y}, M_{1,z})$ is independent on $V$. For small modulations of $V$ around a reference bias $V_0$, we can also expand $H_0(V)$ in powers of $\delta V=V-V_0$:
\begin{equation}
H_0(V)=H_0(V_0)-e\delta V D_1+{\cal O}(\delta V^2)\,
\label{eqpotential}
\end{equation}
where $D_1(\vec{r})=\partial V_t(V, \vec{r})/\partial V|_{V=V_0}$ is the derivative of the total potential $V_t(V, \vec{r})$ in the device with respect to the gate voltage $V$. 

We label $E_{n,\sigma}$ and $\ket{n,\sigma}$ the eigenenergies and eigenstates of $H$, with $n\ge 0$ and $\sigma\in\{\Uparrow,\Downarrow\}$ a ``pseudo-spin'' index (as spin is not a good quantum number in the presence of SOC). At $\vec{B}=\vec{0}$, $E_{n,\Uparrow}=E_{n,\Downarrow}=E_n$ and $\ket{n,\Uparrow}=T\ket{n,\Downarrow}$ (possibly up to an arbitrary phase), where $T$ is the time-reversal symmetry operator (Kramers degeneracy).

We consider a two-level qubit based on states $\ket{\zero}\equiv\ket{0,\Downarrow}$ and $\ket{\one}\equiv\ket{0,\Uparrow}$.\footnote{The models of section \ref{sectionModels} can however be applied to any pair of Kramers degenerate states.} At finite magnetic field, these two states are split by the Zeeman energy $\Delta E=g^*\mu_B B$, where $\mu_B$ is Bohr's magneton and $g^*$ is the effective gyromagnetic factor (which may depend on the orientation of $\vec{B}$). The qubit is manipulated by a radio-frequency (RF) modulation of the gate voltage $\delta V(t)=V_{\rm ac}\sin(2\pi\nu t+\varphi)$ with amplitude $V_{\rm ac}$, frequency $\nu$ and arbitrary phase $\varphi$. At resonance ($h\nu=\Delta E$), this RF modulation drives coherent oscillations between states $\ket{\zero}$ and $\ket{\one}$, with Rabi frequency:
\begin{equation}
f_R=\frac{e}{h}V_{\rm ac}\left|\bra{\one}D_1\ket{\zero}\right|\,.
\label{eqRabi}
\end{equation}
Note that the above expression for $f_R$ is first order in $V_{\rm ac}$, but all orders in the magnetic field through the dependence of $\ket{\zero}$ and $\ket{\one}$ on $\vec{B}$.

In the following, we first expand Eq. (\ref{eqRabi}) to first order in $\vec{B}$, then discuss the connections with the formalism of Refs. \onlinecite{Crippa18} and \onlinecite{Kato03}.

\subsection{First-order perturbation theory}
\label{subsectionperturbation}

We expand Eq. (\ref{eqRabi}) to first order in $\vec{B}$ using perturbation theory on degenerate states. From now on, we assume that the states $\ket{n,\sigma}$ and the energies $E_{n,\sigma}$ have been computed at $V=V_0$ and $\vec{B}=\vec{0}$. The states $\ket{n,\Uparrow}$ and $\ket{n,\Downarrow}$ being degenerate, they are defined up to an unitary transformation. 

Starting from an arbitrary choice for the degenerate $\ket{0,\sigma}$, the zeroth-order $\ket{\zero_0}$ and $\ket{\one_0}$ states are the eigenstates of the matrix:\footnote{Note that $\bra{0,\Downarrow}\vec{B}\cdot\vec{M}_1\ket{0,\Downarrow}=-\bra{0,\Uparrow}\vec{B}\cdot\vec{M}_1\ket{0,\Uparrow}$ since time-reversal symmetry transforms $\vec{B}$ into $-\vec{B}$.}
\begin{equation}
H_1(\vec{B})=-
\begin{pmatrix}{}
\bra{0,\Uparrow}\vec{B}\cdot\vec{M}_1\ket{0,\Uparrow} & \bra{0,\Uparrow}\vec{B}\cdot\vec{M}_1\ket{0,\Downarrow} \\
\bra{0,\Downarrow}\vec{B}\cdot\vec{M}_1\ket{0,\Uparrow} & \bra{0,\Downarrow}\vec{B}\cdot\vec{M}_1\ket{0,\Downarrow}
\end{pmatrix}\,.
\label{eqMB}
\end{equation}
At this order in perturbation, the Rabi frequency $f_R$ is zero. This is easily evidenced on the spinorial form of $\ket{0,\sigma}$. Assuming $\langle\vec{r}|0,\Downarrow\rangle=u(\vec{r})\ket{\uparrow}+v(\vec{r})\ket{\downarrow}$, where $\ket{\uparrow}$ and $\ket{\downarrow}$ are the physical spin components, $\langle\vec{r}|0,\Uparrow\rangle=\bra{\vec{r}}T\ket{0,\Downarrow}=v^*(\vec{r})\ket{\uparrow}-u^*(\vec{r})\ket{\downarrow}$, so that:
\begin{equation}
\bra{0,\Uparrow}D_1\ket{0,\Downarrow}=\int d^3\vec{r}\,D_1(\vec{r})\left[v(\vec{r})u(\vec{r})-u(\vec{r})v(\vec{r})\right]=0\,.
\label{eqDzero}
\end{equation}
The eigenvectors $\ket{\zero_0}=\alpha\ket{0,\Uparrow}+\beta\ket{0,\Downarrow}$ and $\ket{\one_0}=\beta^*\ket{0,\Uparrow}-\alpha^*\ket{0,\Downarrow}$ of $H_1(\vec{B})$ (with $\alpha$, $\beta$ complex numbers) having the same spinorial form, $\bra{\one_0}D_1\ket{\zero_0}$ is still zero. The electric field can not couple Kramers degenerate states, and time-reversal symmetry must be broken in order to achieve Rabi oscillations.

Time-reversal symmetry can actually be broken by the magnetic field at first (and higher) order in perturbation. To first order in $\vec{B}$,
\begin{subequations}
\begin{align}
\ket{\zero_1}&=\ket{\zero_0}-\sum_{n>0,\sigma}\frac{\bra{n,\sigma}\vec{B}\cdot\vec{M}_1\ket{\zero_0}}{E_0-E_n}\ket{n,\sigma} \\
\ket{\one_1}&=\ket{\one_0}-\sum_{n>0,\sigma}\frac{\bra{n,\sigma}\vec{B}\cdot\vec{M}_1\ket{\one_0}}{E_0-E_n}\ket{n,\sigma}\,,
\end{align}
\end{subequations}
so that, to first order in $\vec{B}$ again,
\begin{align}
f_R&=\frac{e}{h}V_{\rm ac}\left|\bra{\one_1}D_1\ket{\zero_1}\right| \nonumber \\
&=\frac{e}{h}BV_{\rm ac}\left|\sum_{n>0,\sigma}\frac{\bra{\one_0}D_1\ket{n,\sigma}\bra{n,\sigma}\vec{b}\cdot\vec{M}_1\ket{\zero_0}}{E_0-E_n}\right. \nonumber \\
&+\left.\sum_{n>0,\sigma}\frac{\bra{\one_0}\vec{b}\cdot\vec{M}_1\ket{n,\sigma}\bra{n,\sigma}D_1\ket{\zero_0}}{E_0-E_n}\right|\,,
\label{eqRabipert}
\end{align}
where $\vec{b}=\vec{B}/B$ is the unit vector pointing along the magnetic field.

The above equation provides a decomposition and interpretation of the Rabi frequency in terms of the energy level structure of the ``unperturbed'' dot at zero magnetic field. Namely, the magnetic field admixes excited states $\ket{n,\sigma^\prime}$ into the ground-state (qubit) levels $\ket{0,\sigma}$; as these states may be coupled by finite dipole matrix elements $\bra{n,\sigma^\prime}D_1\ket{0,\sigma}$, this allows for electrically driven oscillations between $\ket{\zero_0}$ and $\ket{\one_0}$. 

We will discuss the nature of the excitations involved in a \RED{specific example (hole qubit on SOI) in section \ref{sectionSOI}}. We next discuss the relations between Eq. (\ref{eqRabipert}) and the $g$-matrix formalism of Ref. \onlinecite{Crippa18}.

\subsection{The $g$-matrix formalism}

To first order in the magnetic field, the electronic structure of the qubit can be described by the following, generic two-level Hamiltonian:\cite{Weil-Bolton,Crippa18}
\begin{equation}
H(V,\vec{B})=\frac{1}{2}\mu_B\vec{\sigma}\cdot\hat{g}(V)\vec{B}\,,
\label{eqgtensor}
\end{equation}
where $\vec{\sigma}=(\sigma_1,\sigma_2,\sigma_3)$ is the ``vector'' of Pauli matrices, and $\hat{g}$ is a real $3\times 3$ matrix. In the following, the symbol $\cdot$ is used to denote the bilinear map $\vec{u}\cdot\vec{v}=u_1v_1+u_2v_2+u_3v_3$ irrespective of the nature of the three components $u_i$ and $v_i$ of $\vec{u}$ and $\vec{v}$.\footnote{We add a hat on top of all real $3\times3$ matrices, in order to distinguish these matrices from those of the operators, such as the Pauli matrices, acting in the Hilbert space of the qubit. The notations are slightly different from Ref. \onlinecite{Crippa18}, where $\cdot$ was used to denote real $3\times 3$ matrix/matrix and matrix/vector multiplications.} Any linear-in-$B$-and-$V_{\rm ac}$ Hamiltonian can be mapped onto Eq. (\ref{eqgtensor}) up to an irrelevant diagonal energy shift. 

For a given $V$, the first-order energies and zeroth-order states can indeed be obtained from the diagonalization of $H_1(\vec{B})$. Identification of Eqs. (\ref{eqMB}) and (\ref{eqgtensor}) hence provides an explicit expression for the $g$-matrix:
\begin{align}
&\hat{g}(V_0)=-\frac{2}{\mu_B}\times \nonumber \\
&\begin{pmatrix}{}
{\rm Re}\bra{\Downarrow}M_{1,x}\ket{\Uparrow} & {\rm Re}\bra{\Downarrow}M_{1,y}\ket{\Uparrow} & {\rm Re}\bra{\Downarrow}M_{1,z}\ket{\Uparrow} \\
{\rm Im}\bra{\Downarrow}M_{1,x}\ket{\Uparrow} & {\rm Im}\bra{\Downarrow}M_{1,y}\ket{\Uparrow} & {\rm Im}\bra{\Downarrow}M_{1,z}\ket{\Uparrow} \\
\bra{\Uparrow}M_{1,x}\ket{\Uparrow} & \bra{\Uparrow}M_{1,y}\ket{\Uparrow} & \bra{\Uparrow}M_{1,z}\ket{\Uparrow} \\
\end{pmatrix}\,,
\label{eqgmatrix}
\end{align}
where $\ket{\Uparrow}$ is a shorthand notation for $\ket{0,\Uparrow}$ and $\ket{\Downarrow}$ is a shorthand notation for $\ket{0,\Downarrow}$. Note that the $g$-matrix depends on the choice of axes $\vec{x}$, $\vec{y}$, $\vec{z}$ for the magnetic field and on the choice of basis set $\{\ket{0,\Uparrow},\ket{0,\Downarrow}\}$ for the Hilbert space of the qubit. It can be brought to a symmetric or even diagonal form (the $g$-tensor) with an appropriate choice of $\vec{x}$, $\vec{y}$, $\vec{z}$ and $\{\ket{0,\Uparrow},\ket{0,\Downarrow}\}$ (see later discussion). 

Eq. (\ref{eqgtensor}) can also be written:
\begin{equation}
H=\frac{1}{2}g^*\mu_B B(\vec{u}\cdot{\vec{\sigma}})=\frac{1}{2}g^*\mu_B B \sigma_{\vec{u}}
\end{equation}
where $\vec{u}=\hat{g}\vec{b}/|\hat{g}\vec{b}|$ is a unit vector and:
\begin{equation}
g^*=|\hat{g}\vec{b}|
\label{eqgstar}
\end{equation}
is the effective gyromagnetic factor. Indeed, the eigenenergies of $H$ are $E=\pm\frac{1}{2}g^*\mu_B B$, so that the Zeeman splitting is $\Delta E=g^*\mu_B B$.

The dependence of the $\hat{g}$ matrix on $V$ is embedded in the $\ket{0,\sigma}$ states. In order to discuss Rabi oscillations in this framework, we expand $\hat{g}(V)$ to first order in $\delta V$ around the reference bias $V=V_0$:
\begin{equation}
\hat{g}(V)=\hat{g}(V_0)+\delta V\hat{g}^\prime(V_0)\,,
\label{eqgprime}
\end{equation}
where $\hat{g}^\prime=\partial\hat{g}/\partial V$ may be expressed from the $\partial\ket{0,\sigma}/\partial V$'s (see Appendix \ref{appendixEquivalence}). We then introduce the Larmor vector $\hbar\vec{\Omega}=\frac{1}{2}\mu_B\hat{g}(V_0)\vec{B}$ and its derivative $\hbar\vec{\Omega}^\prime=\frac{1}{2}\mu_B\hat{g}^\prime(V_0)\vec{B}$, so that:
\begin{equation}
H(V_0+\delta V,\vec{B})=\hbar|\vec{\Omega}|\sigma_{\vec{\omega}}+\hbar|\vec{\Omega}^\prime|\delta V\sigma_{\vec{\omega}^\prime}\,,
\end{equation}
with $\vec{\omega}=\vec{\Omega}/|\vec{\Omega}|$ and $\vec{\omega}^\prime=\vec{\Omega}^\prime/|\vec{\Omega}^\prime|$. Splitting $\vec{\Omega}^\prime=\vec{\Omega}^\prime_\parallel+\vec{\Omega}^\prime_\perp$ into components parallel and perpendicular to $\vec{\Omega}$,
\begin{equation}
H(V_0+\delta V,\vec{B})=\hbar|\vec{\Omega}+\vec{\Omega}^\prime_\parallel\delta V|\sigma_{\vec{\omega}}+\hbar|\vec{\Omega}^\prime_\perp|\delta V\sigma_{\vec{\omega}^\prime_\perp}\,. \\
\end{equation}
$\vec{\Omega}^\prime_\parallel$ characterizes the gate-driven modulations of the Larmor (precession) frequency, while $\vec{\Omega}^\prime_\perp$ characterizes the gate-driven modulations of the precession axis. For a RF signal $\delta V(t)=V_{\rm ac}\sin(|\vec{\Omega}|t+\varphi)$ resonant with the transition between the eigenstates of $H(V_0,\vec{B})$, the Rabi frequency $f_R$ hence reads:\cite{Kato03,Ares13}
\begin{align}
f_R&=\frac{1}{2\pi}|\vec{\Omega}^\prime_\perp|V_{\rm ac} \nonumber \\
&=\frac{1}{2\pi}|\vec{\omega}\times\vec{\Omega}^\prime|V_{\rm ac} \nonumber \\
&=\frac{\mu_B B V_{\rm ac}}{2hg^*}\Big|[\hat{g}(V_0)\vec{b}]\times[\hat{g}^\prime(V_0)\vec{b}]\Big|\,.
\label{eqRabig}
\end{align}
The matrices $\hat{g}(V_0)$ and $\hat{g}^\prime(V_0)$ therefore provide a comprehensive picture of the Larmor and Rabi frequencies of the qubit for arbitrary orientations of the magnetic field. Since Eqs. (\ref{eqRabipert}) and (\ref{eqRabig}) are both valid to first order in $\vec{B}$ and $V_{\rm ac}$, they must be equivalent. This is actually shown in Appendix \ref{appendixEquivalence}. 

To conclude this paragraph, we would like to discuss the relations between the above $g$-matrix and the so-called $g$-tensor. We remind that the $g$-matrix can always be factorized as:\cite{Chibotaru08,Weil-Bolton}
\begin{equation}
\hat{g}=\hat{U}\hat{g}_d{^t}\hat{V}\,,
\label{eqfactorizationg}
\end{equation}
where $\hat{U}$ and $\hat{V}$ are unitary $3\times 3$ matrices with unit determinant and $\hat{g}_d={\rm diag}(g_1, g_2, g_3)$ is a diagonal matrix (singular value decomposition). $g_1$, $g_2$, $g_3$ are the principal $g$-factors; the columns of $\hat{V}$ define the principal magnetic axes $\{\vec{X},\vec{Y},\vec{Z}\}$, while the columns of $\hat{U}$ define three new pseudo-spin matrices $\vec{\sigma}^\prime={^t}\hat{U}\vec{\sigma}$. Yet there always exists\footnote{Let $R$ be an unitary transformation in the $\{\ket{0,\Uparrow},\ket{0,\Downarrow}\}$ subspace. $R$ can be cast in the form
\begin{equation*}
R=\begin{pmatrix}{}
\alpha e^{i\theta} & -\beta^* \\
\beta e^{i\theta} & \alpha^* 
\end{pmatrix}
\end{equation*}
with $|\alpha|^2+|\beta|^2=1$. In the basis set $\{\ket{0,\Uparrow^\prime},\ket{0,\Downarrow^\prime}\}=\{R\ket{0,\Uparrow},R\ket{0,\Downarrow}\}$, the Hamiltonian reads $H^\prime=\frac{1}{2}\mu_B\vec{\sigma}^\prime\cdot\hat{g}\vec{B}$, with $\sigma^\prime_i=R^\dag\sigma_i R$. Yet $\vec{\sigma}^\prime=\hat{U}\vec{\sigma}$, with:
\begin{equation*}
\hat{U}=\begin{pmatrix}{}
{\rm Re}[(\alpha^2-\beta^2)e^{i\theta}] & {\rm Im}[(\alpha^2-\beta^2)e^{i\theta}] & 2{\rm Re}(\alpha^*\beta) \\
-{\rm Im}[(\alpha^2+\beta^2)e^{i\theta}] & {\rm Re}[(\alpha^2+\beta^2)e^{i\theta}] & 2{\rm Im}(\alpha^*\beta) \\
-2{\rm Re}(\alpha\beta e^{i\theta}) & -2{\rm Im}(\alpha\beta e^{i\theta}) & |\alpha|^2-|\beta|^2
\end{pmatrix}\,.
\end{equation*}
Hence $H^\prime=\frac{1}{2}\mu_B\vec{\sigma}\cdot\hat{g}^\prime\vec{B}$, with $\hat{g}^\prime={^t}\hat{U}\hat{g}$. The $\hat{U}$ matrix is unitary with determinant $+1$. Therefore, any rotation of the $\{\ket{0,\Uparrow},\ket{0,\Downarrow}\}$ basis set results in a corresponding rotation of the $g$-matrix. Conversely, any unitary $3\times3$ matrix $\hat{U}$ with determinant $+1$ can be mapped onto the above equation, and associated with a unitary transformation $R$ in the $\{\ket{0,\Uparrow},\ket{0,\Downarrow}\}$ subspace.} 
a unitary transformation $R$ in the Hilbert space of the qubit such that $\vec{\sigma}^\prime=(R\sigma_1R^\dagger,R\sigma_2R^\dagger,R\sigma_3R^\dagger)$. Therefore, $\hat{U}$ defines a basis set $\{R\ket{0,\Uparrow},R\ket{0,\Downarrow}\}$ of the Hilbert space, and $\hat{V}$ defines a frame for the magnetic field in which the $g$-matrix is diagonal. In this basis set, the Pauli matrices $\sigma_1\equiv\sigma_X$, $\sigma_2\equiv\sigma_Y$ and $\sigma_3\equiv\sigma_Z$ can be identified as the effective pseudo-spin operators along the  principal magnetic axes. The $g$-matrix then remains symmetric upon concomitant rotations of the magnetic axes and pseudo-spin quantization axes ($\vec{B}\to\hat{A}\vec{B}$, $\vec{\sigma}\to\hat{A}\vec{\sigma}$, and $\hat{g}\to\hat{A}\hat{g}{^t}\hat{A}$, with $\hat{A}$ unitary), and is then known as the $g$-tensor. Yet it is important to realize that $\hat{g}^\prime(V_0)$ might be non-symmetric even if $\hat{g}(V_0)$ is (because the matrices $\hat{U}$ and $\hat{V}$ that diagonalize $\hat{g}$ can depend on $V$). We also emphasize that $f_R$ is independent on the choice of gauge for $\vec{M}_1$ and on the choice of magnetic axes $\vec{x}$, $\vec{y}$, $\vec{z}$ and basis set $\{\ket{0,\Uparrow},\ket{0,\Downarrow}\}$,\footnote{As discussed in the previous note, any change of basis set $\{\ket{0,\Uparrow},\ket{0,\Downarrow}\}$, characterized by a complex, unitary $2\times2$ matrix $R$, results in a transformation $\hat{g}\to{^t}\hat{U}\hat{g}$ of the $\hat{g}$ matrix, where $\hat{U}$ is a real, unitary $3\times 3$ matrix (and likewise for $\hat{g}^\prime$). Since $|[{^t}\hat{U}\hat{g}(V_0)\vec{b}]\times[{^t}\hat{U}\hat{g}^\prime(V_0)\vec{b}]|=|[\hat{g}(V_0)\vec{b}]\times[\hat{g}^\prime(V_0)\vec{b}]|$, $f_R$ is invariant under that transformation. Also, a change of gauge for the vector potential results in an unitary transform on $\vec{M}_1$, hence equivalently in a rotation of the basis set $\{\ket{0,\Uparrow},\ket{0,\Downarrow}\}$, which, as discussed above, leaves $f_R$ invariant.} although $\hat{g}(V_0)$ and $\hat{g}^\prime(V_0)$ do depend on these choices.\cite{Crippa18}

\subsection{Discussion}
\label{subsectionDiscussion}

We would first like to emphasize that spin-orbit coupling, although not explicit in the above equations, is the driving force for the Rabi oscillations. Indeed, in the absence of a mechanism coupling the real space motion of the carrier with its spin, $H(V,\vec{B})$ is diagonal in spin whatever $V$ and $\vec{B}$, so that the gate can not couple opposite spins (in particular, $\hat{g}=g_0\hat{I}$ and $\hat{g}^\prime=0$ for all $V$, where $\hat{I}$ is the $3\times 3$ identity matrix and $g_0=2.0023$ is the bare gyromagnetic factor of an electron). The spin-orbit coupling can be ``intrinsic'' to the materials or ``extrinsic'' (resulting from inhomogeneous magnetic fields\cite{Pioro-Ladriere08,Kawakami14}). The above $g$-matrix formalism primarily applies to intrinsic spin-orbit coupling, but may be extended to extrinsic spin-orbit coupling taking care of the additional contribution from the inhomogeneous magnetic field (which may give rise to a Zeeman splitting at $\vec{B}=\vec{0}$) as an effective field $\vec{B}_{\rm in}$. \RED{The $g$-matrix formalism may also be adapted to the description of singlet-triplet qubits.\cite{Jock18}}

As hinted before, the perturbation series, Eq. (\ref{eqRabipert}), provides a meaningful interpretation of the Rabi frequency in terms of the matrix elements of the electric and magnetic fields between the qubit and excited states of the system. It can help, for example, to identify the states at $\vec{B}=\vec{0}$ that form the minimal basis capturing the physics of the device at finite $\vec{B}$. It is not, however, a particularly ``compact'' model for the control of the qubit, as it may require many matrix elements as input.

\RED{Equations (\ref{eqgstar}) and (\ref{eqRabig})} provide such a minimal model for the control of the qubit, characterized by the $2\times9$ elements of $\hat{g}$ and $\hat{g}^\prime$. The $g$-matrix and its derivative do, however, give much less insights into the physics of the device at the microscopic scale.

From a modeling perspective, the $g$-matrix formalism provides a very efficient way to compute the maps of Larmor and Rabi frequencies as a function of the orientation of the magnetic field. Indeed, the $g$-matrix $\hat{g}(V_0)$ can be readily calculated from the wave functions at zero magnetic field using Eq. (\ref{eqgmatrix}). Its derivative $\hat{g}^\prime(V_0)$ can then be obtained from finite differences at points $V=V_0$ and $V=V_0\pm\delta V$. Care must however be taken to align the basis set $\ket{0,\sigma}(V_0\pm\delta V)$ onto the basis set $\ket{0,\sigma}(V_0)$ (see Appendix \ref{appendixCalculation} for details). The map of Rabi frequency can, therefore, be constructed from three electronic structure calculations at zero field, one at $V=V_0$ and two at $V=V_0\pm\delta V$ -- whereas the use of Eq. (\ref{eqRabi}) calls for a specific electronic structure calculation for each different magnitude and orientation of $\vec{B}$. The validity of the linear-in-$B$-and-$V_{\rm ac}$ approximation can (and should) be assessed by a comparison with Eq. (\ref{eqRabi}) for a few magnetic fields.

We would finally like to stress that the experimental characterization of $\hat{g}^\prime$ \RED{can be complex as some transformations of $\hat{g}$ do not result in variations of the Zeeman splitting $\Delta E$.\cite{Crippa18} Indeed, the $g$-matrix or tensor is usually reconstructed from the measurement of:}
\begin{equation}
\Delta E^2=\mu_B^2|\hat{g}\vec{B}|^2=\mu_B^2\vec{B}\cdot\hat{G}\vec{B}\,,
\end{equation}
where $\hat{G}={^t}\hat{g}\hat{g}$ is the symmetric Zeeman tensor. The eigenvalues of $\hat{G}$ are the square of the principal $g$-factors $g_1^2$, $g_2^2$, $g_3^2$, and its eigenvectors are the principal magnetic axes $\vec{X}$, $\vec{Y}$, $\vec{Z}$ [as $\hat{G}=\hat{V}\hat{g}_d^2{^t}\hat{V}$ following Eq. (\ref{eqfactorizationg})]. The iso-surfaces of $\Delta E^2$ are, therefore, ellipsoids in the principal magnetic axes frame: 
\begin{equation}
\Delta E^2(B_1\vec{X}+B_2\vec{Y}+B_3\vec{Z})=\mu_B^2\left(g_1^2B_1^2+g_2^2B_2^2+g_3^2B_3^2\right)\,.
\end{equation}
The six independent elements of the symmetric Zeeman tensor $\hat{G}$ can hence be assessed from the measurement of the Zeeman splitting for at least six orientations of $\vec{B}$. However, $\hat{g}$ is defined by $\hat{G}$ only up to a unitary transformation [because $\hat{G}={^t}\hat{g}\hat{g}={^t}(\hat{A}\hat{g})(\hat{A}\hat{g})$ for any unitary matrix $\hat{A}$]. This results from the fact that a rotation of the basis set $\{\ket{0,\Uparrow},\ket{0,\Downarrow}\}$ leaves the Zeeman splitting (but not the $\hat{g}$-matrix) invariant. The basis set where $\hat{g}=\hat{g}_d$ is diagonal remains, therefore, unknown. To highlight the implications, we may write the expressions of $\hat{g}^\prime$ and $\hat{G}^\prime$ in this (yet implicit) basis set and in the principal magnetic axes at $V=V_0$. Then $\hat{U}=\hat{V}=\hat{I}$,
${^t}\hat{V}^\prime=-\hat{V}^\prime$ since ${^t}\hat{V}\hat{V}=\hat{I}$, ${^t}\hat{U}^\prime=-\hat{U}^\prime$, so that:
\begin{equation}
\hat{G}^\prime=\hat{V}^\prime\hat{g}_d^2-\hat{g}_d^2\hat{V}^\prime+2\hat{g}_d\hat{g}_d^\prime
\end{equation}
and:
\begin{align}
\hat{g}^\prime&=\hat{U}^\prime\hat{g}_d-\hat{g}_d\hat{V}^\prime+\hat{g}_d^\prime
\nonumber \\
&=\frac{1}{2}\hat{g}_d^{-1}\hat{G}^\prime-\frac{1}{2}\hat{g}_d^{-1}\left(\hat{g}_d^2\hat{V}^\prime+\hat{V}^\prime\hat{g}_d^2\right)+\hat{U}^\prime\hat{g}_d\,.
\label{eqgder}
\end{align}
It is clear, therefore, that the electrical modulations of $\hat{U}$ contribute to $\hat{g}^\prime$, but not to $\hat{G}^\prime$. Likewise, the effects of the electrical modulations of $\hat{V}$ are not completely captured by $\hat{G}^\prime$. When the system is sufficiently symmetric so that only the principal $g$-factors are modulated by the electric field, $\hat{g}^\prime=\hat{g}_d^\prime=\frac{1}{2}\hat{g}_d^{-1}\hat{G}^\prime$ can be completely reconstructed from the measurement of the Zeeman tensor ($\hat{G}$) and of its dependence on gate voltage ($\hat{G}^\prime$). This is the conventional $g$-TMR scenario introduced in Ref. \onlinecite{Kato03}. However, $\hat{g}^\prime$ can not, in general, be reconstructed from $\hat{G}$ and $\hat{G}^\prime$ once the principal magnetic axes ($\hat{V}$) and diagonal basis set ($\hat{U}$) depend on gate voltage. We may consequently split $\hat{g}^\prime=\hat{g}_{\rm TMR}^\prime+\hat{g}_{\rm IZR}^\prime$ in two parts, where $\hat{g}_{\rm TMR}^\prime=\hat{g}_d^{-1}\hat{G}^\prime/2$ is a generalized $g$-TMR matrix that can be drawn from the variations of the Zeeman tensor, and $\hat{g}_{\rm IZR}^\prime=-\hat{g}_d^{-1}(\hat{g}_d^2\hat{V}^\prime+\hat{V}^\prime\hat{g}_d^2)/2+\hat{U}^\prime\hat{g}_d$ is an ``iso-Zeeman'' EDSR (IZ-EDSR) matrix that leaves no fingerprints on the electrical dependence of the Zeeman splittings. The latter must, therefore, be reconstructed from the measurement of the Rabi frequency for a few orientations of the magnetic field. As a matter of fact, $\hat{g}_{\rm IZR}$ can be factorized as $\hat{g}_{\rm IZR}^\prime=\hat{g}_d^{-1}\hat{A}$ where $\hat{A}$ is an anti-symmetric matrix, which leaves only three degrees of freedom (out of 9) in $\hat{g}^\prime$ that can not be inferred from the measurement of the Zeeman tensor (see Ref. \onlinecite{Crippa18} for details and an application).\footnote{In principle, both $\hat{g}_d$ and $\hat{V}$ (hence their derivatives) can be completely characterized as a function of gate voltage by the measurement and diagonalization of the symmetric Zeeman tensor $\hat{G}$. We may, therefore, adopt a slightly more general (and physical) definition of $\hat{g}_{\rm TMR}^\prime=\hat{g}_d^\prime-\hat{g}_d\hat{V}^\prime$ as the contribution from the electrical modulations $\hat{g}_d^\prime$ and $\hat{V}^\prime$, and of $\hat{g}_{\rm IZR}^\prime=\hat{U}^\prime\hat{g}_d$ as the contribution from the electrical modulations $\hat{U}^\prime$. However, from an experimental point of view, it may be extremely difficult to reconstruct $\hat{V}^\prime$ from the dependence of the eigenvectors of $\hat{G}$ on gate voltage. Indeed, as part of the modulations of $\hat{V}$ [second and third terms of Eq. (\ref{eqgder})] only make a second or higher-order contribution to $\hat{G}$, the measurement of $\hat{G}$ must be extremely accurate in order to completely capture the dependence of $\hat{V}$ on gate voltage.} 

\section{Effects of symmetries on the $g$-matrix and Rabi frequency}
\label{sectionSymmetries}

\begin{figure}
\includegraphics[width=0.75\columnwidth]{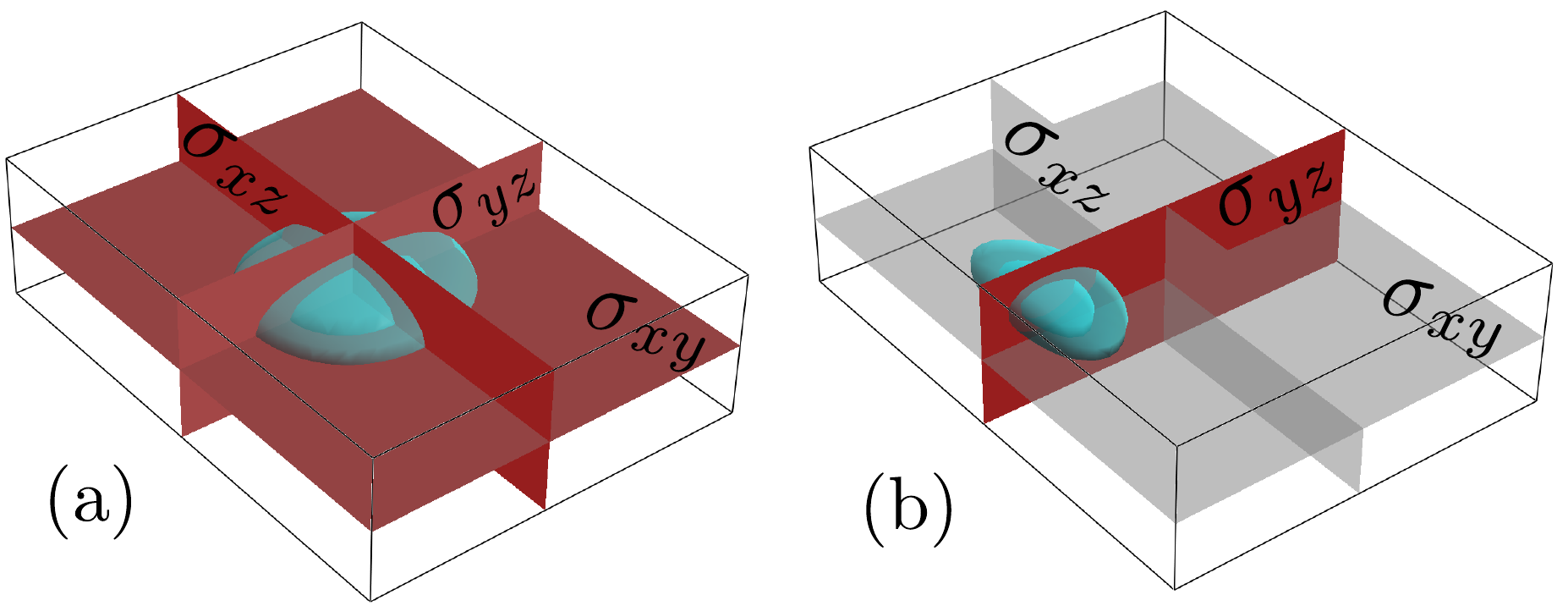}
\caption{\RED{Illustration of the symmetries considered in this work on a particle-in-a-box model. The iso-surfaces of the squared wave function of a hypothetical qubit are plotted in blue. (a) The box (and squared wave function) have three mirror planes, $\sigma_{yz}$, $\sigma_{xz}$ and $\sigma_{xy}$ (in red) ; (b) The application of a static electric field in the $(yz)$ plane however breaks the $\sigma_{xz}$ and $\sigma_{xy}$ planes.}}
\label{figSymmetry}
\end{figure}

In this section, we discuss the impact of symmetries on the shape of $\hat{g}$ and $\hat{g}^\prime$, and on the anisotropy of the Rabi frequency. \RED{Such symmetry considerations can help the analysis of both experimental and computational data.}

We focus on the effects of mirror planes as their enumeration completely explains the trends shown in recent experiments\cite{Crippa18} and in \RED{section \ref{sectionSOI}}. For that purpose, we introduce the following hierarchy of symmetries at zero magnetic field:
\begin{enumerate}
 \item No particular symmetry.
 \item One mirror plane $\sigma_{yz}$ perpendicular to $x$ (double group $C_s$).
 \item Two mirror planes, $\sigma_{yz}$ and $\sigma_{xz}$. Such a system must also have a twofold rotation axis around $z$ (double group $C_{2v}$).
 \item Three mirror planes, $\sigma_{yz}$, $\sigma_{xz}$ and $\sigma_{xy}$. Such a system must also have twofold rotation axes around $x$, $y$ and $z$, and, most notably, an inversion center (double group $D_{2h}$). 
\end{enumerate}
Note that a device may switch (at least approximately, see later discussion) from one group to an other as a function of gate voltage \RED{(see Fig. \ref{figSymmetry})}.

\subsection{Shape of $\hat{g}$}

Each mirror plane sets specific constraints on the shape of $\hat{g}(V_0)$. The arguments, drawn from group theory,\cite{Cornwell97} are detailed in Appendix \ref{appendixGroupTheory}. We outline the main conclusions here.

From now on, we write down the $g$-matrix and its derivative in the magnetic axes $\{\vec{x}, \vec{y}, \vec{z}\}$. In all above symmetry groups, there exists a basis set $\{\ket{0,\Uparrow},\ket{0,\Downarrow}\}$ for the Kramers doublet in which the mirror planes quench the elements of $\hat{g}(V_0)$ listed in Table \ref{table_r}. In this basis set, the $g$-matrix takes therefore the form given in \ref{table_g}.

\begin{table}
\begin{tabular}{|c|c|c|c|} \hline
$\sigma_{\alpha\beta}$ & $\sigma_{yz}$ & $\sigma_{xz}$ & $\sigma_{xy}$ \\ \hline
$\hat{g}$ &  
$\begin{pmatrix}
\bullet & 0 & 0 \\
0 & \bullet & \bullet \\
0 & \bullet & \bullet
\end{pmatrix}$ & 
$\begin{pmatrix}
\bullet & 0 & \bullet \\
0 & \bullet & 0 \\
\bullet & 0 & \bullet
\end{pmatrix}$ &
$\begin{pmatrix}
\bullet & \bullet & 0 \\
\bullet & \bullet & 0 \\
0 & 0 & \bullet
\end{pmatrix}$
\\ \hline
\end{tabular}
\caption{Constraints on the shape of $\hat{g}$ set by each mirror plane $\sigma_{\alpha\beta}$. Black dots are non-zero matrix elements.}
\label{table_r}
\end{table}

\begin{table}
\begin{tabular}{|c|c|c|c|} \hline
group & $C_s$ & $C_{2v}$ & $D_{2h}$ \\ \hline
$\hat{g}$ &  
$\begin{pmatrix}
\bullet & 0 & 0 \\
0 & \bullet & \bullet \\
0 & \bullet & \bullet
\end{pmatrix}$ &
$\begin{pmatrix}
\bullet & 0 & 0 \\
0 & \bullet & 0 \\
0 & 0 & \bullet
\end{pmatrix}$ &
$\begin{pmatrix}
\bullet & 0 & 0 \\
0 & \bullet & 0\\
0 & 0 & \bullet
\end{pmatrix}$ \\ \hline
\end{tabular}
\caption{Shape of the $g$-matrix in each symmetry group. Black dots are non-zero matrix elements.}
\label{table_g}
\end{table}

The main message from Tables \ref{table_r} and \ref{table_g} is that there is a principal magnetic axis perpendicular to each mirror plane. When there is a single mirror plane $\sigma_{yz}$, $\vec{x}$ is a principal magnetic axis, the two others being orthogonal (but {\it a priori} arbitrary) directions of the $(yz)$ plane. Any additional mirror plane locks all magnetic axes onto $\{\vec{x},\vec{y},\vec{z}\}$.

The analysis of the shape of $\hat{g}^\prime(V_0)$ in the same basis set unveils the anisotropy of the Rabi frequency in each group.

\subsection{Shape of $\hat{g}^\prime$}

Sweeping the gate voltage can break the symmetries of the system. We can, therefore, distinguish at least three different behaviors for a given mirror plane $\sigma_{\alpha\beta}$:
\begin{enumerate}
 \item Changing the gate voltage does not break $\sigma_{\alpha\beta}$ -- At least, the first-order variation of the potential in the device, $D_1(\vec{r})$, is invariant under that transformation [$D_1(\sigma_{\alpha\beta}(\vec{r}))=D_1(\vec{r})$]. Equivalently, the first-order electric field $\vec{E}_1(\vec{r})=-\vec{\nabla}D_1(\vec{r})$ is even under $\sigma_{\alpha\beta}$:
 \begin{equation}
 \vec{E}_1(\sigma_{\alpha\beta}(\vec{r}))=\sigma_{\alpha\beta}(\vec{E}_1(\vec{r}))\,.
 \end{equation}
 In that case, $\sigma_{\alpha\beta}$ sets the same constraints on $\hat{g}^\prime(V_0)$ as on $\hat{g}(V_0)$ (Tables \ref{table_r} and first row of Table \ref{table_r2}). This behavior is for example encountered when the first-order electric field $\vec{E}_1$ is homogeneous in the $(\alpha\beta)$ plane.
 \item Changing the gate voltage does break the mirror plane $\sigma_{\alpha\beta}$ ($D_1(\vec{r})$ is not invariant), but $\vec{E}_1$ is odd under $\sigma_{\alpha\beta}$:
 \begin{equation}
 \vec{E}_1(\sigma_{\alpha\beta}(\vec{r}))=-\sigma_{\alpha\beta}(\vec{E}_1(\vec{r}))\,.
 \end{equation}
 Equivalently, $D_1(\vec{r})$ is ``odd'' under that transformation up to a constant $K$ [$D_1(\sigma_{\alpha\beta}(\vec{r}))+D_1(\vec{r})=K$, as homogeneous shifts of the potential in the device have no effects on the spin]. Then $\sigma_{\alpha\beta}$ sets the constraints on $\hat{g}^\prime(V_0)$ listed in the second row of Table \ref{table_r2}. This behavior is for example encountered when the first-order electric field $\vec{E}_1$ is homogeneous and perpendicular to the $(\alpha\beta)$ plane.
 \item Changing the gate voltage does break the mirror plane $\sigma_{\alpha\beta}$, but $\vec{E}_1(\vec{r})$ does not show any relevant parity under that transformation. Then $\sigma_{\alpha\beta}$ does not, in general, set any condition on $\hat{g}^\prime(V_0)$ (third row of Table \ref{table_r2}).
\end{enumerate}
The actual form of $\hat{g}^\prime(V_0)$ is given by the intersection of the non-zero positions in Table \ref{table_r2}, for all operations of the symmetry group at $V=V_0$. Different operations might have different parities.

\begin{table}
\begin{tabular}{|c|c|c|c|} \hline
 $\sigma_{\alpha\beta}$ & $\sigma_{yz}$ & $\sigma_{xz}$ & $\sigma_{xy}$ \\ \hline
$\vec{E}_1$ even &  
$\begin{pmatrix}
\bullet & 0 & 0 \\
0 & \bullet & \bullet \\
0 & \bullet & \bullet
\end{pmatrix}$ &
$\begin{pmatrix}
\bullet & 0 & \bullet \\
0 & \bullet & 0 \\
\bullet & 0 & \bullet
\end{pmatrix}$ & 
$\begin{pmatrix}
\bullet & \bullet & 0 \\
\bullet & \bullet & 0 \\
0 & 0 & \bullet
\end{pmatrix}$ \\ \hline
$\vec{E}_1$ odd & 
$\begin{pmatrix}
0 & \bullet & \bullet \\
\bullet & 0  & 0 \\
\bullet & 0 & 0
\end{pmatrix}$ & 
$\begin{pmatrix}
0 & \bullet & 0 \\
\bullet & 0 & \bullet \\
0 & \bullet & 0
\end{pmatrix}$ & 
$\begin{pmatrix}
0 & 0 & \bullet \\
0 & 0 & \bullet \\
\bullet & \bullet & 0
\end{pmatrix}$ \\ \hline
Other & 
$\begin{pmatrix}
\bullet & \bullet & \bullet \\
\bullet & \bullet & \bullet \\
\bullet & \bullet & \bullet
\end{pmatrix}$ & 
$\begin{pmatrix}
\bullet & \bullet & \bullet \\
\bullet & \bullet & \bullet \\
\bullet & \bullet & \bullet
\end{pmatrix}$ & 
$\begin{pmatrix}
\bullet & \bullet & \bullet \\
\bullet & \bullet & \bullet \\
\bullet & \bullet & \bullet
\end{pmatrix}$\\ \hline
\end{tabular}
\caption{Constraints on the shape of $\hat{g}^\prime$ set by each mirror plane $\sigma_{\alpha\beta}$, depending whether the first-order electric field, $\vec{E}_1(\vec{r})=-\vec{\nabla}D_1(\vec{r})$, is even [$\vec{E}_1(\sigma_{\alpha\beta}(\vec{r}))=\sigma_{\alpha\beta}(\vec{E}_1(\vec{r}))$], odd [$\vec{E}_1(\sigma_{\alpha\beta}(\vec{r}))=-\sigma_{\alpha\beta}(\vec{E}_1(\vec{r}))$], or does not show any relevant parity under that transformation. Black dots are non-zero matrix elements.}
\label{table_r2}
\end{table}

\subsection{Anisotropy of the Rabi frequency}

Tables \ref{table_g} and \ref{table_r2} can be used to analyze the anisotropy of the Rabi frequency, by substituting the form of $\hat{g}$ and $\hat{g}^\prime$ into Eq. (\ref{eqRabig}). As an illustration, we discuss below two cases relevant \RED{for section \ref{sectionSOI} (and depicted in Fig. \ref{figSymmetry})}.

Let us first consider a device with a single mirror plane $\sigma_{yz}$ at bias $V=V_0$ (double group $C_s$). Sweeping the gate voltage around that point does not break the mirror plane. Then, according to Tables \ref{table_g} and \ref{table_r2}, in a suitable basis $\{\ket{0,\Uparrow},\ket{0,\Downarrow}\}$ for the Kramers doublet:
\begin{equation}
\hat{g}(V_0)=
\begin{pmatrix}
g_{11} & 0 & 0 \\
0 & g_{22} & g_{23} \\
0 & g_{32} & g_{33}
\end{pmatrix};
\hat{g}^\prime(V_0)=
\begin{pmatrix}
g^\prime_{11} & 0 & 0 \\
0 & g^\prime_{22} & g^\prime_{23} \\
0 & g^\prime_{32} & g^\prime_{33}
\end{pmatrix}\,.
\end{equation}
If $\vec{B}=B\vec{x}$, $\hat{g}(V_0)\vec{b}=g_{11}\vec{x}$, $\hat{g}^\prime(V_0)\vec{b}=g^\prime_{11}\vec{x}$, so that $f_R\propto|[\hat{g}(V_0)\vec{b}]\times[\hat{g}^\prime(V_0)\vec{b}]|=0$. There are no Rabi oscillations when the magnetic field is perpendicular to the mirror plane. 

Let us now assume that the same device has two extra mirror planes $\sigma_{xz}$ and $\sigma_{xy}$ at an other bias $V=V_0^\prime$ (double group $D_{2h}$). Sweeping the gate voltage around $V_0^\prime$ breaks these mirror planes, but the electric field $\vec{E}_1$ is (at least approximately) homogeneous and parallel to $\vec{y}$, hence even under $\sigma_{yz}$ and $\sigma_{xy}$ but odd under $\sigma_{xz}$. Then,
\begin{equation}
\hat{g}(V_0)=
\begin{pmatrix}
g_{11} & 0 & 0 \\
0 & g_{22} & 0 \\
0 & 0 & g_{33}
\end{pmatrix}
\text{ but }
\hat{g}^\prime(V_0)\sim
\begin{pmatrix}
0 & 0 & 0 \\
0 & 0 & 0 \\
0 & 0 & 0
\end{pmatrix}\,.
\end{equation}
There are no Rabi oscillations at all at this bias point (at least with frequency proportional to $B$ and $V_{\rm ac}$). This is reminiscent of the well-known fact\cite{Winkler03} that the inversion symmetry (implied by the three mirror planes at $V_0^\prime$) is hampering the action of spin-orbit coupling. We will actually highlight such a behavior in the next section. 

\section{Application to a hole qubit}
\label{sectionSOI}

\RED{The equations and arguments of sections \ref{sectionModels} and \ref{sectionSymmetries} hold in both electron and hole spin qubits as long as the Rabi frequency is proportional to $B$ and $V_{\rm ac}$. They apply, therefore, to most III-V qubits, to hole qubits in silicon, but not (for example) to electron qubits in silicon operating near the spin-valley mixing point\cite{Corna18,Bourdet18} (where the Rabi frequency is non-linear in $\vec{B}$). The $g$-matrix formalism can be used as a framework to interpret experimental data,\cite{Crippa18} and/or to compute Larmor and Rabi frequencies from a microscopic model for the qubit. The enumeration of symmetries can, in particular, provide meaningful insights into the operation of the device. As an illustration, we apply our methodology to a hole qubit on SOI.\cite{Maurand16,Crippa18} We first describe the device, then analyze the orientational dependence of the Rabi frequency (and the role of symmetries), and discuss the mechanisms behind the electrical dependence of the $g$-factors. We finally compare the calculated and experimental data.}

\begin{figure}
\includegraphics[width=0.85\columnwidth]{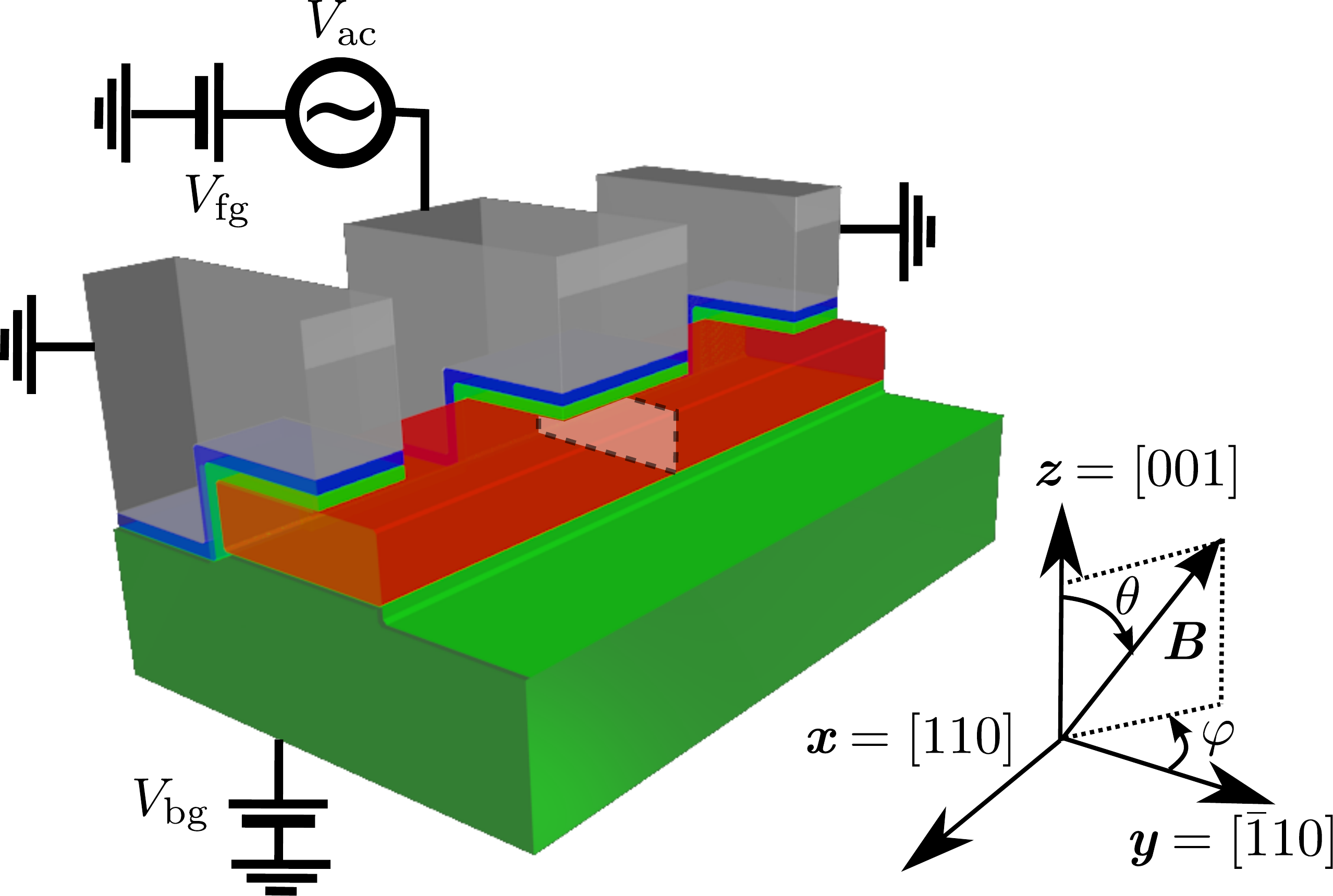}
\caption{Schematics of the device. In red, the $[110]$-oriented, $10\text{\ nm}\times 30\text{\ nm}$ silicon channel on top of the  25 nm thick buried oxide (green) and doped silicon back gate beneath. The 30 nm long front gate (gray) overlaps around half of the channel; it is insulated from the channel by 2 nm of SiO$_2$ and 2 nm of HfO$_2$ (green and blue). The two other lateral gates (also gray) mimic neighboring qubits. They are biased at $V=0$ V unless specified otherwise. The $x$ axis is parallel to the nanowire and the $z$ axis is perpendicular to the substrate. The orientation of the magnetic field $\vec{B}$ is characterized by the polar angle $\theta$ and azimuthal angle $\varphi$ defined on this figure (same angles as in Ref. \onlinecite{Crippa18}). The shaded area in the channel is the $(yz)$ plane where the wave functions of Fig. \ref{figWfnHole} are plotted.}
\label{figDevice}
\end{figure}

\subsection{SOI device and models}

\BLUE{The device, shown in Fig. \ref{figDevice}, is based on a $[110]$-oriented, rectangular nanowire channel with width $W=30$ nm [$(\bar{1}10)$ facets] and height $H=10$ nm [$(001)$ facets]. It is separated from the silicon substrate beneath by a 25 nm thick buried oxide (BOX). This substrate is biased at a potential $V_{\rm bg}$ in order to act as a back gate. A front gate at potential $V_{\rm fg}$ is used to confine the holes in a quantum dot along the channel and to manipulate their spin. This gate is 30 nm long and overlaps around half of the channel (over 20 out of 30 nm). It is insulated from the latter by a 2 nm thick layer of SiO$_2$ and a 2 nm thick layer of HfO$_2$. Two additional gates, laid 30 nm to the left and right of the central dot, mimic neighboring qubits and are biased at potential $V=0$ V unless specified otherwise. The whole device is embedded in Si$_3$N$_4$. In the following, $x\parallel[110]$ is along the nanowire, $y\parallel[\bar{1}10]$ is perpendicular to the nanowire, and $z\parallel[001]$ is perpendicular to the substrate.}

\BLUE{The Rabi frequency is calculated with the methodology introduced in section \ref{sectionModels}. The potential in the device is first computed with a finite volume Poisson solver. The electronic structure of the dot in this potential is then assessed with a six-bands $\vec{k}\cdot\vec{p}$ model\cite{Dresselhaus55,KP09,Luttinger56} discretized on a finite differences mesh (see Appendix \ref{appendixNumerical} for details). This model describes the heavy-hole, light-hole and split-off bands at once and accounts for the ``direct'' Rashba SOC\cite{Kloeffel11,Kloeffel17} that is expected to dominate Rabi oscillations in these devices. This direct Rashba SOC results from the electric and magnetic tunability of the heavy- and light-hole mixing (see later discussion). We assume that the hole wave functions do not penetrate in the oxides and Si$_3$N$_4$ (hard wall boundary conditions at the surface of the channel).}

\subsection{Maps of Rabi frequency}

\begin{figure}
\includegraphics[width=0.95\columnwidth]{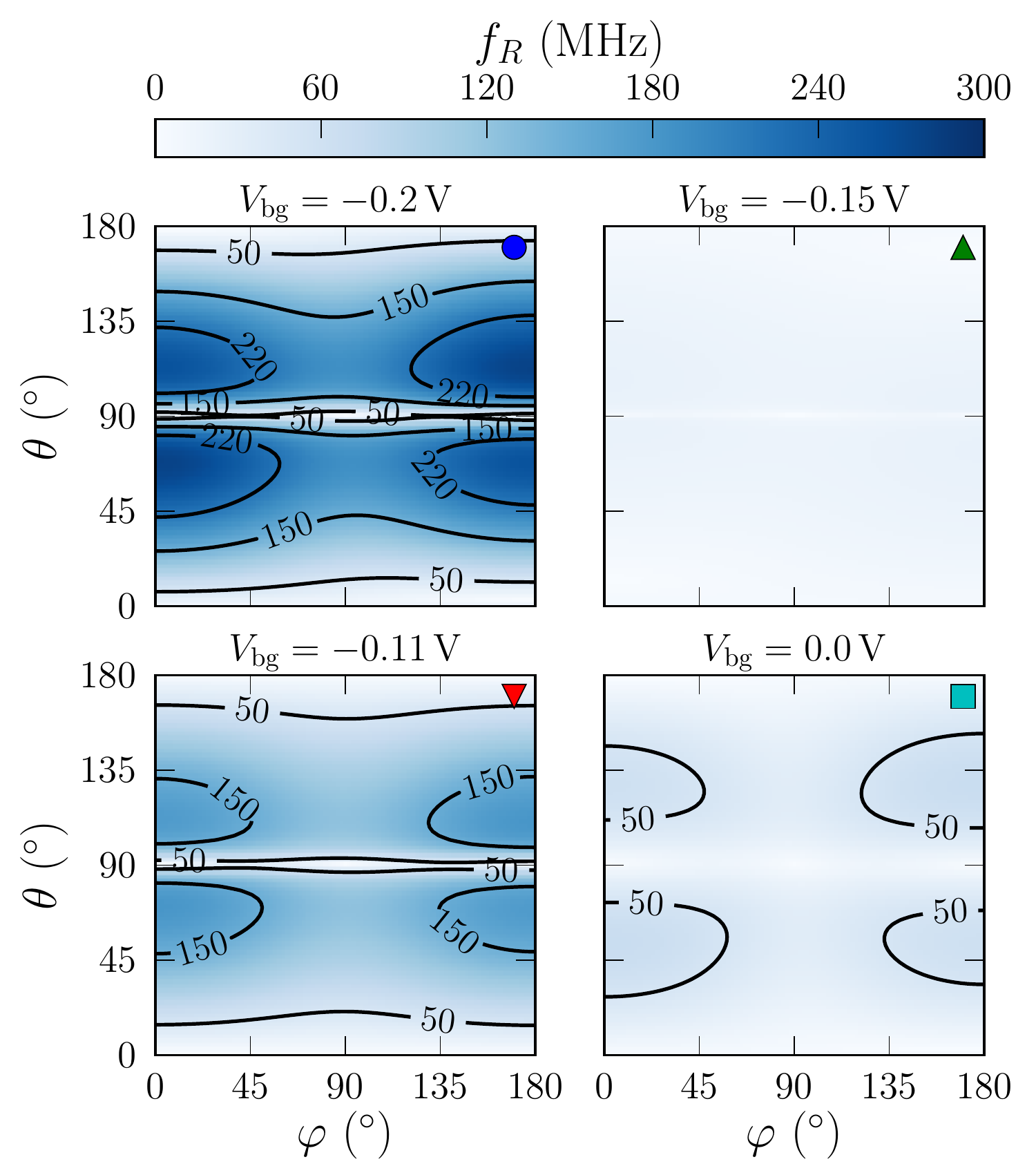}
\caption{Maps of the Rabi frequency as a function of the polar and azimuthal angles of $\vec{B}$ defined on Fig. \ref{figDevice}, for different back gate voltages $V_{\rm bg}$ ($V_{\rm fg}=-0.1$ V, $V_{\rm ac}=1$ mV, $B=1$ T). The Rabi frequency being the same for opposite magnetic fields, the maps are plotted for $0^\circ\le\varphi<180^\circ$. The symbol (circle, triangles and square) on each map uniquely identifies the back gate voltage on Figs. \ref{figMapsRabi}, \ref{figWfnHole}, \ref{figRabiVbg} and \ref{figVt}.}
\label{figMapsRabi} 
\end{figure}

\BLUE{Maps of the calculated Rabi frequency are plotted in Fig. \ref{figMapsRabi} as a function of the polar and azimuthal angles of $\vec{B}$ defined on Fig. \ref{figDevice}, for different back gate voltages. The front gate voltage is $V_{\rm fg}=-0.1$ V and the magnitude of the magnetic field is $B=1$ T. The amplitude of the RF modulation on the front gate is $V_{\rm ac}=1$ mV. The Rabi frequency is calculated from the $g$-matrix and its derivative using Eq. (\ref{eqRabig}). As discussed in section \ref{sectionModels} and Appendix \ref{appendixCalculation}, these matrices can be computed at a given $(V_{\rm fg}, V_{\rm bg})$ from the wave functions of the qubit at zero magnetic field and at three bias points $(V_{\rm fg}-\delta V_{\rm fg}, V_{\rm bg})$, $(V_{\rm fg}, V_{\rm bg})$ and $(V_{\rm fg}+\delta V_{\rm fg}, V_{\rm bg})$ -- there is no need to sweep the orientation of the magnetic field in the $\vec{k}\cdot\vec{p}$ Hamiltonian, which considerably speeds up the calculation.}

\begin{figure}
\includegraphics[width=0.95\columnwidth]{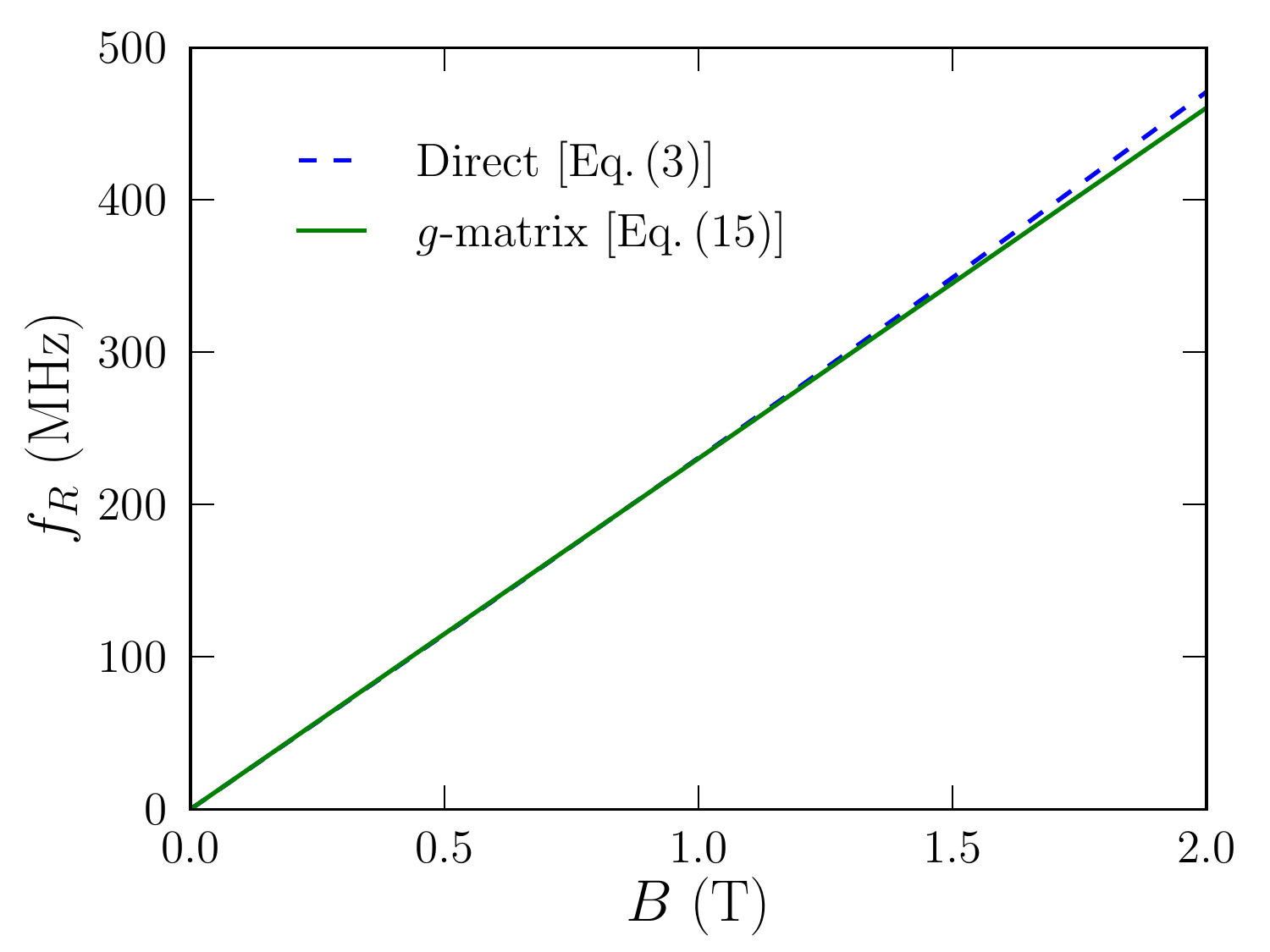}
\caption{Rabi frequency as a function of the magnitude of $\vec{B}\parallel (\vec{y}+\vec{z})$. The $g$-matrix formula [Eq. (\ref{eqRabig})] is compared to the direct evaluation from the wave functions at finite $B$ [Eq. (\ref{eqRabi})]. $V_{\rm fg}=-0.1$V, $V_{\rm bg}=-0.2$V and $V_{\rm ac}=1$ mV.}
\label{figRabiB} 
\end{figure}

\begin{figure}
\includegraphics[width=0.95\columnwidth]{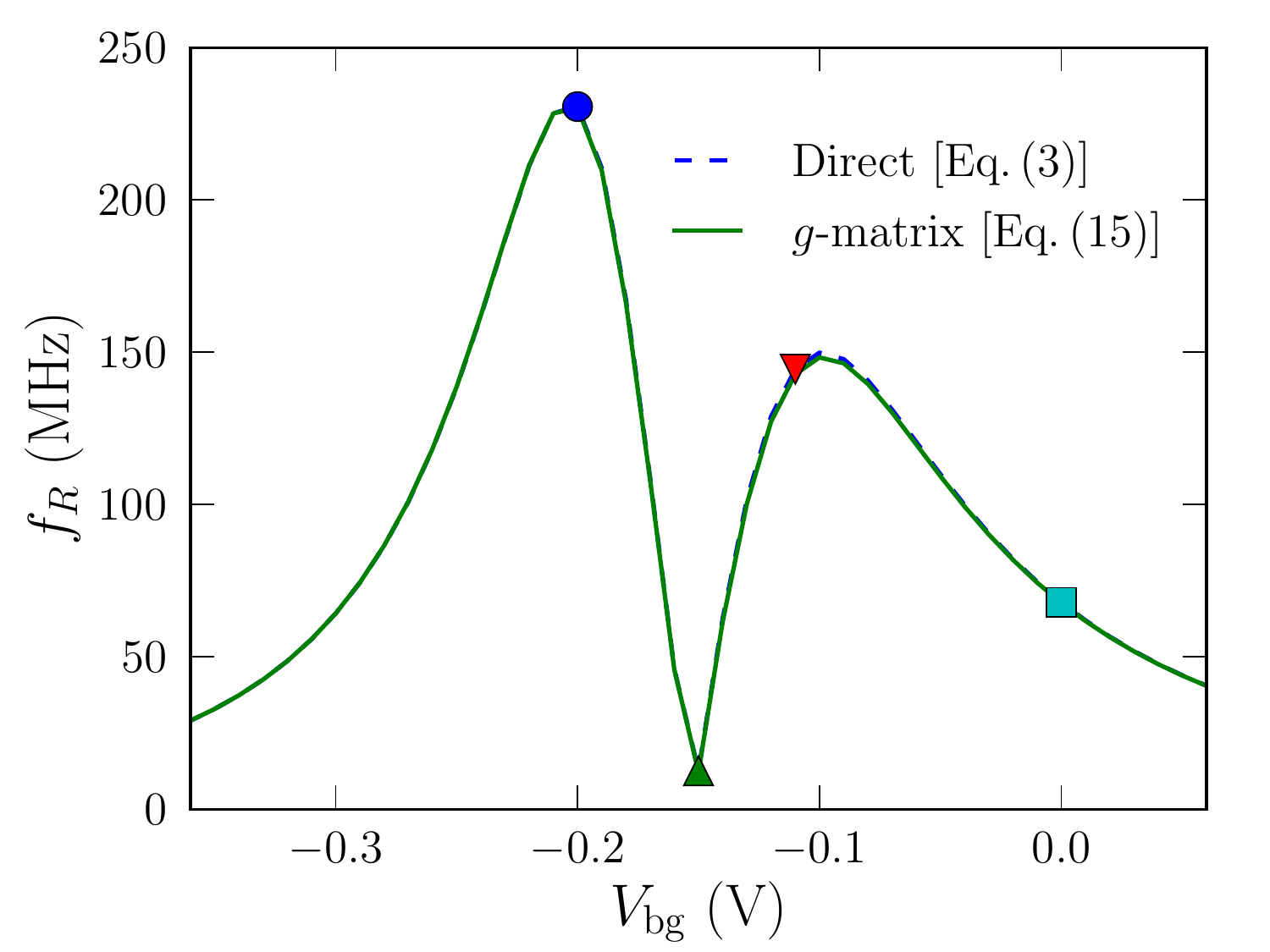}
\caption{Rabi frequency as a function of back gate voltage for $B=1$ T parallel to $\vec{y}+\vec{z}$ ($V_{\rm fg}=-0.1$ V and $V_{\rm ac}=1$ mV). The $g$-matrix formula [Eq. (\ref{eqRabig})] is compared to the direct calculation [Eq. (\ref{eqRabi})]. The maps of Fig. \ref{figMapsRabi} have been plotted at the back gate voltage identified by the corresponding symbol (circle, triangles and square).}
\label{figRabiVbg} 
\end{figure}

\BLUE{In order to assess the validity of the $g$-matrix formalism, we plot in Fig. \ref{figRabiB} the Rabi frequency as a function of the magnitude of $\vec{B}\parallel(\vec{y}+\vec{z})$ ($\theta=45^\circ$, $\varphi=0^\circ$). The $g$-matrix formula [Eq. (\ref{eqRabig})] is compared to the direct \RED{(all orders in $\vec{B}$)} evaluation from the wave functions at finite magnetic field [Eq. (\ref{eqRabi})]. The device operates in the linear-in-$B$ regime where the $g$-matrix formalism holds over a wide range of magnetic fields. This is further supported by Fig. \ref{figRabiVbg}, which shows excellent agreement between the $g$-matrix formula and the direct evaluation at all $V_{\rm bg}$'s and $B=1$ T.}

\BLUE{The maps of Fig. \ref{figMapsRabi} show a complex structure with (quasi-)extinctions along $z$ and in the $(xy)$ plane -- although the Rabi frequency is strictly zero only for $\vec{B}\parallel\vec{x}$. Moreover, the Rabi frequency is very small around back gate voltage $V_{\rm bg}=-0.15$ V. This is also evidenced in Fig. \ref{figRabiVbg}, which displays the Rabi frequency as a function of $V_{\rm bg}$ for $\vec{B}\parallel(\vec{y}+\vec{z})$. There is indeed a sharp dip at $V_{\rm bg}=-0.15\,$V; the Rabi frequency peaks on both sides of this dip and decreases again at large $|V_{\rm bg}|$. It reaches values as large as $\simeq 300$ MHz at $V_{\rm ac}=1$ mV and $B=1$ T.}

\begin{figure}
\includegraphics[width=0.95\columnwidth]{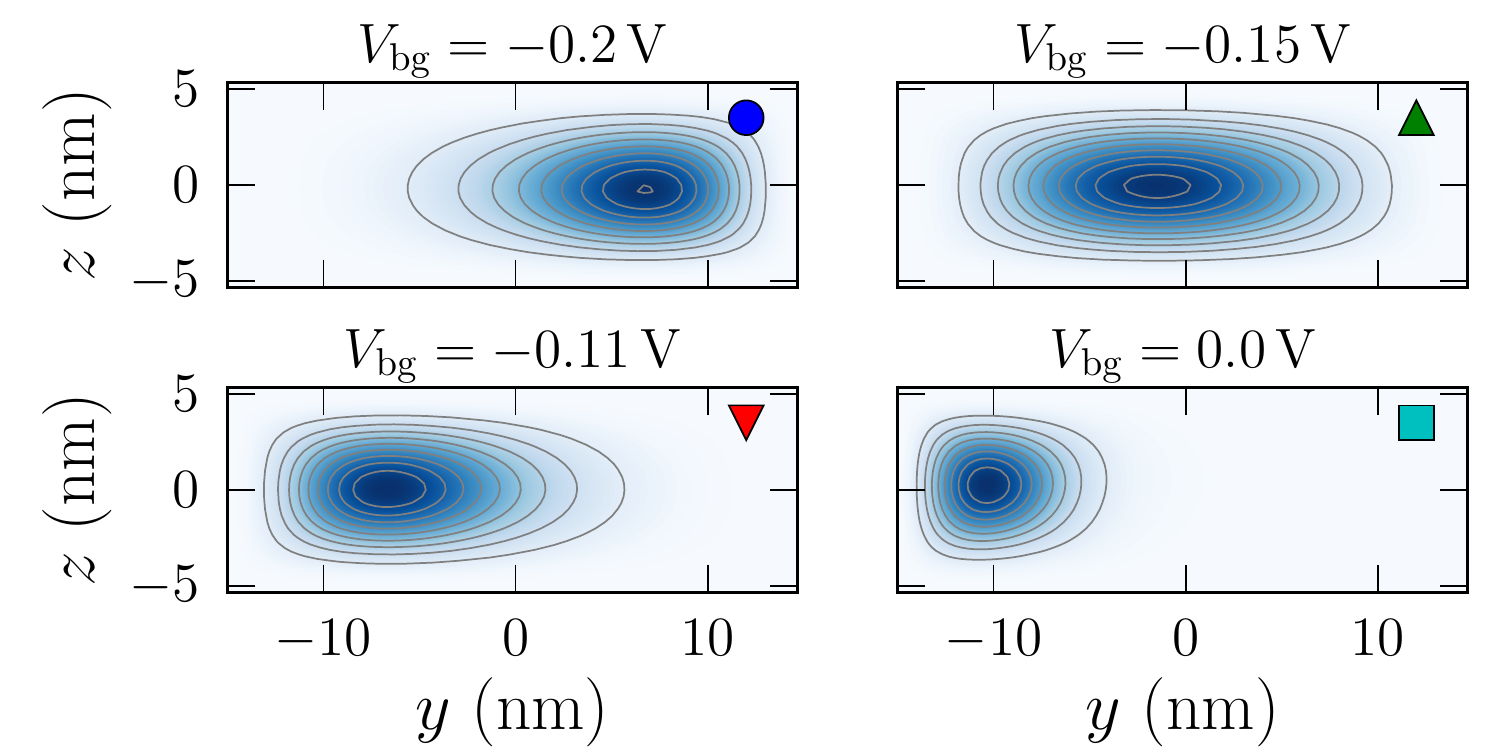}
\caption{Iso-probability surfaces of the ground-state hole wave function at different back gate voltages ($V_{\rm fg}=-0.1$ V). The wave functions are plotted in the $(yz)$ symmetry plane at $x=0$ (shaded area on Fig. \ref{figDevice}). Each plot can be associated with the map of Fig. \ref{figMapsRabi} labeled with the same symbol (circle, triangles and square).}
\label{figWfnHole} 
\end{figure}

\BLUE{In order to understand these trends, we plot the iso-probability surfaces of the ground-state hole wave function at different back gate voltages in Fig. \ref{figWfnHole}. The hole is pushed to the top left corner of the channel at large positive $V_{\rm bg}$, occupies a central position at $V_{\rm bg}\simeq-0.15$ V, and is pulled to the bottom right corner at large negative $V_{\rm bg}$. The back gate therefore controls the position and symmetry of the hole wave function, as it does for electrons.\cite{Bourdet18} The confinement in corner dots is, however, less pronounced for holes than for electrons\cite{Voisin14} due to the different mass anisotropies. We can actually conjecture from Fig. \ref{figWfnHole} that the hole mostly responds to the $y$ component of the electric field of the gates, the polarizability being visibly much smaller along $z$. The extinction of the Rabi frequency corresponds to the back gate voltage where the hole wave function is ``most symmetric''. Note that this control of the symmetry is made easier by the non-planar design of the front gate of this device.}

\begin{figure}
\includegraphics[width=0.95\columnwidth]{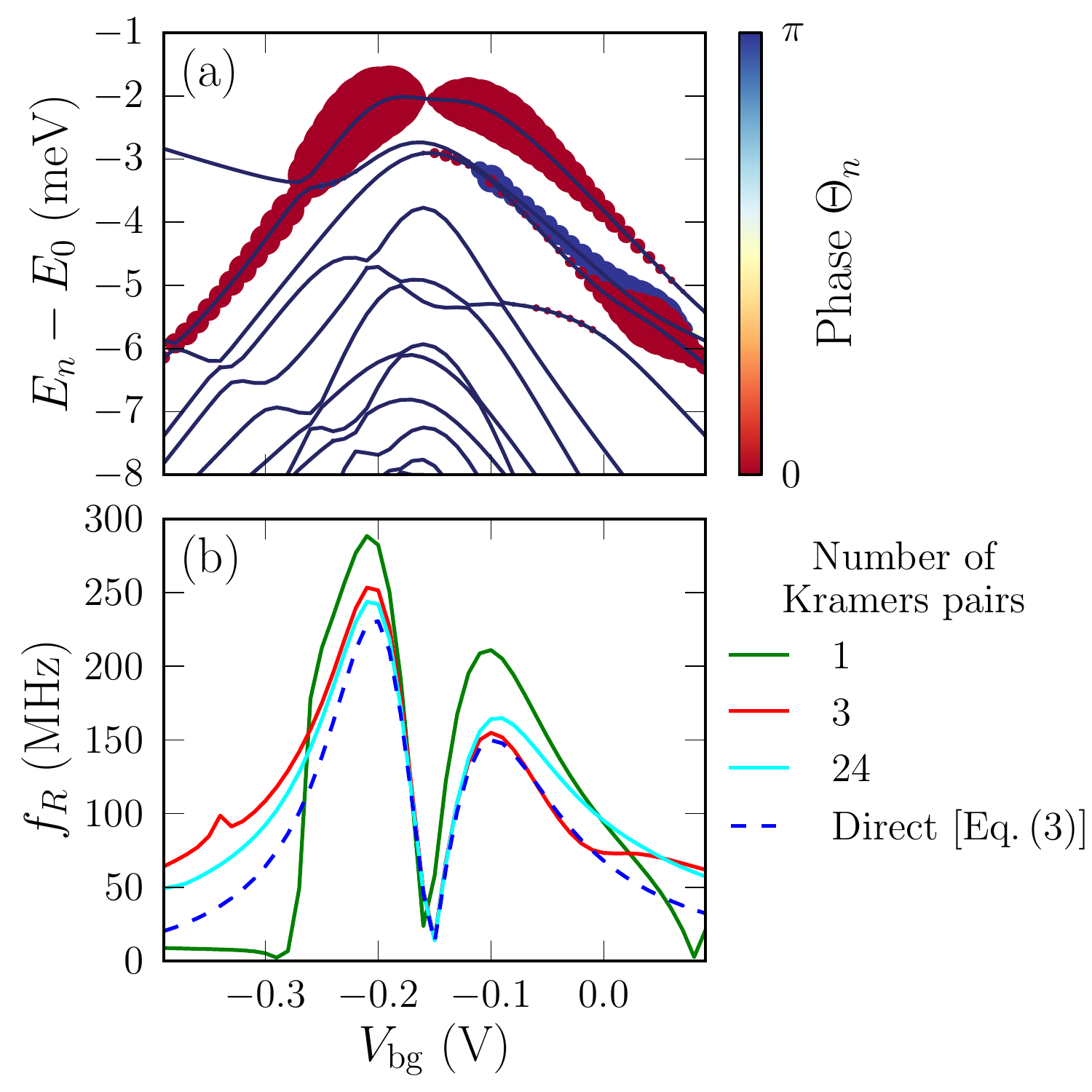}
\caption{(a) Spectrum $E_n-E_0$ of the quantum dot as a function of back gate voltage ($V_{\rm fg}=-0.1$ V, $B=0$ T). We may write the perturbation series for the Rabi frequency, Eq. (\ref{eqRabipert}), as $f_R=|\sum_{n>0}f_{R,n}|$, where $n$ runs over the pairs of Kramers degenerate states. The size of the dots is proportional to the magnitude $|f_{R,n}|$ of the contribution of each pair. The color quantifies the phase of this contribution, $\Theta_n=\arg(f_{R,n})$. (b) Rabi frequency from the perturbation series, Eq. (\ref{eqRabipert}), versus back gate voltage. Summations including 1, 3 and 24 excited pairs of states are compared with Eq. (\ref{eqRabi}) ($V_{\rm fg}=-0.1$ V, $V_{\rm ac}=1$ mV and $\vec{B}\parallel(\vec{y}+\vec{z})$).}
\label{figSpectrum} 
\end{figure}

Strictly speaking, the symmetry group of the system is $C_s$ whatever $V_{\rm bg}$, with one exact symmetry plane $\sigma_{yz}$ perpendicular to the channel (and splitting the gate in two halves). Following section \ref{sectionSymmetries}, we expect the Rabi frequency to be zero for $\vec{B}\parallel\vec{x}$, i.e. along the nanowire ($\theta=90^\circ$ and $\varphi=90^\circ$), which is actually the case. Fig. \ref{figWfnHole} however suggests that the ground-state hole wave function has a horizontal quasi-symmetry plane $\sigma_{xy}$ over a wide range of back gate voltages (weak corner confinement), and an extra vertical quasi-symmetry plane $\sigma_{xz}$ near $V_{\rm bg}=-0.15$ V. This approximate $\sigma_{xy}$ symmetry explains why the Rabi frequency is small for $\vec{B}\parallel\vec{z}$ ($\theta=0$ and $\theta=180^\circ$), while the extra $\sigma_{xz}$ symmetry explains (as further detailed below) the global suppression of the Rabi oscillations at that back gate voltage. 

\begin{figure}
\includegraphics[width=0.95\columnwidth]{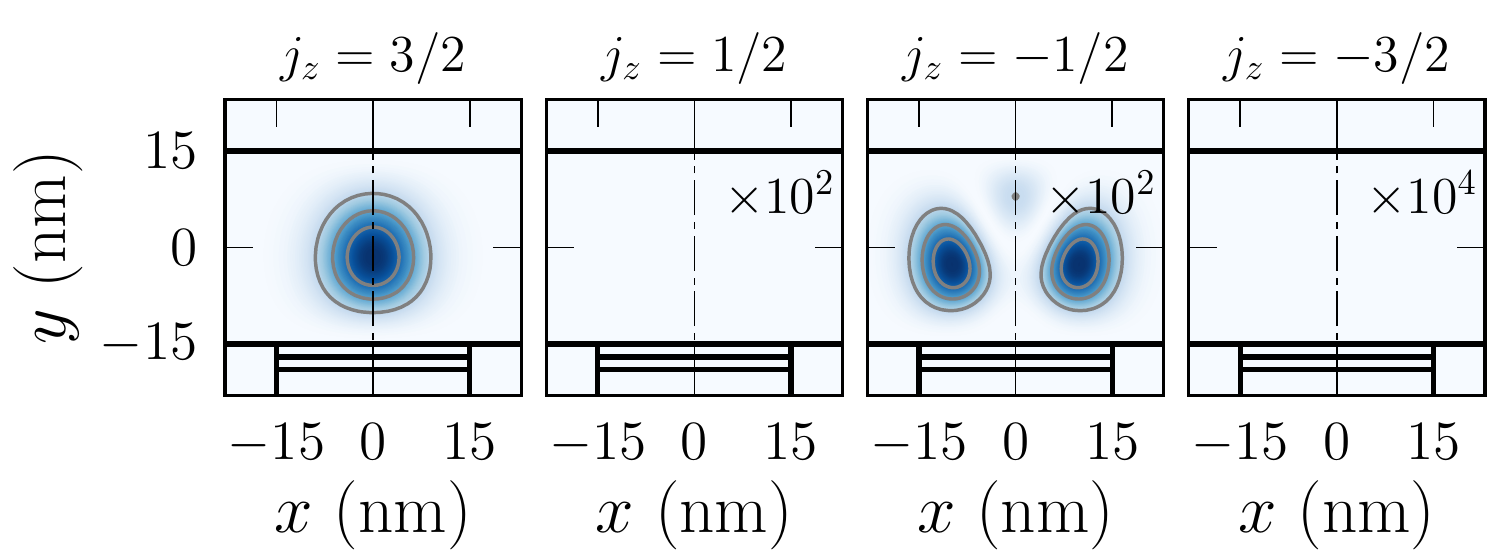}
\includegraphics[width=0.95\columnwidth]{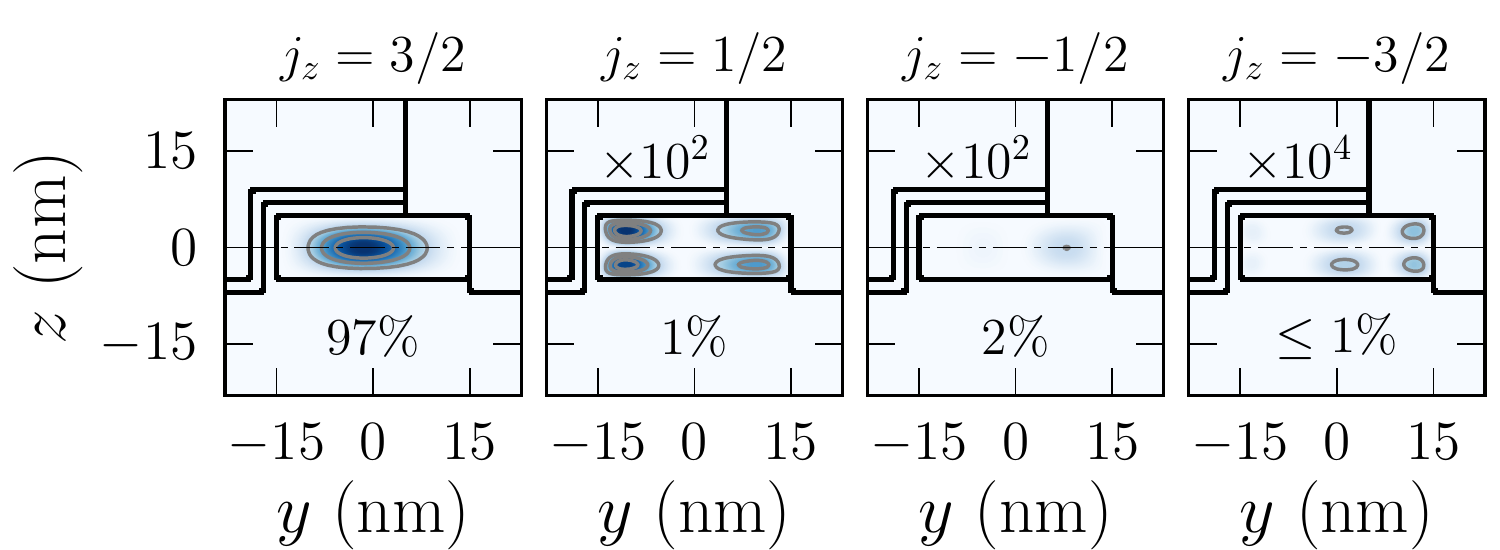}
\caption{Squared heavy-hole ($j_z=\pm\frac{3}{2}$) and light-hole ($j_z=\pm\frac{1}{2}$) envelopes of the qubit ground-state in the $(xy)$ plane at $z=0$ (top) and in the $(zy)$ plan at $x=0$ (bottom) ($V_{\rm fg}=-0.1$ V, $V_{\rm bg}=-0.15$ V). The dash-dotted lines on each plot delineate the position of these planes. Note that this state is degenerate with a partner having time-reversal symmetric envelopes (namely, $j_z=\frac{3}{2}\leftrightarrow j_z=-\frac{3}{2}$, $j_z=\frac{1}{2}\leftrightarrow j_z=-\frac{1}{2}$). The total weight of each envelope is given in the bottom panel. Some squared envelopes have been multiplied by $10^2$ or $10^4$ for clarity.}
\label{figEnvHole1} 
\end{figure}

\begin{figure}
\includegraphics[width=0.95\columnwidth]{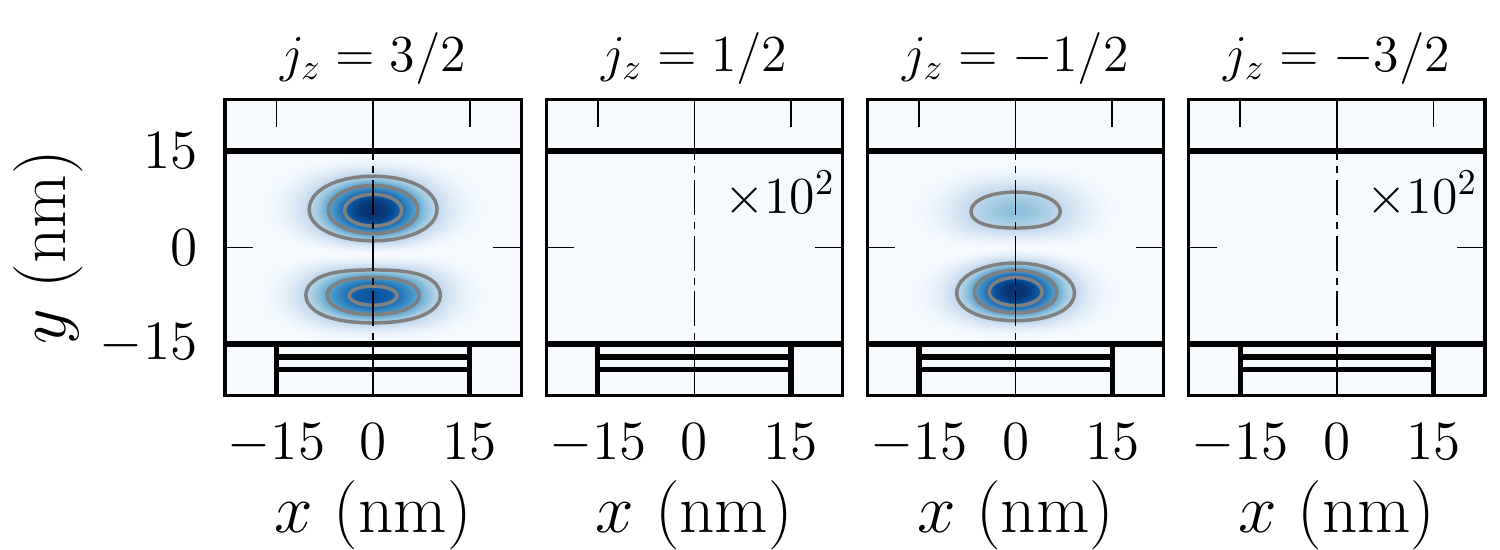}
\includegraphics[width=0.95\columnwidth]{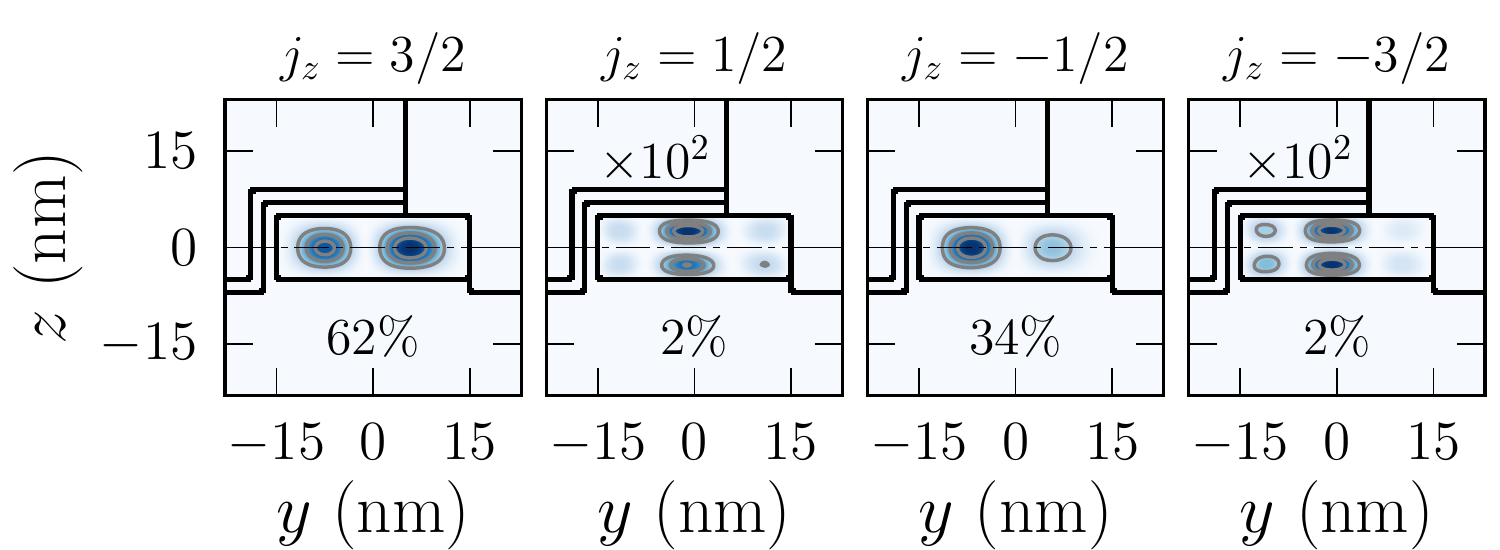}
\caption{Same as Fig. \ref{figEnvHole1}, for the first excited state.}
\label{figEnvHole2} 
\end{figure}

In order to confirm this interpretation, we plot the spectrum $E_n-E_0$ of the quantum dot as a function of $V_{\rm bg}$ in Fig. \ref{figSpectrum}. The contribution of each excited state to the perturbation series, Eq. (\ref{eqRabipert}), is quantified by the thickness and color of the corresponding dots. The perturbation series is dominated by one state (actually the lowest-lying excitation over a wide range of gate voltages), whose splitting with the ground-state shows a strong dependence on $V_{\rm bg}$. This is further highlighted in Fig. \ref{figSpectrum}b, which compares the outcome of Eq. (\ref{eqRabipert}) including the first, first three and first 24 excitations. The perturbation series with the first excitation only indeed captures the main features of the Rabi frequency. The matrix elements $\bra{1,\sigma^\prime}\vec{b}\cdot\vec{M}_1\ket{0,\sigma}$ between the qubit and first excited states in Eq. (\ref{eqRabipert}) vanish near the dip.

\begin{figure}
\includegraphics[width=0.95\columnwidth]{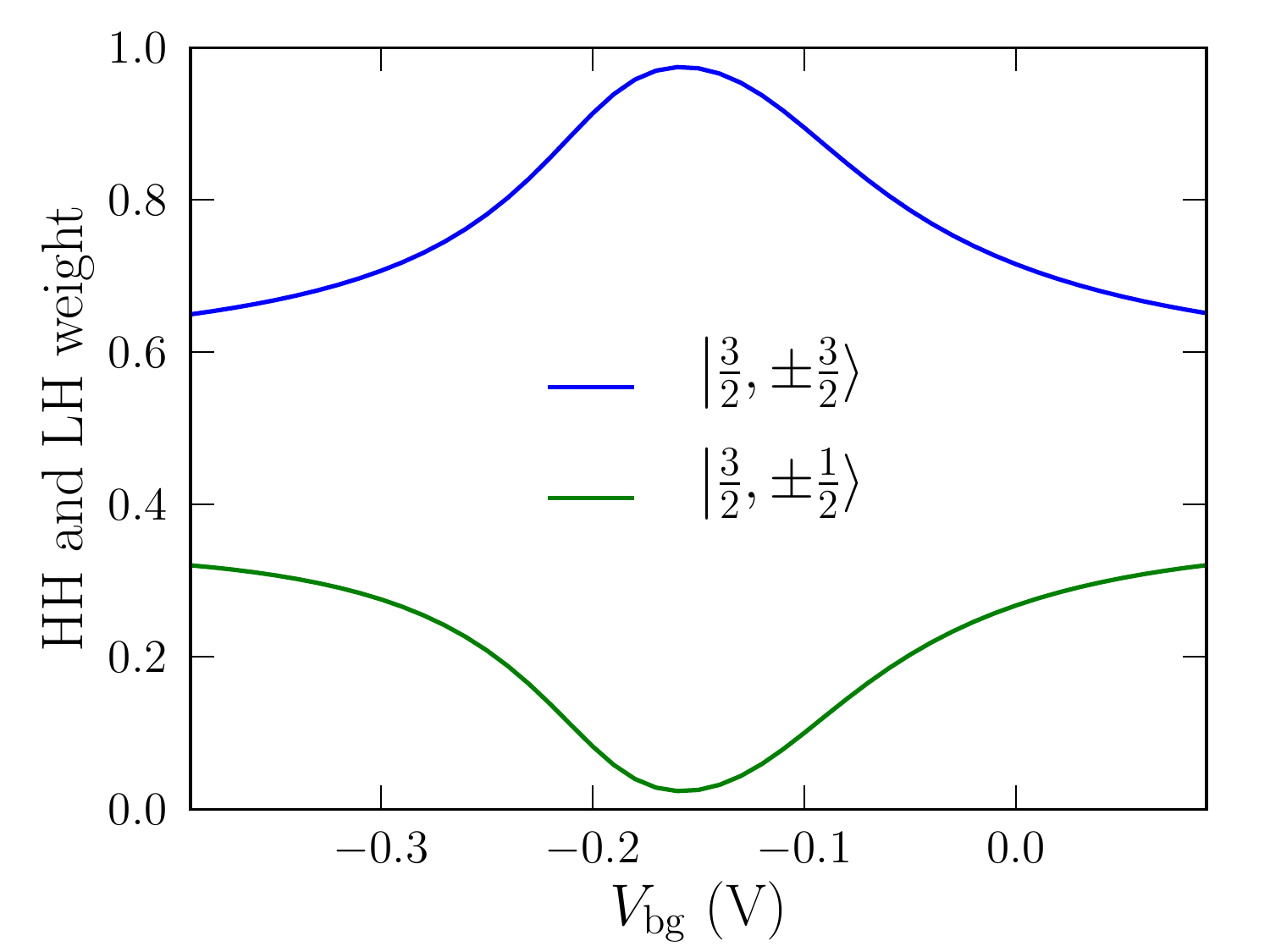}
\caption{Total weight of the heavy (HH, $\ket{\frac{3}{2},\pm\frac{3}{2}}$) and light (LH, $\ket{\frac{3}{2},\pm\frac{1}{2}}$) envelopes in the ground-state (qubit) hole wave functions as a function of $V_{\rm bg}$ ($V_{\rm fg}=-0.1$ V, $B=0$ T). The contribution from split off envelopes is negligible and is not shown.}
\label{figHHLH} 
\end{figure}

The squared heavy-hole ($\ket{\frac{3}{2},\pm\frac{3}{2}}$) and light-hole ($\ket{\frac{3}{2},\pm\frac{1}{2}}$) envelopes of the ground (qubit) and first excited states at $V_{\rm bg}=-0.15$ V are plotted in Figs. \ref{figEnvHole1} and \ref{figEnvHole2}.\footnote{These figures are actually drawn at a small magnetic field $B_z=5$ mT in order to lift Kramers degeneracy and catch the states with maximal $j_z=+\frac{3}{2}$ component.} The total angular momentum is quantized along the strongest confinement axis $\vec{z}$. The split-off envelopes ($\ket{\frac{1}{2},\pm\frac{1}{2}}$) make little contribution to these states. The ground state has a dominant heavy-hole character (as in a silicon thin film) but admixes small light-hole components (Fig. \ref{figHHLH}) due to the extra lateral confinement by the structure and gate fields. The first excited state shows a stronger light-hole character and ``$p_y$''-like heavy-hole envelope (namely, with one nodal plane perpendicular to $y$). When varying $V_{\rm bg}$ around the symmetric position $V_{\rm bg}\simeq-0.15$ V, these two states are mixed by the $y$ component of the gates field. This strengthens heavy-/light-hole mixing in the qubit states, and leads to the strong dependence of the $E_1-E_0$ splitting shown in Fig. \ref{figSpectrum}. 

The same coupling to the $y$ component of the gates field is responsible for the Rabi oscillations driven by the front gate. The excited states with modulations of the heavy-hole envelopes along $z$ are too confined and too far in energy to allow for significant electrical polarizability perpendicular to the substrate. The $z$ component of the electric field, though dominant in many parts of the device, is therefore inefficient in driving Rabi oscillations. 

\begin{figure}
\includegraphics[width=0.95\columnwidth]{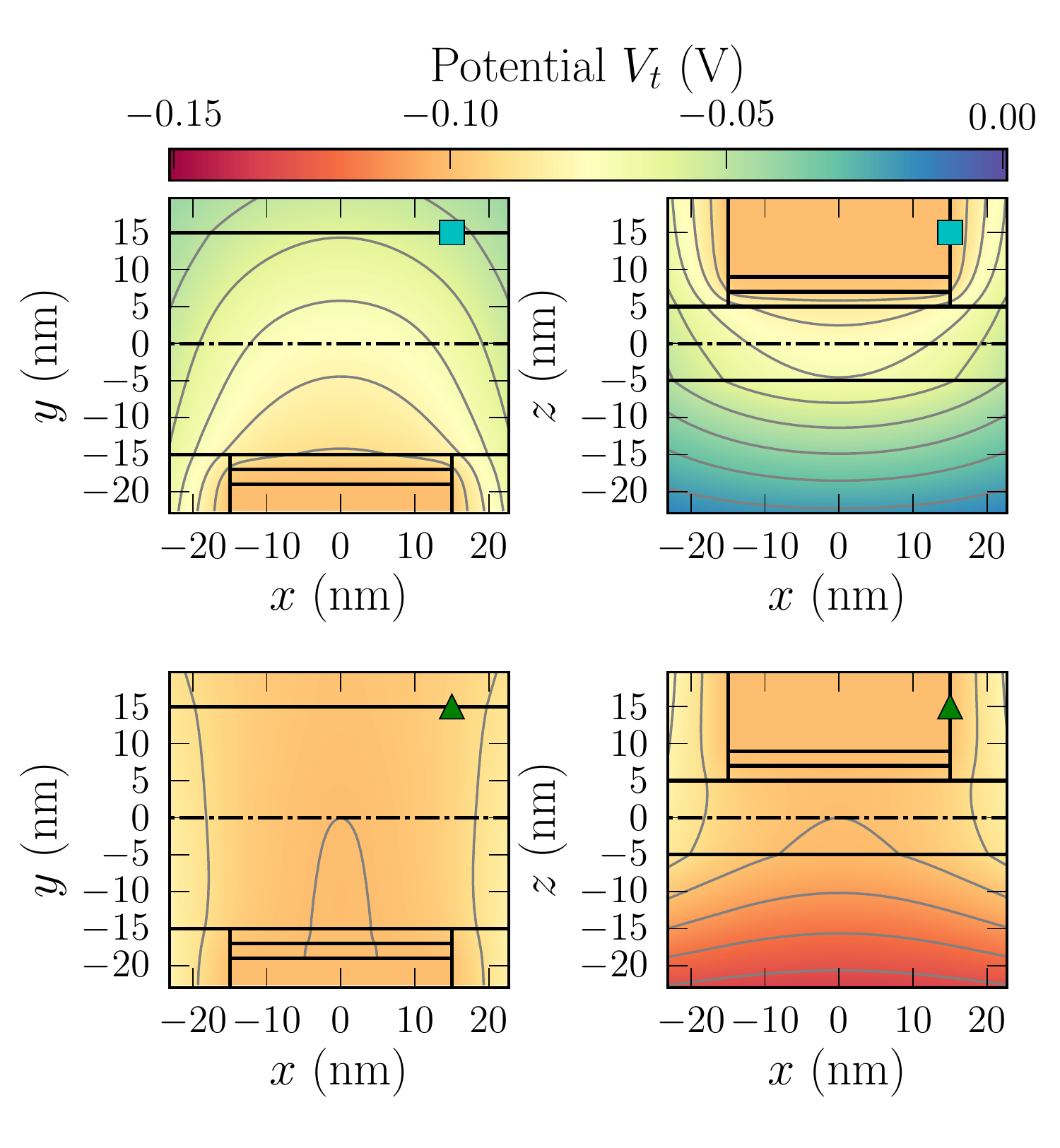}
\caption{The total potential $V_t(\vec{r})$ in the $(xy)$ plane at $z=0$ and in the $(xz)$ plane at $y=0$, at (top) $V_{\rm bg}=0$ V and (down) $V_{\rm bg}=-0.15$ V ($V_{\rm fg}=-0.1$ V). The dash-dotted lines on each plot delineate the position of these planes. Each plot can be associated with the corresponding map of Fig. \ref{figMapsRabi} labeled with the same symbol (triangle and square). The isovalue lines are separated by $10\,$mV.}
\label{figVt} 
\end{figure}

\begin{figure}
\includegraphics[width=0.95\columnwidth]{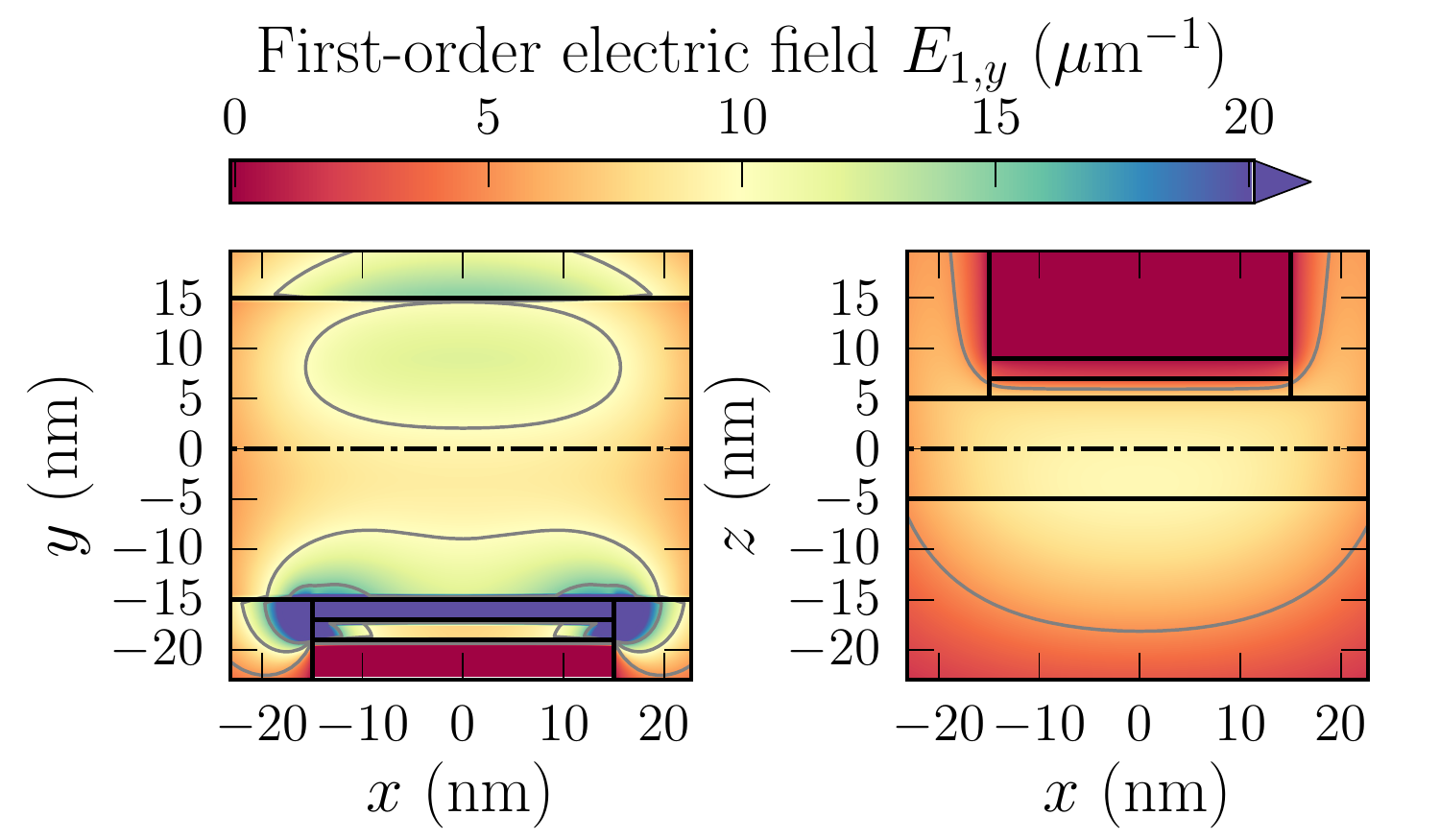}
\caption{The first-order electric field $E_{1,y}$ from the top gate in the $(xy)$ plane at $z=0$ and in the $(xz)$ plane at $y=0$. The dash-dotted lines on each plot delineate the position of these planes.}
\label{figEy} 
\end{figure}

Having identified the main couplings responsible for Rabi oscillations, we can refine our symmetry analysis. The total potential $V_t(\vec{r})$ in the device is plotted in Fig. \ref{figVt} at $V_{\rm bg}=0$ V and $V_{\rm bg}=-0.15$ V. It shows approximate $\sigma_{xy}$ and $\sigma_{xz}$ symmetries at $V_{\rm bg}=-0.15$ V, consistent with Fig. \ref{figWfnHole}. The $\sigma_{xy}$ plane is more noticeably broken, yet this does not result in sizable dissymetries in the ground-state wave function owing, again, to the low polarizability along $z$. Since the Rabi oscillations are driven by the $y$ component of the front gate field, it is sufficient to analyze the parity of $E_{1,y}$ to make use of Table \ref{table_r2}. As the electrostatics is linear in the present description (see Appendix \ref{appendixNumerical}), $D_1(\vec{r})$ is simply the potential in the device with the front gate at a unit voltage and all other gates grounded, so that $E_{1,y}$ is independent on the bias point. Fig. \ref{figVt}b shows that $E_{1,y}$ is pretty homogeneous in a large volume under the gate; it is, therefore, nearly even with respect to $\sigma_{xy}$ and $\sigma_{yz}$, but odd with respect to $\sigma_{xz}$. According to Tables \ref{table_r2}, the matrix $\hat{g}^\prime$, hence the Rabi frequency must indeed be zero at $V_{\rm bg}=-0.15$ V (or almost so, as the $\sigma_{xy}$ and $\sigma_{xz}$ symmetries are only approximate). This highlights the importance of breaking symmetries to maximize the opportunities for Rabi oscillations, as already shown in Refs. \onlinecite{Crippa18} and \onlinecite{Corna18}. We also emphasize that the specific geometry of these SOI devices, with the gate overlapping only the top and lateral facets of the channel, helps to break $\sigma_{xz}$ and to enhance $E_{1,y}$.  \BLUE{Such a qubit may, in principle, be switched between a bias point (e.g., $V_{\rm bg}=-0.2$ V) where it is strongly coupled to the RF electric field for manipulation and a bias point ($V_{\rm bg}=-0.15$ V) where it gets largely decoupled, but is as a consequence more immune to charge and gate noise.\cite{Bourdet18,Kloeffel13}}

Beyond the dip near $V_{\rm bg}=-0.15$ V, the Rabi frequency also decreases at large $|V_{\rm bg}|$ when the hole gets localized on the left or right of the channel. This mostly results from a decrease of the dipole matrix elements $\langle 1,\sigma^\prime|D_1|0,\sigma\rangle$ between the strongly confined ground- and first excited states (as well as from an increase of the corresponding denominator in Eq. (\ref{eqRabipert})), as easily evidenced in a simple triangular well model. Fig. \ref{figRabiVbg} can, therefore, be best described as a single broad peak (primarily shaped by the matrix element(s) of $D_1$), split by a dip near $V_{\rm bg}=-0.15$ V owing to the increase of symmetry at that point (and shaped by the matrix element(s) of $\vec{M}_1$).

The mixing of heavy- and light-hole components in the qubit states, and its electro-magnetic tunability highlighted in Fig. \ref{figHHLH} is the key ingredient of the direct Rashba SOC at play here.\cite{Kloeffel11,Kloeffel17} Indeed, neither the $D_1$ operator (which is diagonal in spin) nor the $\vec{b}\cdot\vec{M}_1$ operator can couple $\ket{\frac{3}{2},+\frac{3}{2}}$ and $\ket{\frac{3}{2},-\frac{3}{2}}$ envelopes and connect pure heavy-hole $\ket{\zero}$ and $\ket{\one}$ states in Eq. (\ref{eqRabipert}) [this can be inferred from the form of Eqs. (\ref{eq6kp}) and (\ref{eqHbloch}) of Appendix \ref{appendixNumerical}]. In essence, the strong atomistic SOC around the silicon nuclei couples the spin of the holes with their momentum. As a consequence, holes with spin parallel to the wave vector tend to show heavier masses in bulk silicon. In nanoscale structures, the mixing between heavy- and light-hole components gives rise to a complex spin distribution in real space. This mixing is efficiently controlled by both electric and magnetic fields; a modulation of the electric field can therefore give rise to a redistribution of spin and ultimately to a rotation under resonant conditions in a finite magnetic field that breaks Kramers degeneracy. The effect of the electric field on the heavy- and light-hole components can be cast as an effective Rashba Hamiltonian; this Rashba effect is ``direct'' in the sense that the electric field couples bands that belong to the same quasi-degenerate manifold, at variance with the Rashba effect in the conduction band of III-V's for example where the electric field couples the conduction band with remote (in particular, valence) bands and which is, therefore, much weaker.

\subsection{Analysis of the $g$-factors}
\label{subsectionganalysis}

\begin{figure}
\includegraphics[width=0.95\columnwidth]{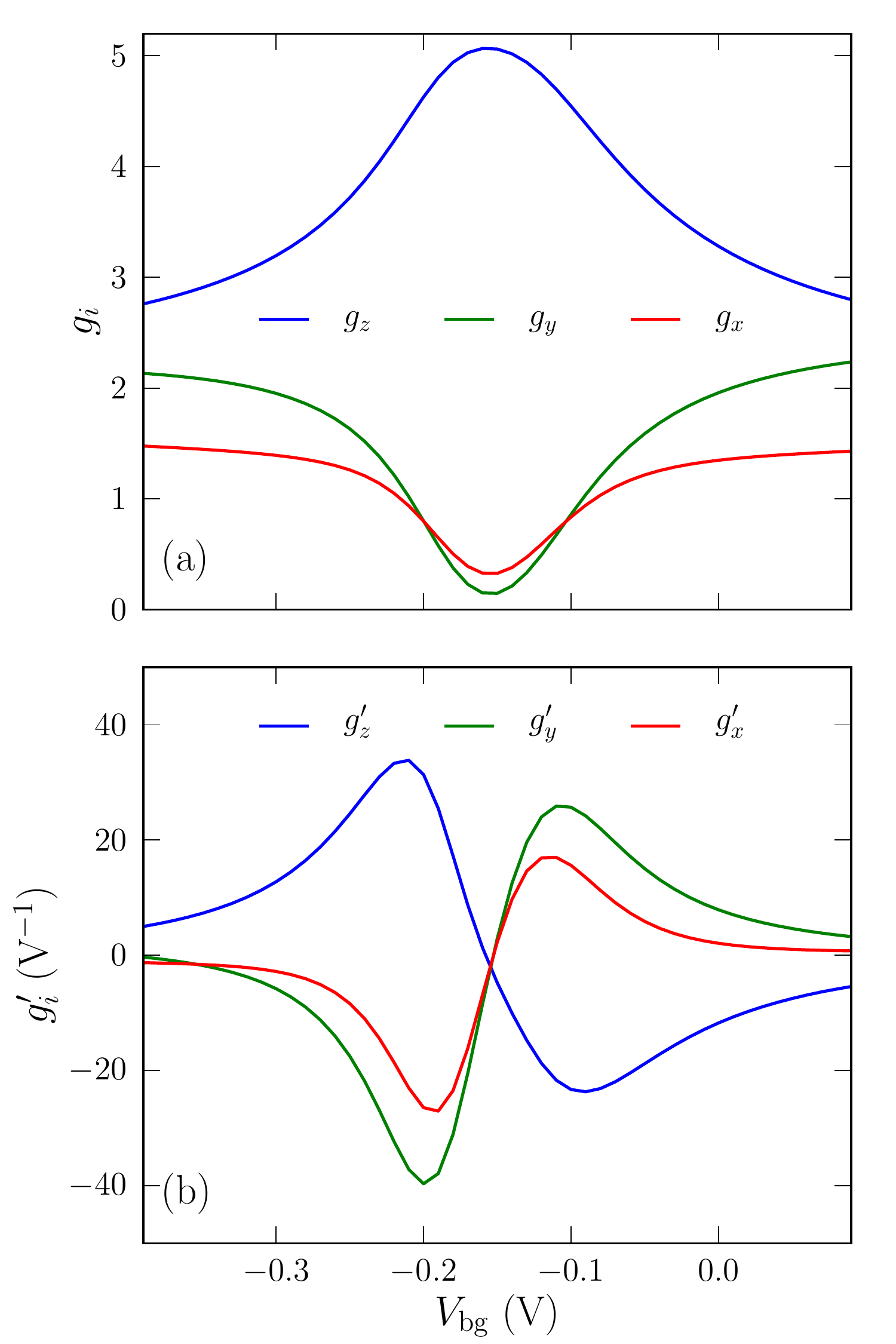}
\caption{(a) The $g$-factors $g_x$, $g_y$, $g_z$ and (b) their derivatives with respect to $V_{\rm fg}$, as a function of the back gate voltage ($V_{\rm fg}=-0.1$ V).}
\label{figgfactors} 
\end{figure}

We now turn to a more quantitative analysis of the $g$-matrix and derivative in order to explain extra features on Fig. \ref{figMapsRabi} such as the quasi-extinction of the Rabi frequency for magnetic fields in the $(xy)$ plane.

As the device has an exact $\sigma_{yz}$ symmetry, $x$ (the axis of the nanowire) is a principal magnetic axis whatever $V_{\rm bg}$. With an additional, weak $\sigma_{xy}$ symmetry in a large range of $V_{\rm bg}$, $y$ and $z$ shall also be approximate principal magnetic axes. The factorization of the $g$-matrix [Eq. (\ref{eqfactorizationg})] confirms that this is indeed the case and yields the principal $g$-factors $g_1\equiv g_x$, $g_2\equiv g_y$, and $g_3\equiv g_z$. They are all plotted as a function of $V_{\rm bg}$ in Fig. \ref{figgfactors}a. We may apply the same transformation that diagonalizes $\hat{g}$ onto $\hat{g}^\prime$ ($\hat{g}^\prime\to{^t}U\hat{g}^\prime V)$. It turns out that the resulting derivative matrix is also (almost) diagonal. The diagonal elements $g_x^\prime$, $g_y^\prime$, and $g_z^\prime$ are plotted as a function of $V_{\rm bg}$ in Fig. \ref{figgfactors}b. 

Neglecting the action of the vector potential on the envelope functions, the magnetic Hamiltonian of $J=3/2$ holes reduces to $H=-2\kappa\mu_B\vec{B}\cdot\vec{J}$, where $\vec{J}$ is the total angular momentum of the Bloch function and $\kappa=-0.42$ [this is Eq. (\ref{eqHbloch}) of Appendix \ref{appendixNumerical} in the $J=3/2$ subspace]. The $g$-factors of a pure heavy-hole ($\ket{\frac{3}{2},\pm\frac{3}{2}}$) doublet are therefore expected to be $g_x=g_y=0$, $g_z=-6\kappa=2.52$, while the $g$-factors of a pure light-hole ($\ket{\frac{3}{2},\pm\frac{1}{2}}$) doublet are expected to be $g_x=g_y=-4\kappa=1.68$, $g_z=-2\kappa=0.84$. Comparison between the heavy-/light-hole mixing data of Fig. \ref{figHHLH} and the $g$ factors of Fig. \ref{figgfactors}a shows qualitative agreement with this trend. $g_z$ is indeed maximal (and $g_x$, $g_y$ minimal) near $V_{\rm bg}=-0.15$ V where the heavy-hole component is the largest. $g_z$ then decreases while $g_x$ and $g_y$ increase as lateral confinement in the gates field enhances heavy-/light-hole mixing. However, the calculated $g$ factors do not match quantitatively the above Hamiltonian. In particular, $g_z$ peaks around 5, far above $6|\kappa|$. These discrepancies result from the action of the magnetic field on the heavy- and light-hole envelopes, as discussed in Ref. \onlinecite{Watzinger16}. Namely, lateral confinement admixes $\ket{\frac{3}{2},\pm\frac{1}{2}}$ envelopes into the mostly $\ket{\frac{3}{2},\pm\frac{3}{2}}$ ground-states [through the ``S'' term of the six-bands $\vec{k}\cdot\vec{p}$ Hamiltonian, see Eq. (\ref{eq6kp}) of Appendix \ref{appendixNumerical} and Fig. \ref{figEnvHole1}]. The mixing coefficients of $\ket{\frac{3}{2},+\frac{1}{2}}$ into $\ket{\frac{3}{2},+\frac{3}{2}}$ and of $\ket{\frac{3}{2},-\frac{1}{2}}$ into $\ket{\frac{3}{2},-\frac{3}{2}}$ do, however, vary in opposite ways with increasing magnetic field owing to the breaking of time-reversal symmetry. This further splits the two states, and makes an extra contribution $\Delta g_z$ to the total $g$-factor. In a model, box-shaped quantum dot with sides $L_z\ll L_x,\,L_y$,\cite{Watzinger16}
\begin{equation}
\Delta g_z=\frac{2^{17}\gamma_3^2}{81\pi^4\left(3\gamma_1+10\gamma_2\right)}\,,
\label{eqdeltag}
\end{equation}
where $\gamma_1$, $\gamma_2$ and $\gamma_3$ are the Luttinger parameters of the host material. In silicon, $\Delta g_z=2.14$ so that $g_z$ is expected to peak around $-6\kappa+\Delta g_z\simeq4.66$ for an almost pure heavy-hole state, in good agreement with the numerical simulations at $V_{\rm bg}=-0.15$ V. Eq. (\ref{eqdeltag}) is strikingly proportional to $\gamma_3$ and independent on the sides $L_x$ and $L_y$ of the dot. $\Delta g_z$ shall, therefore, be weakly dependent on the front and back gate voltages that control the size of the dot. As a matter of fact, setting $\gamma_3=0$ in $\vec{M}_1$ just results in an almost rigid shift of $g_z$.

As expected, all $g_i^\prime$ go through zero near $V_{\rm bg}=-0.15$ V where the heavy-hole/light-hole mixing becomes independent on the electric field to first-order owing to the extra symmetries (see Fig. \ref{figHHLH}).

The fact that the Rabi frequency is much smaller when $\vec{B}$ lies in the $(xy)$ plane (see Fig. \ref{figMapsRabi}) can be traced back to the lack of anisotropy of the $g_i$ and $g_i^\prime$ in that plane ($g_xg_y^\prime-g_x^\prime g_y\ll g_zg_x^\prime-g_z^\prime g_x,\,g_zg_y^\prime-g_z^\prime g_y$). Neglecting $g_xg_y^\prime-g_yg_x^\prime$ in Eq. (\ref{eqRabig}) indeed yields:
\begin{align}
f_R&\simeq\frac{\mu_B B V_{\rm ac}}{2h}|b_z| \nonumber \\
&\times\frac{\sqrt{\left(g_zg_x^\prime-g_z^\prime g_x\right)^2b_x^2+\left(g_zg_y^\prime-g_z^\prime g_y\right)^2b_y^2}}{\sqrt{g_x^2b_x^2+g_y^2b_y^2+g_z^2b_z^2}}\,,
\label{eqapproxfr}
\end{align}
where $b_x$, $b_y$ and $b_z$ are the components of the unit vector $\vec{b}=\vec{B}/B$. The above approximation nicely reproduces the main features of Fig. \ref{figgfactors}. The Luttinger parameter $\gamma_3$ plays again a central role in this equation. An analysis along the lines of Ref. \onlinecite{Watzinger16} indeed suggests that $g_x$ and $g_y$ are, to lowest order, proportional to $\gamma_3$ (as a result of the mixing between $\ket{\frac{3}{2},\pm\frac{3}{2}}$ and $\ket{\frac{3}{2},\mp\frac{1}{2}}$ envelopes through the $R$ term of the 6 bands $\vec{k}\cdot\vec{p}$ Hamiltonian), so that $f_R\propto\gamma_3$. We have indeed verified that $f_R\to0$ when $\gamma_3\to 0$.

Since $\hat{g}^\prime$ remains (almost) diagonal in the basis sets where $\hat{g}$ is, the RF electric field essentially modulates the principal $g$-factors but hardly rotates the matrices $\hat{U}$ and $\hat{V}$ of the singular value decomposition $\hat{g}=\hat{U}\hat{g}_d{^t}\hat{V}$. The dependence of $\hat{g}_d$ on the electric field can practically be reconstructed from the measurement of the Zeeman splitting $\Delta E$ as a function of gate voltage and magnetic field orientation (see discussion in paragraph \ref{subsectionDiscussion}). When the electric field modulates only the principal $g$-factors, as is the case here, the Rabi oscillations can be interpreted as conventional $g$-TMR (that is, can be unequivocally related to the electrical tunability of the Zeeman splitting). Such modulations of $\hat{g}_d$ must result from changes in the shape of the wave function in a strongly anharmonic potential (as shown in Fig. \ref{figWfnHole}). Indeed, a mere translation of the wave function by the electric field in a quasi-harmonic potential can not give rise to variations of the Zeeman splitting in a homogeneous magnetic field.\cite{Golovach06} It may, however, restructure the phase of the wave function, as the vector potential breaks translational symmetry. The effects of such a motion appear in $\hat{U}^\prime$ and give rise to iso-Zeeman EDSR. We will discuss an example of iso-Zeeman EDSR in the next paragraph.

\subsection{Comparison with experimental data}

We now make a comparison with the experimental data of Ref. \onlinecite{Crippa18}. In that work, the orientational dependence of the $g$-factor and Rabi frequency of a SOI hole qubit have been completely characterized near a single bias point at zero back gate voltage. The geometry and dimensions of the experimental qubit are close to those of Fig. \ref{figDevice}, the main difference being that the gates overlap the channel on all three facets. Disorder is, however, expected to break the symmetry between the two sides of the wire, the hole being ultimately localized on the left or right of the channel.

The $x$ (nanowire) axis was found to be a principal magnetic axis, the two other magnetic axes being close (yet not parallel) to $y$ and $z$. This is consistent with the existence of a symmetry or quasi-symmetry plane perpendicular to $x$. However, the measured $g$ factors $g_x=2.48$, $g_y=2.08$ and $g_z=1.62$ do not match those calculated above. In particular, $g_z$ is the smallest experimental $g$-factor, while it is the largest $g$-factor in the calculations.

The map of Rabi frequency also looks very different. Experimentally, the Rabi oscillations are dominated by an iso-Zeeman EDSR contribution, and the Rabi frequency is maximal when $\vec{B}\parallel\vec{z}$. Theoretically, the Rabi oscillations are dominated by a $g$-TMR contribution, and the Rabi frequency is very small when $\vec{B}\parallel\vec{z}$.

\RED{These two discrepancies are in fact largely unrelated. First of all, Ref. \onlinecite{Crippa18} pointed out that the quantum dot was sitting between two gates ``G1'' and ``G2'', the lever arm parameters of the dot with respect to these gates being almost the same. Therefore, the excitation pattern was different from the one considered up to now: In a dot located between two gates, $\vec{E}_1$ indeed tends to align with the wire axis $\vec{x}$ instead of being perpendicular to it. This, as further discussed below, explains the mismatch between the Rabi frequency, but not between the $g$-factor anisotropies (since the $g$-matrix and $g$-factors do not depend on the RF electric field distribution -- only $\hat{g}^\prime$ does).}

\begin{figure}
\includegraphics[width=0.95\columnwidth]{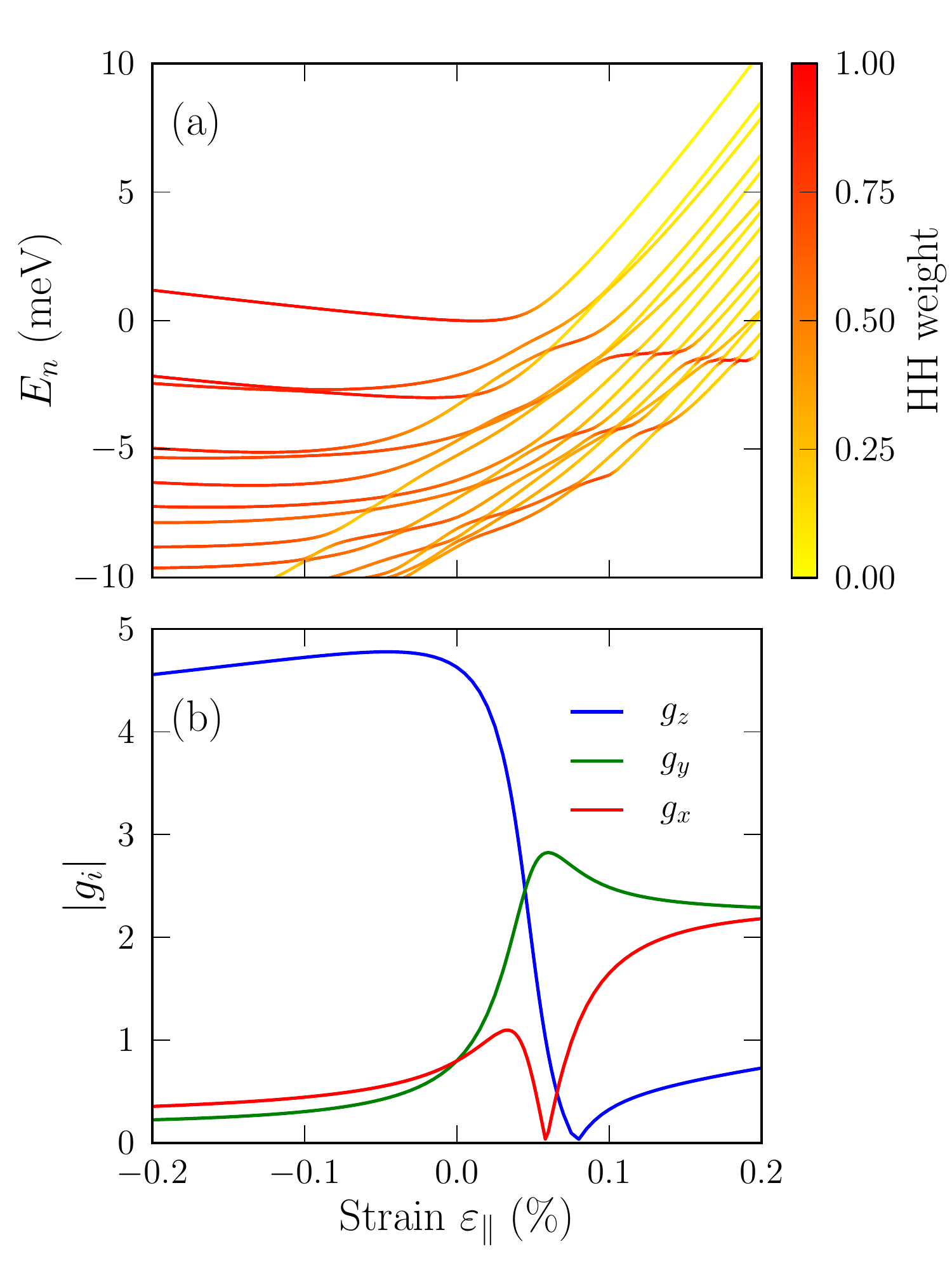}
\caption{\RED{(a) Energies and nature of the highest-lying hole states as a function of the biaxial strain $\varepsilon_{xx}=\varepsilon_{yy}=\varepsilon_{\parallel}$. The heavy-hole character is quantified by the color of the lines. The origin of energies is set to the highest valence band state at $\varepsilon_{\parallel}=0$. (b) Principal $g$-factors of the ground-state hole doublet as a function of the biaxial strain $\varepsilon_{\parallel}$.  $V_{\rm fg}=-0.1$ V and $V_{\rm bg}=-0.2$ V (blue point on Fig. \ref{figMapsRabi}).}}
\label{figBiaxial} 
\end{figure}

The measured $g$-factor anisotropy seems, in fact, more consistent with a mostly light-hole doublet. There was, in particular, a finite number of holes in the dot of Ref. \onlinecite{Crippa18} (estimated between 10 and 30), so that the $g$-factors and Rabi frequencies have been measured on a deeper-lying doublet. Some of the calculated excited states indeed show a strong light-hole character, yet their $g$-factors and Rabi frequency maps hardly match the experimental data (we did not, however, account for Coulomb correlations at this stage). Moreover, a similar $g$-factor anisotropy has been repeatedly measured in other devices with a different number of holes.\cite{Voisin16,Maurand16} We therefore look for a robust mechanism able to explain these observations.

\begin{figure}
\includegraphics[width=0.95\columnwidth]{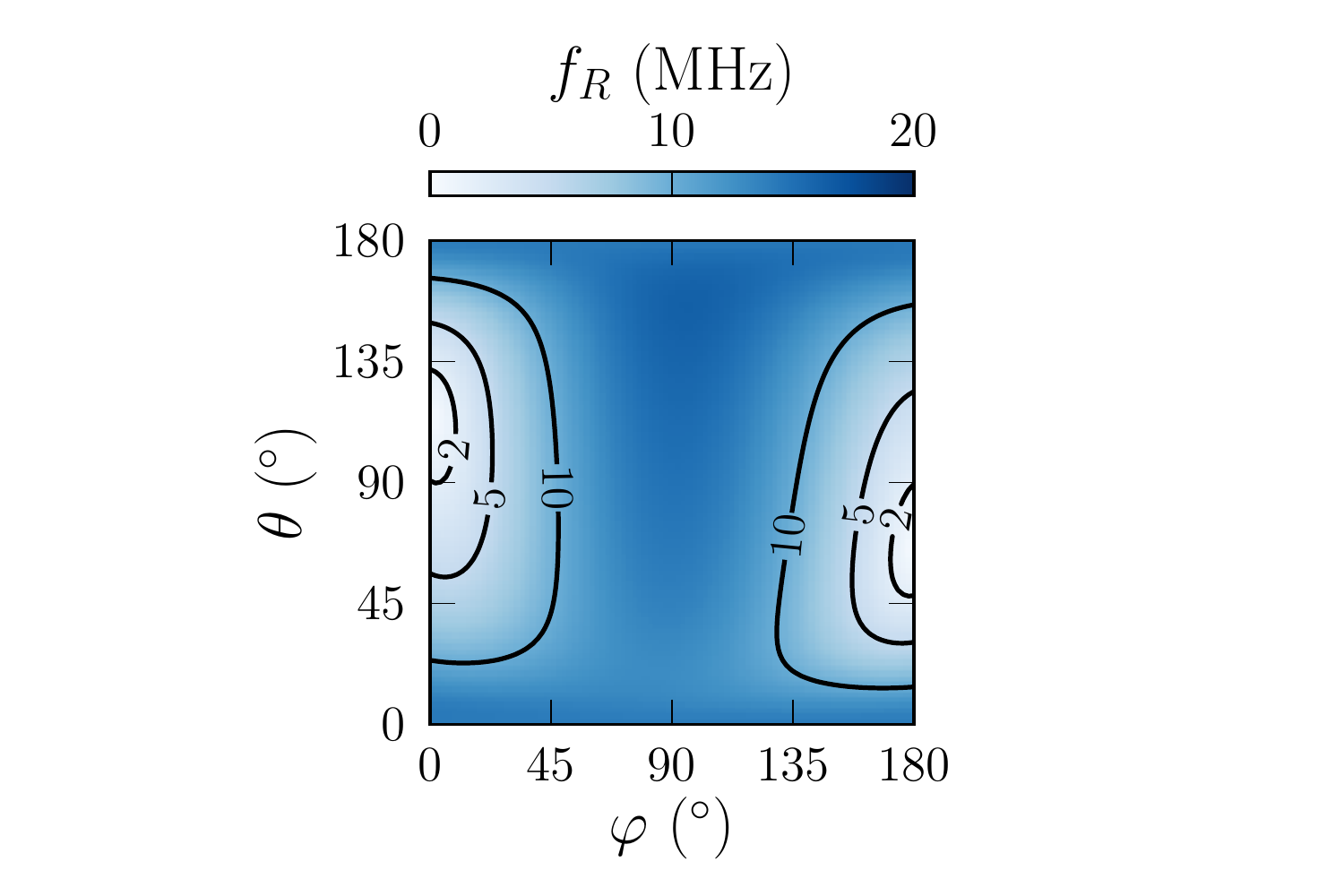}
\caption{Map of the Rabi frequency as a function of the polar and azimuthal angles of $\vec{B}$ defined in Fig. \ref{figDevice} ($V_{\rm ac}=1$ mV). This map is drawn for the ground-state doublet of a biaxially strained qubit ($\varepsilon_{xx}=\varepsilon_{yy}=0.2\%$, $\varepsilon_{zz}=-0.15\%$). The qubit is biased so that the dot lies between the central and leftmost gate of Fig. \ref{figDevice}: $V_{\rm fg}=0.15$ V, $V_{\rm left}=0.15$ V on the leftmost gate, $V_{\rm right}=0.2$ V on the rightmost gate, and $V_{\rm bg}=0$ V. The principal $g$-factors of this doublet are $g_x=2.06$, $g_y=2.41$ and $g_z=0.77$. The map is drawn at constant Zeeman splitting $\Delta E/h=9$ GHz instead of constant magnetic field, as done in Ref. \onlinecite{Crippa18}.}
\label{figMapRabiLH} 
\end{figure}

One possible scenario explaining the measured $g$-factor anisotropy is the introduction of non-intentional strains during processing and/or cooling down to cryogenic temperatures. In-plane tensile strains do, in particular, promote light-hole states at the top of the valence band. \RED{As an illustration, the energy levels and heavy-hole character of the highest-lying hole states is plotted as a function of an in-plane biaxial strain $\varepsilon_{\parallel}$ in Fig. \ref{figBiaxial}a. Namely, the strains within the channel are $\varepsilon_{xx}=\varepsilon_{yy}=\varepsilon_{\parallel}$ and $\varepsilon_{zz}=-2c_{12}\varepsilon_{\parallel}/c_{11}=-0.77\varepsilon_{\parallel}$, where $c_{11}=166$ GPa and $c_{12}=64$ GPa are the elastic constants of silicon. Note that such strains do not break any symmetry of the device (hence do not interfere with the previous discussion on the effects of mirror planes on the Rabi frequency). The character of the topmost hole doublet switches from mostly heavy to mostly light hole for strains as small as $\varepsilon_{\parallel}\simeq 0.1$\%. The $g$-factors of this doublet show correspondingly heavy-hole fingerprints ($g_z>g_x, g_y$) for $\varepsilon_{\parallel}<0$ and light-hole fingerprints ($g_z<g_x, g_y$, like in the experiment of Ref. \onlinecite{Crippa18}) for $\varepsilon_{\parallel}\gtrsim 0.1$\%. At $\varepsilon_{\parallel}\simeq 0.2$\%, more than 10 hole doublets have a dominant light-hole character at the top of the valence band.}

\RED{We plot in Fig. \ref{figMapRabiLH} the Rabi frequency map drawn from the ground-state hole doublet of a biaxially strained device ($\varepsilon_{\parallel}=0.2$\%) biased such that the quantum dot is located between the central and leftmost gate of Fig. \ref{figDevice}, as suggested by the experiment.} As hinted above, the Rabi oscillations become dominated by an iso-Zeeman EDSR contribution, since the RF electric field, mostly aligned along $x$ in this configuration, primarily drives an oscillation of the wavefunction as a whole along that axis. \RED{Therefore the principal $g$-factors $g_x$, $g_y$ and $g_z$ depend much less on gate voltage, unlike the matrix $\hat{U}$. The calculated $g$-factors and Rabi frequency map are in much better qualitative agreement with the experiment, yet do not reach quantitative accuracy, the Rabi frequency being, in particular, too small.} These remaining discrepancies likely result from the presence of other holes in the dot (smoothing the potential), and from the (largely unknown) surface roughness and charge disorder around the qubit.

While the existence of such strains remains to be confirmed, the present results show that qubits can be very responsive to stress.\cite{Mansir18,Pla18} Strain engineering, which has been very successful in classical MOS transistors,\cite{Lee05,Sun07} could therefore open new opportunities for semiconductor qubits as well. 

\section{Conclusions}

We have shown that the $g$-matrix formalism allows for a fast and efficient calculation of the angular maps of Larmor and Rabi frequencies of a qubit, in the linear-in-$B$-and-$V_{\rm ac}$ regime. The electronic structure of the qubit at zero magnetic field indeed embeds all the information relevant for linear response theory. The $g$-matrix formula also provides a ``compact'' model for the control of the qubit, and lends itself to simple symmetry analysis. We have, in particular, discussed the effects of mirror planes on the anisotropy of the $g$-factors and Rabi frequency. The $g$-matrix formalism can be supplemented with a perturbation series analysis that provides additional insights into the physics at the microscopic scale. These models apply to a large variety of electron and hole spin qubits. We have, as an example, considered a hole spin qubit in a silicon-on-insulator nanowire device with both front and back gates.\cite{Crippa18} We have highlighted the complex dependence of the Rabi frequency on the orientation of the magnetic field, which results from the symmetries of the wave functions. These symmetries can be controlled by the front and back gate, thanks to the non-planar front gate design of these devices. In particular, the qubit can be switched from a (low symmetry) bias point where it is strongly coupled to the RF electric field for manipulation, to a (high symmetry) bias point where it is mostly decoupled but less sensitive to electrical noise. The calculated $g$-factors remain, however, qualitatively different from recent experimental data. We suggest that these discrepancies may result from the build-up of non-intentional strains in the silicon quantum dot. This pinpoints the importance of strains in semiconductor qubits, and opens the way for strain engineering in these devices.

\begin{acknowledgments}
We thank Alessandro Crippa, Romain Maurand, Silvano de Franceschi and Louis Hutin for fruitful discussions. This work was supported by the European Union's Horizon 2020 research and innovation program under grant agreement No 688539 MOSQUITO. 
\end{acknowledgments}

\vspace{5cm}

\appendix

\section{Equivalence between the $g$-matrix formalism and the perturbation series}
\label{appendixEquivalence}

To prove the equivalence between Eqs. (\ref{eqRabipert}) and (\ref{eqRabig}), we choose $\vec{z}\parallel\vec{B}$ for a given (yet arbitrary) orientation of $\vec{B}$, and write down $\hat{g}(V_0)$ in the $\{\ket{\zero_0},\ket{\one_0}\}$ basis set. Since $\ket{\zero_0}$ and $\ket{\one_0}$ are eigenstates of $M_{1,z}$, $g_{13}=g_{23}=0$ in Eq. (\ref{eqgmatrix}), so that $\hat{g}\vec{b}=g_{33}\vec{z}$, and $g^*=g_{33}$. Therefore,
\begin{align}
f_R&=\frac{\mu_B B V_{\rm ac}}{2h}\left|\vec{v}\right|\text{ with }\vec{v}=(-g^\prime_{23}, g^\prime_{13}, 0) \nonumber \\
&=\frac{\mu_B B V_{\rm ac}}{2h}\left|g^\prime_{13}+ig^\prime_{23}\right| \nonumber \\
&=\frac{B V_{\rm ac}}{h}\left|\frac{\partial}{\partial V}\bra{\one_0}M_{1,z}\ket{\zero_0}\right|\,.
\label{eqRabieq}
\end{align}
The above derivative can easily be calculated from first-order perturbation theory in the operator $D_1$. Although $\ket{\zero_0}$ and $\ket{\one_0}$ are degenerate eigenstates of $H_0(V_0,\vec{B})$, they are not coupled by the electric field, so that non-degenerate perturbation theory applies for our purpose:
\begin{subequations}
\begin{align}
\frac{\partial}{\partial V}\ket{\zero_0}&=-e\sum_{n>0,\sigma}\frac{\bra{n,\sigma}D_1\ket{\zero_0}}{E_0-E_n}\ket{n,\sigma} \\
\frac{\partial}{\partial V}\ket{\one_0}&=-e\sum_{n>0,\sigma}\frac{\bra{n,\sigma}D_1\ket{\one_0}}{E_0-E_n}\ket{n,\sigma}\,.
\label{eqgpert}
\end{align}
\end{subequations}
We then get:
\begin{align}
\frac{\partial}{\partial V}\bra{\one_0}&M_{1,z}\ket{\zero_0}= \nonumber \\
=&-e\sum_{n>0,\sigma}\frac{\bra{\one_0}M_{1,z}\ket{n,\sigma}\bra{n,\sigma}D_1\ket{\zero_0}}{E_0-E_n} \nonumber \\
&-e\sum_{n>0,\sigma}\frac{\bra{\one_0}D_1\ket{n,\sigma}\bra{n,\sigma}M_{1,z}\ket{\zero_0}}{E_0-E_n}\,.
\label{eqderpsi}
\end{align}
Substitution into Eq. (\ref{eqRabieq}) yields back Eq. (\ref{eqRabipert}) as expected. Note that the wave functions have been developed to first order in the magnetic field in paragraph \ref{subsectionperturbation}, but to first order in the potential here, in order to reach the same first-order-in-$B$-and-$V_{\rm ac}$ formula for the Rabi frequency.

\section{Calculation of the $\hat{g}$ and $\hat{g}^\prime$ matrices}
\label{appendixCalculation}

\BLUE{In this appendix, we discuss the methodology for the numerical calculation of $\hat{g}^\prime$.}

\BLUE{The eigenstates $\{\ket{0,\Uparrow},\ket{0,\Downarrow}\}$ at $V=V_0$ must first be computed with the method of one's choice ($\vec{k}\cdot\vec{p}$, tight-binding...). The $g$-matrix $\hat{g}(V_0)$ in the basis set $\{\ket{0,\Uparrow},\ket{0,\Downarrow}\}$ is then easily obtained from Eqs. (\ref{eqgmatrix}). Doing so again at $V=V_0\pm\delta V$ yields $\hat{g}(V_0\pm\delta V)$ in some basis sets $\{\ket{0_\pm,\Uparrow},\ket{0_\pm,\Downarrow}\}$. $\hat{g}^\prime(V_0)$ may then be tentatively calculated from finite differences:
\begin{equation}
\hat{g}^\prime(V_0)=\frac{\hat{g}(V_0+\delta V)-\hat{g}(V_0-\delta V)}{2\delta V}\,.
\label{eqgprimefd}
\end{equation}
However, the above equation is usually meaningless as $\{\ket{0_\pm,\Uparrow},\ket{0_\pm,\Downarrow}\}$ may differ from $\{\ket{0,\Uparrow},\ket{0,\Downarrow}\}$ by an arbitrary unitary transformation $R_\pm$ on top of the expected gate-driven modulation, Eq. (\ref{eqderpsi}). This equation actually suggests that the appropriate $\ket{0_\pm,\sigma^\prime}$ must fulfill:
\begin{equation}
\begin{pmatrix}
\langle 0_\pm,\Uparrow^\prime|0,\Uparrow\rangle & \langle 0_\pm,\Uparrow^\prime|0,\Downarrow\rangle \\
\langle 0_\pm,\Downarrow^\prime|0,\Uparrow\rangle & \langle 0_\pm,\Downarrow^\prime|0,\Downarrow\rangle
\end{pmatrix}
=\alpha_\pm I\,,
\label{eqPtrans}
\end{equation}
where $I$ is the identity matrix and $0<\alpha_\pm\le1$. We therefore seek the unitary transformation $P_\pm\equiv R_\pm^\dagger$ in the $\{\ket{0_\pm,\Uparrow},\ket{0_\pm,\Downarrow}\}$ subspace such that the states $\ket{0_\pm,\sigma^\prime}=P_\pm\ket{0_\pm,\sigma}$ satisfy the above relations. Solving this problem for Kramers degenerate states uniquely defines $P_\pm$:
\begin{equation}
P_\pm=\beta_\pm
\begin{pmatrix}
\langle 0_\pm,\Uparrow|0,\Uparrow\rangle & \langle 0_\pm,\Downarrow|0,\Uparrow\rangle \\
\langle 0_\pm,\Uparrow|0,\Downarrow\rangle & \langle 0_\pm,\Downarrow|0,\Downarrow\rangle
\end{pmatrix}\,,
\end{equation}
where:
\begin{subequations}
\begin{align}
\beta_\pm^{-2}&=\left|\langle 0_\pm,\Uparrow|0,\Uparrow\rangle\right|^2+\left|\langle 0_\pm,\Downarrow|0,\Uparrow\rangle\right|^2 \\
&=\left|\langle 0_\pm,\Uparrow|0,\Downarrow\rangle\right|^2+\left|\langle 0_\pm,\Downarrow|0,\Downarrow\rangle\right|^2\,. 
\end{align}
\end{subequations}
Finite differences [Eq. (\ref{eqgprimefd})] can then be safely calculated in the basis sets $\{\ket{0_\pm,\Uparrow^\prime},\ket{0_\pm,\Downarrow^\prime}\}$:
\begin{subequations}
\begin{align}
\ket{0_\pm,\Uparrow^\prime}&=\beta_\pm\left(\langle 0_\pm,\Uparrow|0,\Uparrow\rangle\ket{0_\pm,\Uparrow}+\langle 0_\pm,\Downarrow|0,\Uparrow\rangle\ket{0_\pm,\Downarrow}\right)  \\
\ket{0_\pm,\Downarrow^\prime}&=\beta_\pm\left(\langle 0_\pm,\Uparrow|0,\Downarrow\rangle\ket{0_\pm,\Uparrow}+\langle 0_\pm,\Downarrow|0,\Downarrow\rangle\ket{0_\pm,\Downarrow}\right)\,.
\end{align}
\end{subequations}
We practically use $\delta V=1$ mV in section \ref{sectionSOI}.}

\section{Group theory for the $\hat{g}$ and $\hat{g}^\prime$ matrices}
\label{appendixGroupTheory}

In this appendix, we give some details about the application of group theory to the $\hat{g}$ and $\hat{g}^\prime$ matrices.\cite{Cornwell97}

For any operation $R$ that leaves the system invariant, and for any states $\ket{\varphi_1}$ and $\ket{\varphi_2}$,
\begin{align}
\bra{\varphi_1}&H(V,\vec{B})\ket{\varphi_2}= \nonumber \\
&\bra{\Gamma_S(R)\varphi_1}H(V,\hat{\Gamma}_B(R)\vec{B})\ket{\Gamma_S(R)\varphi_2}\,,
\end{align}
where $\Gamma_S(R)$ is the representation of the symmetry group onto the Hilbert space of the qubit and $\hat{\Gamma}_B(R)$ is the representation of the symmetry group onto the magnetic field pseudo-vector space. This implies:
\begin{equation}
H(V,\vec{B})=\Gamma_S^\dagger(R)H(V,\hat{\Gamma}_B(R)\vec{B})\Gamma_S(R)\,.
\label{eqgroup}
\end{equation}
The matrices $\hat{\Gamma}_B(R)$ in the $\{\vec{x},\vec{y},\vec{z}\}$ axes are given in Table \ref{table_repB} for the $\sigma_{yz}$, $\sigma_{xz}$ and $\sigma_{xy}$ operations. With a suitable choice of basis set $\{\ket{0,\Uparrow},\ket{0,\Downarrow}\}$ (which are defined up to an unitary transformation), the matrices $\Gamma_S(R)$ take the form given\cite{Onodera66} in Table \ref{table_repH}.\footnote{The double group $D_{2h}$ has two possible irreducible representations for a Kramers doublet that differ by the sign of $\Gamma_S(\sigma_{xy})$. The conclusions are, therefore, the same whether the Kramers doublet belongs to one or the other irreducible representation.}

\begin{center}
\begin{table}
\begin{tabular}{|c|c|c|c|} \hline
$\sigma_{\alpha\beta}$ & $\sigma_{yz}$ & $\sigma_{xz}$ & $\sigma_{xy}$  \\ \hline
$\Gamma_B(\sigma_{\alpha\beta})$  & 
$\begin{pmatrix}
1 & 0 & 0 \\
0 & -1 & 0 \\
0 & 0 & -1
\end{pmatrix}$ &
$\begin{pmatrix}
-1 & 0 & 0 \\
0 & 1 & 0 \\
0 & 0 & -1
\end{pmatrix}$ &
$\begin{pmatrix}
-1 & 0 & 0 \\
0 & -1 & 0 \\
0 & 0 & 1
\end{pmatrix}$ \\ \hline
\end{tabular}
\caption{Representation $\hat{\Gamma}_B(\sigma_{\alpha\beta})$ in the $\{\vec{x},\vec{y},\vec{z}\}$ axes.}
\label{table_repB}
\end{table}
\end{center}

\begin{table}
\begin{tabular}{|c|c|c|c|} \hline
$\sigma_{\alpha\beta}$ & $\sigma_{yz}$ & $\sigma_{xz}$  & $\sigma_{xy}$ \\ \hline
$\Gamma_S(\sigma_{\alpha\beta})$ & 
$\begin{pmatrix}
0 & -i \\
-i & 0
\end{pmatrix}$ & 
$\begin{pmatrix}
0 & -1 \\
1 & 0
\end{pmatrix}$ & 
$\begin{pmatrix}
-i & 0 \\
0 & i
\end{pmatrix}$ \\ \hline
\end{tabular}
\caption{Representation $\Gamma_S(\sigma_{\alpha\beta})$.}
\label{table_repH}
\end{table}

Using the $g$-matrix form of the Hamiltonian, Eq. (\ref{eqgtensor}), as input for Eq. (\ref{eqgroup}) yields the relation
\begin{equation}
\vec{\sigma}\cdot\hat{g}(V_0)\vec{B}=\left[\Gamma_S^\dagger(R)\vec{\sigma}\Gamma_S(R)\right]\cdot\hat{g} (V_0)\hat{\Gamma}_B(R)\vec{B}
\label{eqgroupg}
\end{equation}
for any symmetry operation $R$ and magnetic field $\vec{B}$, where it is understood that $\Gamma_S^\dagger\vec{\sigma}\Gamma_S=(\Gamma_S^\dagger\sigma_1\Gamma_S, \Gamma_S^\dagger\sigma_2\Gamma_S, \Gamma_S^\dagger\sigma_3\Gamma_S)$. Expanding this relation for the mirror planes $\sigma_{yz}$, $\sigma_{xz}$ and $\sigma_{xy}$ finally provides Tables \ref{table_r} and \ref{table_g}. 

In order to establish the conditions on $\hat{g}^\prime(V_0)$, we must generalize Eqs. (\ref{eqH}) and (\ref{eqpotential}) and consider the Hamiltonian as a function of $\vec{B}$ and as a functional of the total potential $V_t(V,\vec{r})$ in the device. Then,
\begin{align}
\hat{g}^\prime(V_0)=\left.\frac{\partial\hat{g}}{\partial V}\right|_{V=V_0}&=\int d^3\vec{r}\,\frac{\delta\hat{g}}{\delta V_t(\vec{r})}\left.\frac{\partial V_t(V, \vec{r})}{\partial V}\right|_{V=V_0} \nonumber \\
&=\int d^3\vec{r}\,\frac{\delta\hat{g}}{\delta V_t(\vec{r})}D_1(\vec{r})\,,
\label{eqgprimefnd}
\end{align}
where the functional derivative $\delta\hat{g}/\delta V_t(\vec{r})$ is evaluated at potential $V_t(V_0, \vec{r})$. Inserting Eq. (\ref{eqgprimefnd}) into Eqs. (\ref{eqgprime}) and (\ref{eqgroup}) therefore yields:
\begin{align}
&\vec{\sigma}\cdot\left[\int d^3\vec{r}\,\frac{\delta\hat{g}}{\delta V_t(\vec{r})}D_1(\vec{r})\right]\vec{B}= \nonumber \\
&\left[\Gamma_S^\dagger(R)\vec{\sigma}\Gamma_S(R)\right]\cdot\left[\int d^3\vec{r}\,\frac{\delta\hat{g}}{\delta V_t(\vec{r})}D_1(\hat{\Gamma}_R(R)\vec{r})\right]\hat{\Gamma}_B(R)\vec{B}\,,
\label{eqgroupgprime}
\end{align}
where $\hat{\Gamma}_R(R)$ is a representation of the symmetry group in real space (defined by the standard rotation matrices). This relation sets non-trivial conditions on $\hat{g}^\prime(V_0)$ in at least two cases:
\begin{enumerate}
 \item $D_1(\vec{r})=D_1(\hat{\Gamma}_R(R)\vec{r})$ -- namely, $D_1$ is also invariant by the symmetry operation $R$. Eq. (\ref{eqgroupgprime}) is the same as Eq. (\ref{eqgroupg}) with $\hat{g}$ replaced by $\hat{g}^\prime$, so that $R$ sets the same conditions on $\hat{g}(V_0)$ and $\hat{g}^\prime(V_0)$ (first row of Table \ref{table_r2}). Taking the gradient on both sides, $\vec{E}_1(\vec{r})=\hat{\Gamma}_R(R)\vec{E}_1(\hat{\Gamma}_R(R)\vec{r})$ must be even under the transformation $R$.
 \item $D_1(\vec{r})=-D_1(\hat{\Gamma}_R(R)\vec{r})$. Then Eq. (\ref{eqgroupgprime}) imposes that $\hat{g}^\prime(V_0)$ takes the form given in the second row of Table \ref{table_r2}. As the $\hat{g}$ matrix must not change if $D_1(\vec{r})$ is a constant (which translates into $\int d^3\vec{r}\,\delta\hat{g}/\delta V_t(\vec{r})=0$), the above condition can be generalized as $D_1(\vec{r})+D_1(\hat{\Gamma}_R(R)\vec{r})=K$ ($K$ independent on $\vec{r}$). Hence, $\vec{E}_1(\vec{r})=-\hat{\Gamma}_R(R)\vec{E}_1(\hat{\Gamma}_R(R)\vec{r})$ must be odd under the transformation $R$.
\end{enumerate}
These conditions actually result from the interplay between the parities of the magnetic field Hamiltonian (defined in Table \ref{table_repB}) and the parities of the electric field Hamiltonian (defined above). When these parities are incompatible, the contributions from each pair of excited states sum up to zero in Eq. (\ref{eqRabipert}), leaving no connection between the qubit states $\ket{\zero}$ and $\ket{\one}$.

\section{Details about numerical simulations}
\label{appendixNumerical}

\BLUE{The potential in the SOI device of section \ref{sectionSOI} is computed with a finite volume Poisson solver assuming dielectric constants $\varepsilon_{\rm Si}=11.7$, $\varepsilon_{\rm SiO_2}=3.9$, $\varepsilon_{\rm HfO_2}=20$ and $\varepsilon_{\rm Si_3N_4}=7.5$. The wave functions of the qubit in this potential are then computed with a six-bands $\vec{k}\cdot\vec{p}$ model.}

\BLUE{In bulk silicon, the six-bands $\vec{k}\cdot\vec{p}$ Hamiltonian\cite{Dresselhaus55} reads in the $\{\ket{\frac{3}{2},+\frac{3}{2}},\ket{\frac{3}{2},+\frac{1}{2}},\ket{\frac{3}{2},-\frac{1}{2}},\ket{\frac{3}{2},-\frac{3}{2}},\ket{\frac{1}{2},+\frac{1}{2}},\ket{\frac{1}{2},-\frac{1}{2}}\}$ Bloch functions basis set:\cite{KP09}
\begin{widetext}
\begin{equation}
H_{\rm 6kp}=-
\begin{pmatrix}
P+Q & -S & R & 0 & \frac{1}{\sqrt{2}}S & -\sqrt{2}R \\
-S^* & P-Q & 0 & R & \sqrt{2}Q & -\sqrt{\frac{3}{2}}S & \\
R^* & 0 & P-Q & S & -\sqrt{\frac{3}{2}}S^* & -\sqrt{2}Q \\
0 & R^* & S^* & P+Q & \sqrt{2}R^* & \frac{1}{\sqrt{2}}S^* \\
\frac{1}{\sqrt{2}}S^* & \sqrt{2}Q & -\sqrt{\frac{3}{2}}S & \sqrt{2}R & P+\Delta & 0 \\
-\sqrt{2}R^* & -\sqrt{\frac{3}{2}}S^* & -\sqrt{2}Q & \frac{1}{\sqrt{2}}S & 0 & P+\Delta
\end{pmatrix}
\label{eq6kp}
\end{equation}
\end{widetext}
where:
\begin{subequations}
\begin{align}
P&=\frac{\hbar^2}{2m_0}\gamma_1\left(k_x^2+k_y^2+k_z^2\right) \\
Q&=\frac{\hbar^2}{2m_0}\gamma_2\left(k_x^2+k_y^2-2k_z^2\right) \\
R&=\frac{\hbar^2}{2m_0}\sqrt{3}\left[-\gamma_3\left(k_x^2-k_y^2\right)+2i\gamma_2k_xk_y\right] \label{eqR} \\
S&=\frac{\hbar^2}{2m_0}2\sqrt{3}\gamma_3\left(k_x-ik_y\right)k_z \label{eqS} \,.
\end{align}
\end{subequations}
$k_x$, $k_y$ and $k_z$ are the components of the wave vector in the device axis set (defined on Fig. \ref{figDevice}), $\gamma_1$, $\gamma_2$ and $\gamma_3$ are the Luttinger parameters, $m_0$ is the free electron mass, and $\Delta$ is the spin-orbit coupling parameter. In silicon, $\gamma_1=4.285$, $\gamma_2=0.339$, $\gamma_3=1.446$ and $\Delta=44$ meV.}

\BLUE{The equations for the envelope functions obtained after the substitution $\vec{k}\to-i\vec{\nabla}$ are discretized on a finite differences mesh. Periodic boundary conditions are applied along the wire axis $x$, although the dots are effectively decoupled by the lateral gates on Fig. \ref{figDevice}. Hard wall boundary conditions are applied at the surface of the wire (the wave functions do not penetrate in the oxides and Si$_3$N$_4$). The effect of the potential vector $\vec{A}$ on the envelope functions is included through Peierl's substitution as done in the tight-binding approximation.\cite{Vogl95} The effect of the magnetic field on the Bloch functions and spin is described by the following Hamiltonian:\cite{Luttinger56}
\begin{equation}
H_{\rm Bloch}=-(3\kappa+1)\mu_B\vec{B}\cdot\vec{L}+g_0\mu_B\vec{B}\cdot\vec{S}=\mu_B\vec{B}\cdot\vec{K}\,,
\label{eqHbloch}
\end{equation}
where $\vec{L}$ is the (orbital) angular momentum of the Bloch function, $\vec{S}$ its spin, and $\kappa=-0.42$ in silicon. We neglect the effects of the much smaller $\propto q$ term of Ref. \onlinecite{Luttinger56}. For completeness, we give below the expression of the matrices $K_x$, $K_y$, $K_z$ consistent with our choice of phases for the Bloch functions:\footnote{We assume $g_0=2$ in Eq. (\ref{eqHbloch}).}
\begin{subequations}
\begin{equation}
K_x=-
\begin{pmatrix}
0 & \sqrt{3}\kappa & 0 & 0 & -\sqrt{\frac{3}{2}}\kappa^\prime & 0 \\
\sqrt{3}\kappa  & 0 & 2\kappa & 0 & 0 & -\frac{\kappa^\prime}{\sqrt{2}} \\
0 & 2\kappa & 0 & \sqrt{3}\kappa & \frac{\kappa^\prime}{\sqrt{2}} & 0 \\
0 & 0 & \sqrt{3}\kappa & 0 & 0 & \sqrt{\frac{3}{2}}\kappa^\prime \\
-\sqrt{\frac{3}{2}}\kappa^\prime & 0 & \frac{\kappa^\prime}{\sqrt{2}} & 0 & 0 & \kappa^{\prime\prime} \\
0 & -\frac{\kappa^\prime}{\sqrt{2}} & 0 & \sqrt{\frac{3}{2}}\kappa^\prime & \kappa^{\prime\prime} & 0
\end{pmatrix} 
\end{equation}
\begin{equation}
K_y=i
\begin{pmatrix}
0 & \sqrt{3}\kappa & 0 & 0 & -\sqrt{\frac{3}{2}}\kappa^\prime & 0 \\
-\sqrt{3}\kappa  & 0 & 2\kappa & 0 & 0 & -\frac{\kappa^\prime}{\sqrt{2}} \\
0 & -2\kappa & 0 & \sqrt{3}\kappa & -\frac{\kappa^\prime}{\sqrt{2}} & 0 \\
0 & 0 & -\sqrt{3}\kappa & 0 & 0 & -\sqrt{\frac{3}{2}}\kappa^\prime \\
\sqrt{\frac{3}{2}}\kappa^\prime & 0 & \frac{\kappa^\prime}{\sqrt{2}} & 0 & 0 & \kappa^{\prime\prime} \\
0 & \frac{\kappa^\prime}{\sqrt{2}} & 0 & \sqrt{\frac{3}{2}}\kappa^\prime & -\kappa^{\prime\prime} & 0
\end{pmatrix}
\end{equation}
\begin{equation}
K_z=-
\begin{pmatrix}
3\kappa & 0 & 0 & 0 & 0 & 0 \\
0 & \kappa & 0 & 0 & \sqrt{2}\kappa^\prime & 0 \\
0 & 0 & -\kappa & 0 & 0 & \sqrt{2}\kappa^\prime \\
0 & 0 & 0 & -3\kappa & 0 & 0 \\
0 & \sqrt{2}\kappa^\prime & 0 & 0 & \kappa^{\prime\prime}  & 0 \\
0 & 0 & \sqrt{2}\kappa^\prime & 0 & 0 & -\kappa^{\prime\prime}
\end{pmatrix}\,,
\end{equation}
\end{subequations}
with $\kappa^\prime=1+\kappa$ and $\kappa^{\prime\prime}=1+2\kappa$. Note that in the $J=3/2$ subspace (the top left $4\times4$ sub-blocks of $K_x$, $K_y$ and $K_z$), $H_{\rm Bloch}$ is formally equivalent to $-2\kappa\mu_B\vec{B}\cdot\vec{J}$, where $\vec{J}=\vec{L}+\vec{S}$ is the total angular momentum of the Bloch function.\cite{Luttinger56} The eigenstates $\ket{0,\sigma}$ at $V=V_0$ are computed with an iterative Jacobi-Davidson eigensolver.\cite{Sleijpen00,Templates00}}

\BLUE{The operator $\vec{M}_1=-\vec{\nabla}_{\vec{B}}H(V,\vec{B})|_{\vec{B}=\vec{0}}$ is not constructed explicitly. The matrix elements of $M_{1,x}$ for example are evaluated from finite differences as:
\begin{align}
\bra{\varphi_1}M_{1,x}\ket{\varphi_2}=-\frac{1}{2\delta B}&\left[\bra{\varphi_1}H(V,+\delta B\vec{x})\ket{\varphi_2}\right. \nonumber \\
-&\left.\bra{\varphi_1}H(V,-\delta B\vec{x})\ket{\varphi_2}\right]\,,
\label{eqM1}
\end{align}
where $\delta B=0.1$ mT. The value of $V$ is irrelevant as $\vec{M}_1$ is independent on the bias voltage in the present approximation for the electro-magnetic Hamiltonian.}

\BLUE{If the electrostatics is linear (which is the case in section \ref{sectionSOI}), then $D_1(\vec{r})=\partial V_t(V, \vec{r})/\partial V|_{V=V_0}$ is simply the potential created by a unit voltage on the control gate with all other terminals (if any) grounded. $D_1(\vec{r})$ is, notably, also independent on $V_0$ in that case.}


\begin{thebibliography}{64}%
\makeatletter
\providecommand \@ifxundefined [1]{%
 \@ifx{#1\undefined}
}%
\providecommand \@ifnum [1]{%
 \ifnum #1\expandafter \@firstoftwo
 \else \expandafter \@secondoftwo
 \fi
}%
\providecommand \@ifx [1]{%
 \ifx #1\expandafter \@firstoftwo
 \else \expandafter \@secondoftwo
 \fi
}%
\providecommand \natexlab [1]{#1}%
\providecommand \enquote  [1]{``#1''}%
\providecommand \bibnamefont  [1]{#1}%
\providecommand \bibfnamefont [1]{#1}%
\providecommand \citenamefont [1]{#1}%
\providecommand \href@noop [0]{\@secondoftwo}%
\providecommand \href [0]{\begingroup \@sanitize@url \@href}%
\providecommand \@href[1]{\@@startlink{#1}\@@href}%
\providecommand \@@href[1]{\endgroup#1\@@endlink}%
\providecommand \@sanitize@url [0]{\catcode `\\12\catcode `\$12\catcode
  `\&12\catcode `\#12\catcode `\^12\catcode `\_12\catcode `\%12\relax}%
\providecommand \@@startlink[1]{}%
\providecommand \@@endlink[0]{}%
\providecommand \url  [0]{\begingroup\@sanitize@url \@url }%
\providecommand \@url [1]{\endgroup\@href {#1}{\urlprefix }}%
\providecommand \urlprefix  [0]{URL }%
\providecommand \Eprint [0]{\href }%
\providecommand \doibase [0]{http://dx.doi.org/}%
\providecommand \selectlanguage [0]{\@gobble}%
\providecommand \bibinfo  [0]{\@secondoftwo}%
\providecommand \bibfield  [0]{\@secondoftwo}%
\providecommand \translation [1]{[#1]}%
\providecommand \BibitemOpen [0]{}%
\providecommand \bibitemStop [0]{}%
\providecommand \bibitemNoStop [0]{.\EOS\space}%
\providecommand \EOS [0]{\spacefactor3000\relax}%
\providecommand \BibitemShut  [1]{\csname bibitem#1\endcsname}%
\let\auto@bib@innerbib\@empty
%</preamble>
\bibitem [{\citenamefont {Kane}(1998)}]{Kane98}%
  \BibitemOpen
  \bibfield  {author} {\bibinfo {author} {\bibfnamefont {B.~E.}\ \bibnamefont
  {Kane}},\ }\href {\doibase 10.1038/30156} {\bibfield  {journal} {\bibinfo
  {journal} {Nature}\ }\textbf {\bibinfo {volume} {393}},\ \bibinfo {pages}
  {133} (\bibinfo {year} {1998})}\BibitemShut {NoStop}%
\bibitem [{\citenamefont {Loss}\ and\ \citenamefont
  {DiVincenzo}(1998)}]{Loss98}%
  \BibitemOpen
  \bibfield  {author} {\bibinfo {author} {\bibfnamefont {D.}~\bibnamefont
  {Loss}}\ and\ \bibinfo {author} {\bibfnamefont {D.~P.}\ \bibnamefont
  {DiVincenzo}},\ }\href {\doibase 10.1103/PhysRevA.57.120} {\bibfield
  {journal} {\bibinfo  {journal} {Physical Review A}\ }\textbf {\bibinfo
  {volume} {57}},\ \bibinfo {pages} {120} (\bibinfo {year} {1998})}\BibitemShut
  {NoStop}%
\bibitem [{\citenamefont {Petta}\ \emph {et~al.}(2005)\citenamefont {Petta},
  \citenamefont {Johnson}, \citenamefont {Taylor}, \citenamefont {Laird},
  \citenamefont {Yacoby}, \citenamefont {Lukin}, \citenamefont {Marcus},
  \citenamefont {Hanson},\ and\ \citenamefont {Gossard}}]{Petta05}%
  \BibitemOpen
  \bibfield  {author} {\bibinfo {author} {\bibfnamefont {J.~R.}\ \bibnamefont
  {Petta}}, \bibinfo {author} {\bibfnamefont {A.~C.}\ \bibnamefont {Johnson}},
  \bibinfo {author} {\bibfnamefont {J.~M.}\ \bibnamefont {Taylor}}, \bibinfo
  {author} {\bibfnamefont {E.~A.}\ \bibnamefont {Laird}}, \bibinfo {author}
  {\bibfnamefont {A.}~\bibnamefont {Yacoby}}, \bibinfo {author} {\bibfnamefont
  {M.~D.}\ \bibnamefont {Lukin}}, \bibinfo {author} {\bibfnamefont {C.~M.}\
  \bibnamefont {Marcus}}, \bibinfo {author} {\bibfnamefont {M.~P.}\
  \bibnamefont {Hanson}}, \ and\ \bibinfo {author} {\bibfnamefont {A.~C.}\
  \bibnamefont {Gossard}},\ }\href {\doibase 10.1126/science.1116955}
  {\bibfield  {journal} {\bibinfo  {journal} {Science}\ }\textbf {\bibinfo
  {volume} {309}},\ \bibinfo {pages} {2180} (\bibinfo {year}
  {2005})}\BibitemShut {NoStop}%
\bibitem [{\citenamefont {Koppens}\ \emph {et~al.}(2006)\citenamefont
  {Koppens}, \citenamefont {Buizert}, \citenamefont {Tielrooij}, \citenamefont
  {Vink}, \citenamefont {Nowack}, \citenamefont {Meunier}, \citenamefont
  {Kouwenhoven},\ and\ \citenamefont {Vandersypen}}]{Koppens06}%
  \BibitemOpen
  \bibfield  {author} {\bibinfo {author} {\bibfnamefont {F.~H.~L.}\
  \bibnamefont {Koppens}}, \bibinfo {author} {\bibfnamefont {C.}~\bibnamefont
  {Buizert}}, \bibinfo {author} {\bibfnamefont {K.~J.}\ \bibnamefont
  {Tielrooij}}, \bibinfo {author} {\bibfnamefont {I.~T.}\ \bibnamefont {Vink}},
  \bibinfo {author} {\bibfnamefont {K.~C.}\ \bibnamefont {Nowack}}, \bibinfo
  {author} {\bibfnamefont {T.}~\bibnamefont {Meunier}}, \bibinfo {author}
  {\bibfnamefont {L.~P.}\ \bibnamefont {Kouwenhoven}}, \ and\ \bibinfo {author}
  {\bibfnamefont {L.~M.~K.}\ \bibnamefont {Vandersypen}},\ }\href {\doibase
  10.1038/nature05065} {\bibfield  {journal} {\bibinfo  {journal} {Nature}\
  }\textbf {\bibinfo {volume} {442}},\ \bibinfo {pages} {766} (\bibinfo {year}
  {2006})}\BibitemShut {NoStop}%
\bibitem [{\citenamefont {Pla}\ \emph {et~al.}(2012)\citenamefont {Pla},
  \citenamefont {Tan}, \citenamefont {Dehollain}, \citenamefont {Lim},
  \citenamefont {Morton}, \citenamefont {Jamieson}, \citenamefont {Dzurak},\
  and\ \citenamefont {Morello}}]{Pla12}%
  \BibitemOpen
  \bibfield  {author} {\bibinfo {author} {\bibfnamefont {J.~J.}\ \bibnamefont
  {Pla}}, \bibinfo {author} {\bibfnamefont {K.~Y.}\ \bibnamefont {Tan}},
  \bibinfo {author} {\bibfnamefont {J.~P.}\ \bibnamefont {Dehollain}}, \bibinfo
  {author} {\bibfnamefont {W.~H.}\ \bibnamefont {Lim}}, \bibinfo {author}
  {\bibfnamefont {J.~J.~L.}\ \bibnamefont {Morton}}, \bibinfo {author}
  {\bibfnamefont {D.~N.}\ \bibnamefont {Jamieson}}, \bibinfo {author}
  {\bibfnamefont {A.~S.}\ \bibnamefont {Dzurak}}, \ and\ \bibinfo {author}
  {\bibfnamefont {A.}~\bibnamefont {Morello}},\ }\href {\doibase
  10.1038/nature11449} {\bibfield  {journal} {\bibinfo  {journal} {Nature}\
  }\textbf {\bibinfo {volume} {489}},\ \bibinfo {pages} {541} (\bibinfo {year}
  {2012})}\BibitemShut {NoStop}%
\bibitem [{\citenamefont {Hanson}\ \emph {et~al.}(2007)\citenamefont {Hanson},
  \citenamefont {Kouwenhoven}, \citenamefont {Petta}, \citenamefont {Tarucha},\
  and\ \citenamefont {Vandersypen}}]{Hanson07}%
  \BibitemOpen
  \bibfield  {author} {\bibinfo {author} {\bibfnamefont {R.}~\bibnamefont
  {Hanson}}, \bibinfo {author} {\bibfnamefont {L.~P.}\ \bibnamefont
  {Kouwenhoven}}, \bibinfo {author} {\bibfnamefont {J.~R.}\ \bibnamefont
  {Petta}}, \bibinfo {author} {\bibfnamefont {S.}~\bibnamefont {Tarucha}}, \
  and\ \bibinfo {author} {\bibfnamefont {L.~M.~K.}\ \bibnamefont
  {Vandersypen}},\ }\href {\doibase 10.1103/RevModPhys.79.1217} {\bibfield
  {journal} {\bibinfo  {journal} {Review of Modern Physics}\ }\textbf {\bibinfo
  {volume} {79}},\ \bibinfo {pages} {1217} (\bibinfo {year}
  {2007})}\BibitemShut {NoStop}%
\bibitem [{\citenamefont {Zwanenburg}\ \emph {et~al.}(2013)\citenamefont
  {Zwanenburg}, \citenamefont {Dzurak}, \citenamefont {Morello}, \citenamefont
  {Simmons}, \citenamefont {Hollenberg}, \citenamefont {Klimeck}, \citenamefont
  {Rogge}, \citenamefont {Coppersmith},\ and\ \citenamefont
  {Eriksson}}]{Zwanenburg13}%
  \BibitemOpen
  \bibfield  {author} {\bibinfo {author} {\bibfnamefont {F.~A.}\ \bibnamefont
  {Zwanenburg}}, \bibinfo {author} {\bibfnamefont {A.~S.}\ \bibnamefont
  {Dzurak}}, \bibinfo {author} {\bibfnamefont {A.}~\bibnamefont {Morello}},
  \bibinfo {author} {\bibfnamefont {M.~Y.}\ \bibnamefont {Simmons}}, \bibinfo
  {author} {\bibfnamefont {L.~C.~L.}\ \bibnamefont {Hollenberg}}, \bibinfo
  {author} {\bibfnamefont {G.}~\bibnamefont {Klimeck}}, \bibinfo {author}
  {\bibfnamefont {S.}~\bibnamefont {Rogge}}, \bibinfo {author} {\bibfnamefont
  {S.~N.}\ \bibnamefont {Coppersmith}}, \ and\ \bibinfo {author} {\bibfnamefont
  {M.~A.}\ \bibnamefont {Eriksson}},\ }\href {\doibase
  10.1103/RevModPhys.85.961} {\bibfield  {journal} {\bibinfo  {journal} {Review
  of Modern Physics}\ }\textbf {\bibinfo {volume} {85}},\ \bibinfo {pages}
  {961} (\bibinfo {year} {2013})}\BibitemShut {NoStop}%
\bibitem [{\citenamefont {Tyryshkin}\ \emph {et~al.}(2012)\citenamefont
  {Tyryshkin}, \citenamefont {Tojo}, \citenamefont {Morton}, \citenamefont
  {Riemann}, \citenamefont {Abrosimov}, \citenamefont {Becker}, \citenamefont
  {J.}, \citenamefont {Schenkel}, \citenamefont {Thewalt}, \citenamefont
  {Itoh},\ and\ \citenamefont {Lyon}}]{Tyryshkin12}%
  \BibitemOpen
  \bibfield  {author} {\bibinfo {author} {\bibfnamefont {A.~M.}\ \bibnamefont
  {Tyryshkin}}, \bibinfo {author} {\bibfnamefont {S.}~\bibnamefont {Tojo}},
  \bibinfo {author} {\bibfnamefont {J.~J.~L.}\ \bibnamefont {Morton}}, \bibinfo
  {author} {\bibfnamefont {H.}~\bibnamefont {Riemann}}, \bibinfo {author}
  {\bibfnamefont {N.~V.}\ \bibnamefont {Abrosimov}}, \bibinfo {author}
  {\bibfnamefont {P.}~\bibnamefont {Becker}}, \bibinfo {author} {\bibfnamefont
  {P.~H.}\ \bibnamefont {J.}}, \bibinfo {author} {\bibfnamefont
  {T.}~\bibnamefont {Schenkel}}, \bibinfo {author} {\bibfnamefont {M.~L.~W.}\
  \bibnamefont {Thewalt}}, \bibinfo {author} {\bibfnamefont {K.~M.}\
  \bibnamefont {Itoh}}, \ and\ \bibinfo {author} {\bibfnamefont {S.~A.}\
  \bibnamefont {Lyon}},\ }\href {\doibase http://dx.doi.org/10.1038/nmat3182}
  {\bibfield  {journal} {\bibinfo  {journal} {Nature Materials}\ }\textbf
  {\bibinfo {volume} {11}},\ \bibinfo {pages} {143} (\bibinfo {year}
  {2012})}\BibitemShut {NoStop}%
\bibitem [{\citenamefont {Veldhorst}\ \emph {et~al.}(2014)\citenamefont
  {Veldhorst}, \citenamefont {Hwang}, \citenamefont {Yang}, \citenamefont
  {Leenstra}, \citenamefont {de~Ronde}, \citenamefont {Dehollain},
  \citenamefont {Muhonen}, \citenamefont {Hudson}, \citenamefont {Itoh},
  \citenamefont {Morello},\ and\ \citenamefont {Dzurak}}]{Veldhorst14}%
  \BibitemOpen
  \bibfield  {author} {\bibinfo {author} {\bibfnamefont {M.}~\bibnamefont
  {Veldhorst}}, \bibinfo {author} {\bibfnamefont {J.~C.~C.}\ \bibnamefont
  {Hwang}}, \bibinfo {author} {\bibfnamefont {C.~H.}\ \bibnamefont {Yang}},
  \bibinfo {author} {\bibfnamefont {A.~W.}\ \bibnamefont {Leenstra}}, \bibinfo
  {author} {\bibfnamefont {B.}~\bibnamefont {de~Ronde}}, \bibinfo {author}
  {\bibfnamefont {J.~P.}\ \bibnamefont {Dehollain}}, \bibinfo {author}
  {\bibfnamefont {J.~T.}\ \bibnamefont {Muhonen}}, \bibinfo {author}
  {\bibfnamefont {F.~E.}\ \bibnamefont {Hudson}}, \bibinfo {author}
  {\bibfnamefont {K.~M.}\ \bibnamefont {Itoh}}, \bibinfo {author}
  {\bibfnamefont {A.}~\bibnamefont {Morello}}, \ and\ \bibinfo {author}
  {\bibfnamefont {A.~S.}\ \bibnamefont {Dzurak}},\ }\href {\doibase
  10.1038/nnano.2014.216} {\bibfield  {journal} {\bibinfo  {journal} {Nature
  Nanotechnology}\ }\textbf {\bibinfo {volume} {9}},\ \bibinfo {pages} {981}
  (\bibinfo {year} {2014})}\BibitemShut {NoStop}%
\bibitem [{\citenamefont {Veldhorst}\ \emph {et~al.}(2015)\citenamefont
  {Veldhorst}, \citenamefont {Yang}, \citenamefont {Hwang}, \citenamefont
  {Huang}, \citenamefont {Dehollain}, \citenamefont {Muhonen}, \citenamefont
  {Simmons}, \citenamefont {Laucht}, \citenamefont {Hudson}, \citenamefont
  {Itoh}, \citenamefont {Morello},\ and\ \citenamefont
  {Dzurak}}]{Veldhorst15b}%
  \BibitemOpen
  \bibfield  {author} {\bibinfo {author} {\bibfnamefont {M.}~\bibnamefont
  {Veldhorst}}, \bibinfo {author} {\bibfnamefont {C.~H.}\ \bibnamefont {Yang}},
  \bibinfo {author} {\bibfnamefont {J.~C.~C.}\ \bibnamefont {Hwang}}, \bibinfo
  {author} {\bibfnamefont {W.}~\bibnamefont {Huang}}, \bibinfo {author}
  {\bibfnamefont {J.~P.}\ \bibnamefont {Dehollain}}, \bibinfo {author}
  {\bibfnamefont {J.~T.}\ \bibnamefont {Muhonen}}, \bibinfo {author}
  {\bibfnamefont {S.}~\bibnamefont {Simmons}}, \bibinfo {author} {\bibfnamefont
  {A.}~\bibnamefont {Laucht}}, \bibinfo {author} {\bibfnamefont {F.~E.}\
  \bibnamefont {Hudson}}, \bibinfo {author} {\bibfnamefont {K.~M.}\
  \bibnamefont {Itoh}}, \bibinfo {author} {\bibfnamefont {A.}~\bibnamefont
  {Morello}}, \ and\ \bibinfo {author} {\bibfnamefont {A.~S.}\ \bibnamefont
  {Dzurak}},\ }\href {\doibase http://dx.doi.org/10.1038/nature15263}
  {\bibfield  {journal} {\bibinfo  {journal} {Nature}\ }\textbf {\bibinfo
  {volume} {526}},\ \bibinfo {pages} {410} (\bibinfo {year}
  {2015})}\BibitemShut {NoStop}%
\bibitem [{\citenamefont {Takeda}\ \emph {et~al.}(2016)\citenamefont {Takeda},
  \citenamefont {Kamioka}, \citenamefont {Otsuka}, \citenamefont {Yoneda},
  \citenamefont {Nakajima}, \citenamefont {Delbecq}, \citenamefont {Amaha},
  \citenamefont {Allison}, \citenamefont {Kodera}, \citenamefont {Oda},\ and\
  \citenamefont {Tarucha}}]{Takeda16}%
  \BibitemOpen
  \bibfield  {author} {\bibinfo {author} {\bibfnamefont {K.}~\bibnamefont
  {Takeda}}, \bibinfo {author} {\bibfnamefont {J.}~\bibnamefont {Kamioka}},
  \bibinfo {author} {\bibfnamefont {T.}~\bibnamefont {Otsuka}}, \bibinfo
  {author} {\bibfnamefont {J.}~\bibnamefont {Yoneda}}, \bibinfo {author}
  {\bibfnamefont {T.}~\bibnamefont {Nakajima}}, \bibinfo {author}
  {\bibfnamefont {M.~R.}\ \bibnamefont {Delbecq}}, \bibinfo {author}
  {\bibfnamefont {S.}~\bibnamefont {Amaha}}, \bibinfo {author} {\bibfnamefont
  {G.}~\bibnamefont {Allison}}, \bibinfo {author} {\bibfnamefont
  {T.}~\bibnamefont {Kodera}}, \bibinfo {author} {\bibfnamefont
  {S.}~\bibnamefont {Oda}}, \ and\ \bibinfo {author} {\bibfnamefont
  {S.}~\bibnamefont {Tarucha}},\ }\href {\doibase 10.1126/sciadv.1600694}
  {\bibfield  {journal} {\bibinfo  {journal} {Science Advances}\ }\textbf
  {\bibinfo {volume} {2}},\ \bibinfo {pages} {e1600694} (\bibinfo {year}
  {2016})}\BibitemShut {NoStop}%
\bibitem [{\citenamefont {Yoneda}\ \emph {et~al.}(2018)\citenamefont {Yoneda},
  \citenamefont {Takeda}, \citenamefont {Otsuka}, \citenamefont {Nakajima},
  \citenamefont {Delbecq}, \citenamefont {Allison}, \citenamefont {Honda},
  \citenamefont {Kodera}, \citenamefont {Oda}, \citenamefont {Hoshi},
  \citenamefont {Usami}, \citenamefont {Itoh},\ and\ \citenamefont
  {Tarucha}}]{Yoneda18}%
  \BibitemOpen
  \bibfield  {author} {\bibinfo {author} {\bibfnamefont {J.}~\bibnamefont
  {Yoneda}}, \bibinfo {author} {\bibfnamefont {K.}~\bibnamefont {Takeda}},
  \bibinfo {author} {\bibfnamefont {T.}~\bibnamefont {Otsuka}}, \bibinfo
  {author} {\bibfnamefont {T.}~\bibnamefont {Nakajima}}, \bibinfo {author}
  {\bibfnamefont {M.~R.}\ \bibnamefont {Delbecq}}, \bibinfo {author}
  {\bibfnamefont {G.}~\bibnamefont {Allison}}, \bibinfo {author} {\bibfnamefont
  {T.}~\bibnamefont {Honda}}, \bibinfo {author} {\bibfnamefont
  {T.}~\bibnamefont {Kodera}}, \bibinfo {author} {\bibfnamefont
  {S.}~\bibnamefont {Oda}}, \bibinfo {author} {\bibfnamefont {Y.}~\bibnamefont
  {Hoshi}}, \bibinfo {author} {\bibfnamefont {N.}~\bibnamefont {Usami}},
  \bibinfo {author} {\bibfnamefont {K.~M.}\ \bibnamefont {Itoh}}, \ and\
  \bibinfo {author} {\bibfnamefont {S.}~\bibnamefont {Tarucha}},\ }\href
  {\doibase https://doi.org/10.1038/s41565-017-0014-x} {\bibfield  {journal}
  {\bibinfo  {journal} {Nature Nanotechnology}\ }\textbf {\bibinfo {volume}
  {13}},\ \bibinfo {pages} {102} (\bibinfo {year} {2018})}\BibitemShut
  {NoStop}%
\bibitem [{\citenamefont {Watson}\ \emph {et~al.}(2018)\citenamefont {Watson},
  \citenamefont {Philips}, \citenamefont {Kawakami}, \citenamefont {Ward},
  \citenamefont {Scarlino}, \citenamefont {Veldhorst}, \citenamefont {Savage},
  \citenamefont {Lagally}, \citenamefont {Friesen}, \citenamefont
  {Coppersmith}, \citenamefont {Eriksson},\ and\ \citenamefont
  {Vandersypen}}]{Watson18}%
  \BibitemOpen
  \bibfield  {author} {\bibinfo {author} {\bibfnamefont {T.~F.}\ \bibnamefont
  {Watson}}, \bibinfo {author} {\bibfnamefont {S.~G.~J.}\ \bibnamefont
  {Philips}}, \bibinfo {author} {\bibfnamefont {E.}~\bibnamefont {Kawakami}},
  \bibinfo {author} {\bibfnamefont {D.~R.}\ \bibnamefont {Ward}}, \bibinfo
  {author} {\bibfnamefont {P.}~\bibnamefont {Scarlino}}, \bibinfo {author}
  {\bibfnamefont {M.}~\bibnamefont {Veldhorst}}, \bibinfo {author}
  {\bibfnamefont {D.~E.}\ \bibnamefont {Savage}}, \bibinfo {author}
  {\bibfnamefont {M.~G.}\ \bibnamefont {Lagally}}, \bibinfo {author}
  {\bibfnamefont {M.}~\bibnamefont {Friesen}}, \bibinfo {author} {\bibfnamefont
  {S.~N.}\ \bibnamefont {Coppersmith}}, \bibinfo {author} {\bibfnamefont
  {M.~A.}\ \bibnamefont {Eriksson}}, \ and\ \bibinfo {author} {\bibfnamefont
  {L.~M.~K.}\ \bibnamefont {Vandersypen}},\ }\href {\doibase
  10.1038/nature25766} {\bibfield  {journal} {\bibinfo  {journal} {Nature}\
  }\textbf {\bibinfo {volume} {555}},\ \bibinfo {pages} {633} (\bibinfo {year}
  {2018})}\BibitemShut {NoStop}%
\bibitem [{\citenamefont {Maurand}\ \emph {et~al.}(2016)\citenamefont
  {Maurand}, \citenamefont {Jehl}, \citenamefont {Kotekar-Patil}, \citenamefont
  {Corna}, \citenamefont {Bohuslavskyi}, \citenamefont {Lavi\'{e}ville},
  \citenamefont {Hutin}, \citenamefont {Barraud}, \citenamefont {Vinet},
  \citenamefont {Sanquer},\ and\ \citenamefont {de~Franceschi}}]{Maurand16}%
  \BibitemOpen
  \bibfield  {author} {\bibinfo {author} {\bibfnamefont {R.}~\bibnamefont
  {Maurand}}, \bibinfo {author} {\bibfnamefont {X.}~\bibnamefont {Jehl}},
  \bibinfo {author} {\bibfnamefont {D.}~\bibnamefont {Kotekar-Patil}}, \bibinfo
  {author} {\bibfnamefont {A.}~\bibnamefont {Corna}}, \bibinfo {author}
  {\bibfnamefont {H.}~\bibnamefont {Bohuslavskyi}}, \bibinfo {author}
  {\bibfnamefont {R.}~\bibnamefont {Lavi\'{e}ville}}, \bibinfo {author}
  {\bibfnamefont {L.}~\bibnamefont {Hutin}}, \bibinfo {author} {\bibfnamefont
  {S.}~\bibnamefont {Barraud}}, \bibinfo {author} {\bibfnamefont
  {M.}~\bibnamefont {Vinet}}, \bibinfo {author} {\bibfnamefont
  {M.}~\bibnamefont {Sanquer}}, \ and\ \bibinfo {author} {\bibfnamefont
  {S.}~\bibnamefont {de~Franceschi}},\ }\href {\doibase 10.1038/ncomms13575}
  {\bibfield  {journal} {\bibinfo  {journal} {Nature Communications}\ }\textbf
  {\bibinfo {volume} {7}},\ \bibinfo {pages} {13575} (\bibinfo {year}
  {2016})}\BibitemShut {NoStop}%
\bibitem [{\citenamefont {Crippa}\ \emph {et~al.}(2018)\citenamefont {Crippa},
  \citenamefont {Maurand}, \citenamefont {Bourdet}, \citenamefont
  {Kotekar-Patil}, \citenamefont {Amisse}, \citenamefont {Jehl}, \citenamefont
  {Sanquer}, \citenamefont {Laviéville}, \citenamefont {Bohuslavskyi},
  \citenamefont {Hutin}, \citenamefont {Barraud}, \citenamefont {Vinet},
  \citenamefont {Niquet},\ and\ \citenamefont {{De Franceschi}}}]{Crippa18}%
  \BibitemOpen
  \bibfield  {author} {\bibinfo {author} {\bibfnamefont {A.}~\bibnamefont
  {Crippa}}, \bibinfo {author} {\bibfnamefont {R.}~\bibnamefont {Maurand}},
  \bibinfo {author} {\bibfnamefont {L.}~\bibnamefont {Bourdet}}, \bibinfo
  {author} {\bibfnamefont {D.}~\bibnamefont {Kotekar-Patil}}, \bibinfo {author}
  {\bibfnamefont {A.}~\bibnamefont {Amisse}}, \bibinfo {author} {\bibfnamefont
  {X.}~\bibnamefont {Jehl}}, \bibinfo {author} {\bibfnamefont {M.}~\bibnamefont
  {Sanquer}}, \bibinfo {author} {\bibfnamefont {R.}~\bibnamefont
  {Laviéville}}, \bibinfo {author} {\bibfnamefont {H.}~\bibnamefont
  {Bohuslavskyi}}, \bibinfo {author} {\bibfnamefont {L.}~\bibnamefont {Hutin}},
  \bibinfo {author} {\bibfnamefont {S.}~\bibnamefont {Barraud}}, \bibinfo
  {author} {\bibfnamefont {M.}~\bibnamefont {Vinet}}, \bibinfo {author}
  {\bibfnamefont {Y.-M.}\ \bibnamefont {Niquet}}, \ and\ \bibinfo {author}
  {\bibfnamefont {S.}~\bibnamefont {{De Franceschi}}},\ }\href {\doibase
  10.1103/PhysRevLett.120.137702} {\bibfield  {journal} {\bibinfo  {journal}
  {Physical Review Letters}\ }\textbf {\bibinfo {volume} {120}},\ \bibinfo
  {pages} {137702} (\bibinfo {year} {2018})}\BibitemShut {NoStop}%
\bibitem [{\citenamefont {Koppens}\ \emph {et~al.}(2005)\citenamefont
  {Koppens}, \citenamefont {Folk}, \citenamefont {Elzerman}, \citenamefont
  {Hanson}, \citenamefont {van Beveren}, \citenamefont {Vink}, \citenamefont
  {Tranitz}, \citenamefont {Wegscheider}, \citenamefont {Kouwenhoven},\ and\
  \citenamefont {Vandersypen}}]{Koppens05}%
  \BibitemOpen
  \bibfield  {author} {\bibinfo {author} {\bibfnamefont {F.~H.~L.}\
  \bibnamefont {Koppens}}, \bibinfo {author} {\bibfnamefont {J.~A.}\
  \bibnamefont {Folk}}, \bibinfo {author} {\bibfnamefont {J.~M.}\ \bibnamefont
  {Elzerman}}, \bibinfo {author} {\bibfnamefont {R.}~\bibnamefont {Hanson}},
  \bibinfo {author} {\bibfnamefont {L.~H.~W.}\ \bibnamefont {van Beveren}},
  \bibinfo {author} {\bibfnamefont {I.~T.}\ \bibnamefont {Vink}}, \bibinfo
  {author} {\bibfnamefont {H.~P.}\ \bibnamefont {Tranitz}}, \bibinfo {author}
  {\bibfnamefont {W.}~\bibnamefont {Wegscheider}}, \bibinfo {author}
  {\bibfnamefont {L.~P.}\ \bibnamefont {Kouwenhoven}}, \ and\ \bibinfo {author}
  {\bibfnamefont {L.~M.~K.}\ \bibnamefont {Vandersypen}},\ }\href {\doibase
  10.1126/science.1113719} {\bibfield  {journal} {\bibinfo  {journal}
  {Science}\ }\textbf {\bibinfo {volume} {309}},\ \bibinfo {pages} {1346}
  (\bibinfo {year} {2005})}\BibitemShut {NoStop}%
\bibitem [{\citenamefont {Rashba}\ and\ \citenamefont
  {Sheka}(1991)}]{Rashba91}%
  \BibitemOpen
  \bibfield  {author} {\bibinfo {author} {\bibfnamefont {E.~I.}\ \bibnamefont
  {Rashba}}\ and\ \bibinfo {author} {\bibfnamefont {V.~I.}\ \bibnamefont
  {Sheka}},\ }in\ \href {\doibase
  https://doi.org/10.1016/B978-0-444-88535-7.50011-X} {\emph {\bibinfo
  {booktitle} {Landau Level Spectroscopy}}},\ \bibinfo {series} {Modern
  Problems in Condensed Matter Sciences}, Vol.~\bibinfo {volume} {27},\
  \bibinfo {editor} {edited by\ \bibinfo {editor} {\bibfnamefont
  {G.}~\bibnamefont {Landwehr}}\ and\ \bibinfo {editor} {\bibfnamefont {E.~I.}\
  \bibnamefont {Rashba}}}\ (\bibinfo  {publisher} {Elsevier},\ \bibinfo
  {address} {Amsterdam},\ \bibinfo {year} {1991})\ p.\ \bibinfo {pages}
  {131}\BibitemShut {NoStop}%
\bibitem [{\citenamefont {Rashba}(2008)}]{Rashba08}%
  \BibitemOpen
  \bibfield  {author} {\bibinfo {author} {\bibfnamefont {E.~I.}\ \bibnamefont
  {Rashba}},\ }\href {\doibase 10.1103/PhysRevB.78.195302} {\bibfield
  {journal} {\bibinfo  {journal} {Physical Review B}\ }\textbf {\bibinfo
  {volume} {78}},\ \bibinfo {pages} {195302} (\bibinfo {year}
  {2008})}\BibitemShut {NoStop}%
\bibitem [{\citenamefont {Winkler}(2003)}]{Winkler03}%
  \BibitemOpen
  \bibfield  {author} {\bibinfo {author} {\bibfnamefont {R.}~\bibnamefont
  {Winkler}},\ }\href {\doibase 10.1007/b13586} {\emph {\bibinfo {title}
  {Spin-orbit coupling in two-dimensional electron and hole systems}}}\
  (\bibinfo  {publisher} {Springer},\ \bibinfo {address} {Berlin},\ \bibinfo
  {year} {2003})\BibitemShut {NoStop}%
\bibitem [{\citenamefont {Kato}\ \emph {et~al.}(2003)\citenamefont {Kato},
  \citenamefont {Myers}, \citenamefont {Driscoll}, \citenamefont {Gossard},
  \citenamefont {Levy},\ and\ \citenamefont {Awschalom}}]{Kato03}%
  \BibitemOpen
  \bibfield  {author} {\bibinfo {author} {\bibfnamefont {Y.}~\bibnamefont
  {Kato}}, \bibinfo {author} {\bibfnamefont {R.~C.}\ \bibnamefont {Myers}},
  \bibinfo {author} {\bibfnamefont {D.~C.}\ \bibnamefont {Driscoll}}, \bibinfo
  {author} {\bibfnamefont {A.~C.}\ \bibnamefont {Gossard}}, \bibinfo {author}
  {\bibfnamefont {J.}~\bibnamefont {Levy}}, \ and\ \bibinfo {author}
  {\bibfnamefont {D.~D.}\ \bibnamefont {Awschalom}},\ }\href {\doibase
  10.1126/science.1080880} {\bibfield  {journal} {\bibinfo  {journal}
  {Science}\ }\textbf {\bibinfo {volume} {299}},\ \bibinfo {pages} {1201}
  (\bibinfo {year} {2003})}\BibitemShut {NoStop}%
\bibitem [{\citenamefont {Nowack}\ \emph {et~al.}(2007)\citenamefont {Nowack},
  \citenamefont {Koppens}, \citenamefont {Nazarov},\ and\ \citenamefont
  {Vandersypen}}]{Nowack07}%
  \BibitemOpen
  \bibfield  {author} {\bibinfo {author} {\bibfnamefont {K.~C.}\ \bibnamefont
  {Nowack}}, \bibinfo {author} {\bibfnamefont {F.~H.~L.}\ \bibnamefont
  {Koppens}}, \bibinfo {author} {\bibfnamefont {Y.~V.}\ \bibnamefont
  {Nazarov}}, \ and\ \bibinfo {author} {\bibfnamefont {L.~M.~K.}\ \bibnamefont
  {Vandersypen}},\ }\href {\doibase 10.1126/science.1148092} {\bibfield
  {journal} {\bibinfo  {journal} {Science}\ }\textbf {\bibinfo {volume}
  {318}},\ \bibinfo {pages} {1430} (\bibinfo {year} {2007})}\BibitemShut
  {NoStop}%
\bibitem [{\citenamefont {Nadj-Perge}\ \emph {et~al.}(2010)\citenamefont
  {Nadj-Perge}, \citenamefont {Frolov}, \citenamefont {Bakkers},\ and\
  \citenamefont {Kouwenhoven}}]{NadjPerge10}%
  \BibitemOpen
  \bibfield  {author} {\bibinfo {author} {\bibfnamefont {S.}~\bibnamefont
  {Nadj-Perge}}, \bibinfo {author} {\bibfnamefont {S.~M.}\ \bibnamefont
  {Frolov}}, \bibinfo {author} {\bibfnamefont {E.~P. A.~M.}\ \bibnamefont
  {Bakkers}}, \ and\ \bibinfo {author} {\bibfnamefont {L.~P.}\ \bibnamefont
  {Kouwenhoven}},\ }\href {\doibase 10.1038/nature09682} {\bibfield  {journal}
  {\bibinfo  {journal} {Nature}\ }\textbf {\bibinfo {volume} {468}},\ \bibinfo
  {pages} {1084} (\bibinfo {year} {2010})}\BibitemShut {NoStop}%
\bibitem [{\citenamefont {van~den Berg}\ \emph {et~al.}(2013)\citenamefont
  {van~den Berg}, \citenamefont {Nadj-Perge}, \citenamefont {Pribiag},
  \citenamefont {Plissard}, \citenamefont {Bakkers}, \citenamefont {Frolov},\
  and\ \citenamefont {Kouwenhoven}}]{vandenBerg13}%
  \BibitemOpen
  \bibfield  {author} {\bibinfo {author} {\bibfnamefont {J.~W.~G.}\
  \bibnamefont {van~den Berg}}, \bibinfo {author} {\bibfnamefont
  {S.}~\bibnamefont {Nadj-Perge}}, \bibinfo {author} {\bibfnamefont {V.~S.}\
  \bibnamefont {Pribiag}}, \bibinfo {author} {\bibfnamefont {S.~R.}\
  \bibnamefont {Plissard}}, \bibinfo {author} {\bibfnamefont {E.~P. A.~M.}\
  \bibnamefont {Bakkers}}, \bibinfo {author} {\bibfnamefont {S.~M.}\
  \bibnamefont {Frolov}}, \ and\ \bibinfo {author} {\bibfnamefont {L.~P.}\
  \bibnamefont {Kouwenhoven}},\ }\href {\doibase
  10.1103/PhysRevLett.110.066806} {\bibfield  {journal} {\bibinfo  {journal}
  {Physical Review Letters}\ }\textbf {\bibinfo {volume} {110}},\ \bibinfo
  {pages} {066806} (\bibinfo {year} {2013})}\BibitemShut {NoStop}%
\bibitem [{\citenamefont {Corna}\ \emph {et~al.}(2018)\citenamefont {Corna},
  \citenamefont {Bourdet}, \citenamefont {Maurand}, \citenamefont {Crippa},
  \citenamefont {Kotekar-Patil}, \citenamefont {Bohuslavskyi}, \citenamefont
  {Lavi\'eville}, \citenamefont {Hutin}, \citenamefont {Barraud}, \citenamefont
  {Jehl}, \citenamefont {Vinet}, \citenamefont {De~Franceschi}, \citenamefont
  {Niquet},\ and\ \citenamefont {Sanquer}}]{Corna18}%
  \BibitemOpen
  \bibfield  {author} {\bibinfo {author} {\bibfnamefont {A.}~\bibnamefont
  {Corna}}, \bibinfo {author} {\bibfnamefont {L.}~\bibnamefont {Bourdet}},
  \bibinfo {author} {\bibfnamefont {R.}~\bibnamefont {Maurand}}, \bibinfo
  {author} {\bibfnamefont {A.}~\bibnamefont {Crippa}}, \bibinfo {author}
  {\bibfnamefont {D.}~\bibnamefont {Kotekar-Patil}}, \bibinfo {author}
  {\bibfnamefont {H.}~\bibnamefont {Bohuslavskyi}}, \bibinfo {author}
  {\bibfnamefont {R.}~\bibnamefont {Lavi\'eville}}, \bibinfo {author}
  {\bibfnamefont {L.}~\bibnamefont {Hutin}}, \bibinfo {author} {\bibfnamefont
  {S.}~\bibnamefont {Barraud}}, \bibinfo {author} {\bibfnamefont
  {X.}~\bibnamefont {Jehl}}, \bibinfo {author} {\bibfnamefont {M.}~\bibnamefont
  {Vinet}}, \bibinfo {author} {\bibfnamefont {S.}~\bibnamefont
  {De~Franceschi}}, \bibinfo {author} {\bibfnamefont {Y.-M.}\ \bibnamefont
  {Niquet}}, \ and\ \bibinfo {author} {\bibfnamefont {M.}~\bibnamefont
  {Sanquer}},\ }\href {\doibase 10.1038/s41534-018-0059-1} {\bibfield
  {journal} {\bibinfo  {journal} {npj Quantum Information}\ }\textbf {\bibinfo
  {volume} {4}},\ \bibinfo {pages} {6} (\bibinfo {year} {2018})}\BibitemShut
  {NoStop}%
\bibitem [{\citenamefont {Pingenot}\ \emph {et~al.}(2011)\citenamefont
  {Pingenot}, \citenamefont {Pryor},\ and\ \citenamefont
  {Flatt\'e}}]{Pingenot11}%
  \BibitemOpen
  \bibfield  {author} {\bibinfo {author} {\bibfnamefont {J.}~\bibnamefont
  {Pingenot}}, \bibinfo {author} {\bibfnamefont {C.~E.}\ \bibnamefont {Pryor}},
  \ and\ \bibinfo {author} {\bibfnamefont {M.~E.}\ \bibnamefont {Flatt\'e}},\
  }\href {\doibase 10.1103/PhysRevB.84.195403} {\bibfield  {journal} {\bibinfo
  {journal} {Physical Review B}\ }\textbf {\bibinfo {volume} {84}},\ \bibinfo
  {pages} {195403} (\bibinfo {year} {2011})}\BibitemShut {NoStop}%
\bibitem [{\citenamefont {Schroer}\ \emph {et~al.}(2011)\citenamefont
  {Schroer}, \citenamefont {Petersson}, \citenamefont {Jung},\ and\
  \citenamefont {Petta}}]{Schroer11}%
  \BibitemOpen
  \bibfield  {author} {\bibinfo {author} {\bibfnamefont {M.~D.}\ \bibnamefont
  {Schroer}}, \bibinfo {author} {\bibfnamefont {K.~D.}\ \bibnamefont
  {Petersson}}, \bibinfo {author} {\bibfnamefont {M.}~\bibnamefont {Jung}}, \
  and\ \bibinfo {author} {\bibfnamefont {J.~R.}\ \bibnamefont {Petta}},\ }\href
  {\doibase 10.1103/PhysRevLett.107.176811} {\bibfield  {journal} {\bibinfo
  {journal} {Physical Review Letters}\ }\textbf {\bibinfo {volume} {107}},\
  \bibinfo {pages} {176811} (\bibinfo {year} {2011})}\BibitemShut {NoStop}%
\bibitem [{\citenamefont {Takahashi}\ \emph {et~al.}(2013)\citenamefont
  {Takahashi}, \citenamefont {Deacon}, \citenamefont {Oiwa}, \citenamefont
  {Shibata}, \citenamefont {Hirakawa},\ and\ \citenamefont
  {Tarucha}}]{Takahashi13}%
  \BibitemOpen
  \bibfield  {author} {\bibinfo {author} {\bibfnamefont {S.}~\bibnamefont
  {Takahashi}}, \bibinfo {author} {\bibfnamefont {R.~S.}\ \bibnamefont
  {Deacon}}, \bibinfo {author} {\bibfnamefont {A.}~\bibnamefont {Oiwa}},
  \bibinfo {author} {\bibfnamefont {K.}~\bibnamefont {Shibata}}, \bibinfo
  {author} {\bibfnamefont {K.}~\bibnamefont {Hirakawa}}, \ and\ \bibinfo
  {author} {\bibfnamefont {S.}~\bibnamefont {Tarucha}},\ }\href {\doibase
  10.1103/PhysRevB.87.161302} {\bibfield  {journal} {\bibinfo  {journal}
  {Physical Review B}\ }\textbf {\bibinfo {volume} {87}},\ \bibinfo {pages}
  {161302} (\bibinfo {year} {2013})}\BibitemShut {NoStop}%
\bibitem [{\citenamefont {Ares}\ \emph {et~al.}(2013)\citenamefont {Ares},
  \citenamefont {Katsaros}, \citenamefont {Golovach}, \citenamefont {Zhang},
  \citenamefont {Prager}, \citenamefont {Glazman}, \citenamefont {Schmidt},\
  and\ \citenamefont {De~Franceschi}}]{Ares13}%
  \BibitemOpen
  \bibfield  {author} {\bibinfo {author} {\bibfnamefont {N.}~\bibnamefont
  {Ares}}, \bibinfo {author} {\bibfnamefont {G.}~\bibnamefont {Katsaros}},
  \bibinfo {author} {\bibfnamefont {V.~N.}\ \bibnamefont {Golovach}}, \bibinfo
  {author} {\bibfnamefont {J.~J.}\ \bibnamefont {Zhang}}, \bibinfo {author}
  {\bibfnamefont {A.}~\bibnamefont {Prager}}, \bibinfo {author} {\bibfnamefont
  {L.~I.}\ \bibnamefont {Glazman}}, \bibinfo {author} {\bibfnamefont {O.~G.}\
  \bibnamefont {Schmidt}}, \ and\ \bibinfo {author} {\bibfnamefont
  {S.}~\bibnamefont {De~Franceschi}},\ }\href {\doibase 10.1063/1.4858959}
  {\bibfield  {journal} {\bibinfo  {journal} {Applied Physics Letters}\
  }\textbf {\bibinfo {volume} {103}},\ \bibinfo {pages} {263113} (\bibinfo
  {year} {2013})}\BibitemShut {NoStop}%
\bibitem [{\citenamefont {Voisin}\ \emph {et~al.}(2016)\citenamefont {Voisin},
  \citenamefont {Maurand}, \citenamefont {Barraud}, \citenamefont {Vinet},
  \citenamefont {Jehl}, \citenamefont {Sanquer}, \citenamefont {Renard},\ and\
  \citenamefont {De~Franceschi}}]{Voisin16}%
  \BibitemOpen
  \bibfield  {author} {\bibinfo {author} {\bibfnamefont {B.}~\bibnamefont
  {Voisin}}, \bibinfo {author} {\bibfnamefont {R.}~\bibnamefont {Maurand}},
  \bibinfo {author} {\bibfnamefont {S.}~\bibnamefont {Barraud}}, \bibinfo
  {author} {\bibfnamefont {M.}~\bibnamefont {Vinet}}, \bibinfo {author}
  {\bibfnamefont {X.}~\bibnamefont {Jehl}}, \bibinfo {author} {\bibfnamefont
  {M.}~\bibnamefont {Sanquer}}, \bibinfo {author} {\bibfnamefont
  {J.}~\bibnamefont {Renard}}, \ and\ \bibinfo {author} {\bibfnamefont
  {S.}~\bibnamefont {De~Franceschi}},\ }\href {\doibase
  10.1021/acs.nanolett.5b02920} {\bibfield  {journal} {\bibinfo  {journal}
  {Nano Letters}\ }\textbf {\bibinfo {volume} {16}},\ \bibinfo {pages} {88}
  (\bibinfo {year} {2016})}\BibitemShut {NoStop}%
\bibitem [{\citenamefont {Rashba}\ and\ \citenamefont
  {Efros}(2003)}]{Rashba03}%
  \BibitemOpen
  \bibfield  {author} {\bibinfo {author} {\bibfnamefont {E.~I.}\ \bibnamefont
  {Rashba}}\ and\ \bibinfo {author} {\bibfnamefont {A.~L.}\ \bibnamefont
  {Efros}},\ }\href {\doibase 10.1103/PhysRevLett.91.126405} {\bibfield
  {journal} {\bibinfo  {journal} {Physical Review Letters}\ }\textbf {\bibinfo
  {volume} {91}},\ \bibinfo {pages} {126405} (\bibinfo {year}
  {2003})}\BibitemShut {NoStop}%
\bibitem [{\citenamefont {Golovach}\ \emph {et~al.}(2006)\citenamefont
  {Golovach}, \citenamefont {Borhani},\ and\ \citenamefont
  {Loss}}]{Golovach06}%
  \BibitemOpen
  \bibfield  {author} {\bibinfo {author} {\bibfnamefont {V.~N.}\ \bibnamefont
  {Golovach}}, \bibinfo {author} {\bibfnamefont {M.}~\bibnamefont {Borhani}}, \
  and\ \bibinfo {author} {\bibfnamefont {D.}~\bibnamefont {Loss}},\ }\href
  {\doibase 10.1103/PhysRevB.74.165319} {\bibfield  {journal} {\bibinfo
  {journal} {Physical Review B}\ }\textbf {\bibinfo {volume} {74}},\ \bibinfo
  {pages} {165319} (\bibinfo {year} {2006})}\BibitemShut {NoStop}%
\bibitem [{\citenamefont {Flindt}\ \emph {et~al.}(2006)\citenamefont {Flindt},
  \citenamefont {S\o{}rensen},\ and\ \citenamefont {Flensberg}}]{Flindt06}%
  \BibitemOpen
  \bibfield  {author} {\bibinfo {author} {\bibfnamefont {C.}~\bibnamefont
  {Flindt}}, \bibinfo {author} {\bibfnamefont {A.~S.}\ \bibnamefont
  {S\o{}rensen}}, \ and\ \bibinfo {author} {\bibfnamefont {K.}~\bibnamefont
  {Flensberg}},\ }\href {\doibase 10.1103/PhysRevLett.97.240501} {\bibfield
  {journal} {\bibinfo  {journal} {Physical Review Letters}\ }\textbf {\bibinfo
  {volume} {97}},\ \bibinfo {pages} {240501} (\bibinfo {year}
  {2006})}\BibitemShut {NoStop}%
\bibitem [{Note1()}]{Note1}%
  \BibitemOpen
  \bibinfo {note} {The models of section \ref {sectionModels} can however be
  applied to any pair of Kramers degenerate states.}\BibitemShut {Stop}%
\bibitem [{Note2()}]{Note2}%
  \BibitemOpen
  \bibinfo {note} {Note that $\protect \bra {0,\delimiter "322B37F }\protect
  \bm {B}\cdot \protect \bm {M}_1\protect \ket {0,\delimiter "322B37F
  }=-\protect \bra {0,\delimiter "322A37E }\protect \bm {B}\cdot \protect \bm
  {M}_1\protect \ket {0,\delimiter "322A37E }$ since time-reversal symmetry
  transforms $\protect \bm {B}$ into $-\protect \bm {B}$.}\BibitemShut {Stop}%
\bibitem [{\citenamefont {Weil}\ and\ \citenamefont
  {Bolton}(2007)}]{Weil-Bolton}%
  \BibitemOpen
  \bibfield  {author} {\bibinfo {author} {\bibfnamefont {J.~A.}\ \bibnamefont
  {Weil}}\ and\ \bibinfo {author} {\bibfnamefont {J.~R.}\ \bibnamefont
  {Bolton}},\ }\href {\doibase 10.1002/0470084987} {\emph {\bibinfo {title}
  {Electron paramagnetic resonance: elementary theory and practical
  applications}}}\ (\bibinfo  {publisher} {John Wiley \& Sons},\ \bibinfo
  {address} {Hoboken},\ \bibinfo {year} {2007})\BibitemShut {NoStop}%
\bibitem [{Note3()}]{Note3}%
  \BibitemOpen
  \bibinfo {note} {We add a hat on top of all real $3\times 3$ matrices, in
  order to distinguish these matrices from those of the operators, such as the
  Pauli matrices, acting in the Hilbert space of the qubit. The notations are
  slightly different from Ref. \protect \rev@citealpnum {Crippa18}, where
  $\cdot $ was used to denote real $3\times 3$ matrix/matrix and matrix/vector
  multiplications.}\BibitemShut {Stop}%
\bibitem [{\citenamefont {Chibotaru}\ \emph {et~al.}(2008)\citenamefont
  {Chibotaru}, \citenamefont {Ceulemans},\ and\ \citenamefont
  {Bolvin}}]{Chibotaru08}%
  \BibitemOpen
  \bibfield  {author} {\bibinfo {author} {\bibfnamefont {L.~F.}\ \bibnamefont
  {Chibotaru}}, \bibinfo {author} {\bibfnamefont {A.}~\bibnamefont
  {Ceulemans}}, \ and\ \bibinfo {author} {\bibfnamefont {H.}~\bibnamefont
  {Bolvin}},\ }\href@noop {} {\bibfield  {journal} {\bibinfo  {journal}
  {Physical Review Letters}\ }\textbf {\bibinfo {volume} {101}},\ \bibinfo
  {pages} {033003} (\bibinfo {year} {2008})}\BibitemShut {NoStop}%
\bibitem [{Note4()}]{Note4}%
  \BibitemOpen
  \bibinfo {note} {Let $R$ be an unitary transformation in the $\protect
  \{\protect \ket {0,\delimiter "322A37E },\protect \ket {0,\delimiter "322B37F
  }\protect \}$ subspace. $R$ can be cast in the form \begin {equation*}
  R=\begin {pmatrix}{} \alpha e^{i\theta } & -\beta ^* \\ \beta e^{i\theta } &
  \alpha ^* \end {pmatrix} \end {equation*} with $|\alpha |^2+|\beta |^2=1$. In
  the basis set $\protect \{\protect \ket {0,\delimiter "322A37E ^\prime
  },\protect \ket {0,\delimiter "322B37F ^\prime }\protect \}=\protect
  \{R\protect \ket {0,\delimiter "322A37E },R\protect \ket {0,\delimiter
  "322B37F }\protect \}$, the Hamiltonian reads $H^\prime =\protect \frac
  {1}{2}\mu _B\protect \bm {\sigma }^\prime \cdot \protect \mathaccentV
  {hat}05E{g}\protect \bm {B}$, with $\sigma ^\prime _i=R^\protect \dag \sigma
  _i R$. Yet $\protect \bm {\sigma }^\prime =\protect \mathaccentV
  {hat}05E{U}\protect \bm {\sigma }$, with: \begin {equation*} \protect
  \mathaccentV {hat}05E{U}=\begin {pmatrix}{} {\protect \rm Re}[(\alpha
  ^2-\beta ^2)e^{i\theta }] & {\protect \rm Im}[(\alpha ^2-\beta ^2)e^{i\theta
  }] & 2{\protect \rm Re}(\alpha ^*\beta ) \\ -{\protect \rm Im}[(\alpha
  ^2+\beta ^2)e^{i\theta }] & {\protect \rm Re}[(\alpha ^2+\beta ^2)e^{i\theta
  }] & 2{\protect \rm Im}(\alpha ^*\beta ) \\ -2{\protect \rm Re}(\alpha \beta
  e^{i\theta }) & -2{\protect \rm Im}(\alpha \beta e^{i\theta }) & |\alpha
  |^2-|\beta |^2 \end {pmatrix}\protect \tmspace +\thinmuskip {.1667em}. \end
  {equation*} Hence $H^\prime =\protect \frac {1}{2}\mu _B\protect \bm {\sigma
  }\cdot \protect \mathaccentV {hat}05E{g}^\prime \protect \bm {B}$, with
  $\protect \mathaccentV {hat}05E{g}^\prime ={^t}\protect \mathaccentV
  {hat}05E{U}\protect \mathaccentV {hat}05E{g}$. The $\protect \mathaccentV
  {hat}05E{U}$ matrix is unitary with determinant $+1$. Therefore, any rotation
  of the $\protect \{\protect \ket {0,\delimiter "322A37E },\protect \ket
  {0,\delimiter "322B37F }\protect \}$ basis set results in a corresponding
  rotation of the $g$-matrix. Conversely, any unitary $3\times 3$ matrix
  $\protect \mathaccentV {hat}05E{U}$ with determinant $+1$ can be mapped onto
  the above equation, and associated with a unitary transformation $R$ in the
  $\protect \{\protect \ket {0,\delimiter "322A37E },\protect \ket
  {0,\delimiter "322B37F }\protect \}$ subspace.}\BibitemShut {Stop}%
\bibitem [{Note5()}]{Note5}%
  \BibitemOpen
  \bibinfo {note} {As discussed in the previous note, any change of basis set
  $\protect \{\protect \ket {0,\delimiter "322A37E },\protect \ket
  {0,\delimiter "322B37F }\protect \}$, characterized by a complex, unitary
  $2\times 2$ matrix $R$, results in a transformation $\protect \mathaccentV
  {hat}05E{g}\to {^t}\protect \mathaccentV {hat}05E{U}\protect \mathaccentV
  {hat}05E{g}$ of the $\protect \mathaccentV {hat}05E{g}$ matrix, where
  $\protect \mathaccentV {hat}05E{U}$ is a real, unitary $3\times 3$ matrix
  (and likewise for $\protect \mathaccentV {hat}05E{g}^\prime $). Since
  $|[{^t}\protect \mathaccentV {hat}05E{U}\protect \mathaccentV
  {hat}05E{g}(V_0)\protect \bm {b}]\times [{^t}\protect \mathaccentV
  {hat}05E{U}\protect \mathaccentV {hat}05E{g}^\prime (V_0)\protect \bm
  {b}]|=|[\protect \mathaccentV {hat}05E{g}(V_0)\protect \bm {b}]\times
  [\protect \mathaccentV {hat}05E{g}^\prime (V_0)\protect \bm {b}]|$, $f_R$ is
  invariant under that transformation. Also, a change of gauge for the vector
  potential results in an unitary transform on $\protect \bm {M}_1$, hence
  equivalently in a rotation of the basis set $\protect \{\protect \ket
  {0,\delimiter "322A37E },\protect \ket {0,\delimiter "322B37F }\protect \}$,
  which, as discussed above, leaves $f_R$ invariant.}\BibitemShut {Stop}%
\bibitem [{\citenamefont {Pioro-Ladri\`{e}re}\ \emph
  {et~al.}(2008)\citenamefont {Pioro-Ladri\`{e}re}, \citenamefont {Obata},
  \citenamefont {Tokura}, \citenamefont {Shin}, \citenamefont {Kubo},
  \citenamefont {Yoshida}, \citenamefont {Taniyama},\ and\ \citenamefont
  {Tarucha}}]{Pioro-Ladriere08}%
  \BibitemOpen
  \bibfield  {author} {\bibinfo {author} {\bibfnamefont {M.}~\bibnamefont
  {Pioro-Ladri\`{e}re}}, \bibinfo {author} {\bibfnamefont {T.}~\bibnamefont
  {Obata}}, \bibinfo {author} {\bibfnamefont {Y.}~\bibnamefont {Tokura}},
  \bibinfo {author} {\bibfnamefont {Y.-S.}\ \bibnamefont {Shin}}, \bibinfo
  {author} {\bibfnamefont {T.}~\bibnamefont {Kubo}}, \bibinfo {author}
  {\bibfnamefont {K.}~\bibnamefont {Yoshida}}, \bibinfo {author} {\bibfnamefont
  {T.}~\bibnamefont {Taniyama}}, \ and\ \bibinfo {author} {\bibfnamefont
  {S.}~\bibnamefont {Tarucha}},\ }\href {\doibase 10.1038/nphys1053} {\bibfield
   {journal} {\bibinfo  {journal} {Nature Physics}\ }\textbf {\bibinfo {volume}
  {4}},\ \bibinfo {pages} {776} (\bibinfo {year} {2008})}\BibitemShut {NoStop}%
\bibitem [{\citenamefont {Kawakami}\ \emph {et~al.}(2014)\citenamefont
  {Kawakami}, \citenamefont {Scarlino}, \citenamefont {Ward}, \citenamefont
  {Braakman}, \citenamefont {Savage}, \citenamefont {Lagally}, \citenamefont
  {Friesen}, \citenamefont {Coppersmith}, \citenamefont {Eriksson},\ and\
  \citenamefont {Vandersypen}}]{Kawakami14}%
  \BibitemOpen
  \bibfield  {author} {\bibinfo {author} {\bibfnamefont {E.}~\bibnamefont
  {Kawakami}}, \bibinfo {author} {\bibfnamefont {P.}~\bibnamefont {Scarlino}},
  \bibinfo {author} {\bibfnamefont {D.~R.}\ \bibnamefont {Ward}}, \bibinfo
  {author} {\bibfnamefont {F.~R.}\ \bibnamefont {Braakman}}, \bibinfo {author}
  {\bibfnamefont {D.~E.}\ \bibnamefont {Savage}}, \bibinfo {author}
  {\bibfnamefont {M.~G.}\ \bibnamefont {Lagally}}, \bibinfo {author}
  {\bibfnamefont {M.}~\bibnamefont {Friesen}}, \bibinfo {author} {\bibfnamefont
  {S.~N.}\ \bibnamefont {Coppersmith}}, \bibinfo {author} {\bibfnamefont
  {M.~A.}\ \bibnamefont {Eriksson}}, \ and\ \bibinfo {author} {\bibfnamefont
  {L.~M.~K.}\ \bibnamefont {Vandersypen}},\ }\href {\doibase
  10.1038/nnano.2014.153} {\bibfield  {journal} {\bibinfo  {journal} {Nature
  Nanotechnology}\ }\textbf {\bibinfo {volume} {9}},\ \bibinfo {pages} {666}
  (\bibinfo {year} {2014})}\BibitemShut {NoStop}%
\bibitem [{\citenamefont {{Jock Ryan M.}}\ \emph {et~al.}(2018)\citenamefont
  {{Jock Ryan M.}}, \citenamefont {{Jacobson N. Tobias}}, \citenamefont
  {{Harvey-Collard Patrick}}, \citenamefont {{Mounce Andrew M.}}, \citenamefont
  {{Srinivasa Vanita}}, \citenamefont {{Ward Dan R.}}, \citenamefont {{Anderson
  John}}, \citenamefont {{Manginell Ron}}, \citenamefont {{Wendt Joel R.}},
  \citenamefont {{Rudolph Martin}}, \citenamefont {{Pluym Tammy}},
  \citenamefont {{Gamble John King}}, \citenamefont {{Baczewski Andrew D.}},
  \citenamefont {{Witzel Wayne M.}},\ and\ \citenamefont {{Carroll Malcolm
  S.}}}]{Jock18}%
  \BibitemOpen
  \bibfield  {author} {\bibinfo {author} {\bibnamefont {{Jock Ryan M.}}},
  \bibinfo {author} {\bibnamefont {{Jacobson N. Tobias}}}, \bibinfo {author}
  {\bibnamefont {{Harvey-Collard Patrick}}}, \bibinfo {author} {\bibnamefont
  {{Mounce Andrew M.}}}, \bibinfo {author} {\bibnamefont {{Srinivasa Vanita}}},
  \bibinfo {author} {\bibnamefont {{Ward Dan R.}}}, \bibinfo {author}
  {\bibnamefont {{Anderson John}}}, \bibinfo {author} {\bibnamefont {{Manginell
  Ron}}}, \bibinfo {author} {\bibnamefont {{Wendt Joel R.}}}, \bibinfo {author}
  {\bibnamefont {{Rudolph Martin}}}, \bibinfo {author} {\bibnamefont {{Pluym
  Tammy}}}, \bibinfo {author} {\bibnamefont {{Gamble John King}}}, \bibinfo
  {author} {\bibnamefont {{Baczewski Andrew D.}}}, \bibinfo {author}
  {\bibnamefont {{Witzel Wayne M.}}}, \ and\ \bibinfo {author} {\bibnamefont
  {{Carroll Malcolm S.}}},\ }\href {\doibase
  https://doi.org/10.1038/s41467-018-04200-0} {\bibfield  {journal} {\bibinfo
  {journal} {Nature Communications}\ }\textbf {\bibinfo {volume} {9}},\
  \bibinfo {pages} {1768} (\bibinfo {year} {2018})}\BibitemShut {NoStop}%
\bibitem [{Note6()}]{Note6}%
  \BibitemOpen
  \bibinfo {note} {In principle, both $\protect \mathaccentV {hat}05E{g}_d$ and
  $\protect \mathaccentV {hat}05E{V}$ (hence their derivatives) can be
  completely characterized as a function of gate voltage by the measurement and
  diagonalization of the symmetric Zeeman tensor $\protect \mathaccentV
  {hat}05E{G}$. We may, therefore, adopt a slightly more general (and physical)
  definition of $\protect \mathaccentV {hat}05E{g}_{\protect \rm TMR}^\prime
  =\protect \mathaccentV {hat}05E{g}_d^\prime -\protect \mathaccentV
  {hat}05E{g}_d\protect \mathaccentV {hat}05E{V}^\prime $ as the contribution
  from the electrical modulations $\protect \mathaccentV {hat}05E{g}_d^\prime $
  and $\protect \mathaccentV {hat}05E{V}^\prime $, and of $\protect
  \mathaccentV {hat}05E{g}_{\protect \rm IZR}^\prime =\protect \mathaccentV
  {hat}05E{U}^\prime \protect \mathaccentV {hat}05E{g}_d$ as the contribution
  from the electrical modulations $\protect \mathaccentV {hat}05E{U}^\prime $.
  However, from an experimental point of view, it may be extremely difficult to
  reconstruct $\protect \mathaccentV {hat}05E{V}^\prime $ from the dependence
  of the eigenvectors of $\protect \mathaccentV {hat}05E{G}$ on gate voltage.
  Indeed, as part of the modulations of $\protect \mathaccentV {hat}05E{V}$
  [second and third terms of Eq. (\ref {eqgder})] only make a second or
  higher-order contribution to $\protect \mathaccentV {hat}05E{G}$, the
  measurement of $\protect \mathaccentV {hat}05E{G}$ must be extremely accurate
  in order to completely capture the dependence of $\protect \mathaccentV
  {hat}05E{V}$ on gate voltage.}\BibitemShut {Stop}%
\bibitem [{\citenamefont {Cornwell}(1997)}]{Cornwell97}%
  \BibitemOpen
  \bibinfo {editor} {\bibfnamefont {J.}~\bibnamefont {Cornwell}},\ ed.,\
  \href@noop {} {\emph {\bibinfo {title} {Group Theory in Physics}}}\ (\bibinfo
   {publisher} {Academic Press},\ \bibinfo {address} {San Diego},\ \bibinfo
  {year} {1997})\BibitemShut {NoStop}%
\bibitem [{\citenamefont {Bourdet}\ and\ \citenamefont
  {Niquet}(2018)}]{Bourdet18}%
  \BibitemOpen
  \bibfield  {author} {\bibinfo {author} {\bibfnamefont {L.}~\bibnamefont
  {Bourdet}}\ and\ \bibinfo {author} {\bibfnamefont {Y.-M.}\ \bibnamefont
  {Niquet}},\ }\href {\doibase 10.1103/PhysRevB.97.155433} {\bibfield
  {journal} {\bibinfo  {journal} {Physical Review B}\ }\textbf {\bibinfo
  {volume} {97}},\ \bibinfo {pages} {155433} (\bibinfo {year}
  {2018})}\BibitemShut {NoStop}%
\bibitem [{\citenamefont {Dresselhaus}\ \emph {et~al.}(1955)\citenamefont
  {Dresselhaus}, \citenamefont {Kip},\ and\ \citenamefont
  {Kittel}}]{Dresselhaus55}%
  \BibitemOpen
  \bibfield  {author} {\bibinfo {author} {\bibfnamefont {G.}~\bibnamefont
  {Dresselhaus}}, \bibinfo {author} {\bibfnamefont {A.~F.}\ \bibnamefont
  {Kip}}, \ and\ \bibinfo {author} {\bibfnamefont {C.}~\bibnamefont {Kittel}},\
  }\href {\doibase 10.1103/PhysRev.98.368} {\bibfield  {journal} {\bibinfo
  {journal} {Physical Review}\ }\textbf {\bibinfo {volume} {98}},\ \bibinfo
  {pages} {368} (\bibinfo {year} {1955})}\BibitemShut {NoStop}%
\bibitem [{\citenamefont {Lew Yan~Voon}\ and\ \citenamefont
  {Willatzen}(2009)}]{KP09}%
  \BibitemOpen
  \bibfield  {author} {\bibinfo {author} {\bibfnamefont {L.~C.}\ \bibnamefont
  {Lew Yan~Voon}}\ and\ \bibinfo {author} {\bibfnamefont {M.}~\bibnamefont
  {Willatzen}},\ }\href {\doibase 10.1007/978-3-540-92872-0} {\emph {\bibinfo
  {title} {The k p Method}}}\ (\bibinfo  {publisher} {Springer},\ \bibinfo
  {address} {Berlin},\ \bibinfo {year} {2009})\BibitemShut {NoStop}%
\bibitem [{\citenamefont {Luttinger}(1956)}]{Luttinger56}%
  \BibitemOpen
  \bibfield  {author} {\bibinfo {author} {\bibfnamefont {J.~M.}\ \bibnamefont
  {Luttinger}},\ }\href {\doibase 10.1103/PhysRev.102.1030} {\bibfield
  {journal} {\bibinfo  {journal} {Physical Review}\ }\textbf {\bibinfo {volume}
  {102}},\ \bibinfo {pages} {1030} (\bibinfo {year} {1956})}\BibitemShut
  {NoStop}%
\bibitem [{\citenamefont {Kloeffel}\ \emph {et~al.}(2011)\citenamefont
  {Kloeffel}, \citenamefont {Trif},\ and\ \citenamefont {Loss}}]{Kloeffel11}%
  \BibitemOpen
  \bibfield  {author} {\bibinfo {author} {\bibfnamefont {C.}~\bibnamefont
  {Kloeffel}}, \bibinfo {author} {\bibfnamefont {M.}~\bibnamefont {Trif}}, \
  and\ \bibinfo {author} {\bibfnamefont {D.}~\bibnamefont {Loss}},\ }\href
  {\doibase 10.1103/PhysRevB.84.195314} {\bibfield  {journal} {\bibinfo
  {journal} {Physical Review B}\ }\textbf {\bibinfo {volume} {84}},\ \bibinfo
  {pages} {195314} (\bibinfo {year} {2011})}\BibitemShut {NoStop}%
\bibitem [{\citenamefont {Kloeffel}\ \emph {et~al.}(2018)\citenamefont
  {Kloeffel}, \citenamefont {Ran\ifmmode \check{c}\else
  \v{c}\fi{}i\ifmmode~\acute{c}\else \'{c}\fi{}},\ and\ \citenamefont
  {Loss}}]{Kloeffel17}%
  \BibitemOpen
  \bibfield  {author} {\bibinfo {author} {\bibfnamefont {C.}~\bibnamefont
  {Kloeffel}}, \bibinfo {author} {\bibfnamefont {M.~J.}\ \bibnamefont
  {Ran\ifmmode \check{c}\else \v{c}\fi{}i\ifmmode~\acute{c}\else \'{c}\fi{}}},
  \ and\ \bibinfo {author} {\bibfnamefont {D.}~\bibnamefont {Loss}},\ }\href
  {\doibase 10.1103/PhysRevB.97.235422} {\bibfield  {journal} {\bibinfo
  {journal} {Physical Review B}\ }\textbf {\bibinfo {volume} {97}},\ \bibinfo
  {pages} {235422} (\bibinfo {year} {2018})}\BibitemShut {NoStop}%
\bibitem [{\citenamefont {Voisin}\ \emph {et~al.}(2014)\citenamefont {Voisin},
  \citenamefont {Nguyen}, \citenamefont {Renard}, \citenamefont {Jehl},
  \citenamefont {Barraud}, \citenamefont {Triozon}, \citenamefont {Vinet},
  \citenamefont {Duchemin}, \citenamefont {Niquet}, \citenamefont
  {de~Franceschi},\ and\ \citenamefont {Sanquer}}]{Voisin14}%
  \BibitemOpen
  \bibfield  {author} {\bibinfo {author} {\bibfnamefont {B.}~\bibnamefont
  {Voisin}}, \bibinfo {author} {\bibfnamefont {V.-H.}\ \bibnamefont {Nguyen}},
  \bibinfo {author} {\bibfnamefont {J.}~\bibnamefont {Renard}}, \bibinfo
  {author} {\bibfnamefont {X.}~\bibnamefont {Jehl}}, \bibinfo {author}
  {\bibfnamefont {S.}~\bibnamefont {Barraud}}, \bibinfo {author} {\bibfnamefont
  {F.}~\bibnamefont {Triozon}}, \bibinfo {author} {\bibfnamefont
  {M.}~\bibnamefont {Vinet}}, \bibinfo {author} {\bibfnamefont
  {I.}~\bibnamefont {Duchemin}}, \bibinfo {author} {\bibfnamefont {Y.~M.}\
  \bibnamefont {Niquet}}, \bibinfo {author} {\bibfnamefont {S.}~\bibnamefont
  {de~Franceschi}}, \ and\ \bibinfo {author} {\bibfnamefont {M.}~\bibnamefont
  {Sanquer}},\ }\href {\doibase 10.1021/nl500299h} {\bibfield  {journal}
  {\bibinfo  {journal} {Nano Letters}\ }\textbf {\bibinfo {volume} {14}},\
  \bibinfo {pages} {2094} (\bibinfo {year} {2014})}\BibitemShut {NoStop}%
\bibitem [{Note7()}]{Note7}%
  \BibitemOpen
  \bibinfo {note} {These figures are actually drawn at a small magnetic field
  $B_z=5$ mT in order to lift Kramers degeneracy and catch the states with
  maximal $j_z=+\protect \frac {3}{2}$ component.}\BibitemShut {Stop}%
\bibitem [{\citenamefont {Kloeffel}\ \emph {et~al.}(2013)\citenamefont
  {Kloeffel}, \citenamefont {Trif}, \citenamefont {Stano},\ and\ \citenamefont
  {Loss}}]{Kloeffel13}%
  \BibitemOpen
  \bibfield  {author} {\bibinfo {author} {\bibfnamefont {C.}~\bibnamefont
  {Kloeffel}}, \bibinfo {author} {\bibfnamefont {M.}~\bibnamefont {Trif}},
  \bibinfo {author} {\bibfnamefont {P.}~\bibnamefont {Stano}}, \ and\ \bibinfo
  {author} {\bibfnamefont {D.}~\bibnamefont {Loss}},\ }\href {\doibase
  10.1103/PhysRevB.88.241405} {\bibfield  {journal} {\bibinfo  {journal}
  {Physical Review B}\ }\textbf {\bibinfo {volume} {88}},\ \bibinfo {pages}
  {241405} (\bibinfo {year} {2013})}\BibitemShut {NoStop}%
\bibitem [{\citenamefont {Watzinger}\ \emph {et~al.}(2016)\citenamefont
  {Watzinger}, \citenamefont {Kloeffel}, \citenamefont {Vukusic}, \citenamefont
  {Rossell}, \citenamefont {Sessi}, \citenamefont {Kukucka}, \citenamefont
  {Kirchschlager}, \citenamefont {Lausecker}, \citenamefont {Truhlar},
  \citenamefont {Glaser}, \citenamefont {Rastelli}, \citenamefont {Fuhrer},
  \citenamefont {Loss},\ and\ \citenamefont {Katsaros}}]{Watzinger16}%
  \BibitemOpen
  \bibfield  {author} {\bibinfo {author} {\bibfnamefont {H.}~\bibnamefont
  {Watzinger}}, \bibinfo {author} {\bibfnamefont {C.}~\bibnamefont {Kloeffel}},
  \bibinfo {author} {\bibfnamefont {L.}~\bibnamefont {Vukusic}}, \bibinfo
  {author} {\bibfnamefont {M.~D.}\ \bibnamefont {Rossell}}, \bibinfo {author}
  {\bibfnamefont {V.}~\bibnamefont {Sessi}}, \bibinfo {author} {\bibfnamefont
  {J.}~\bibnamefont {Kukucka}}, \bibinfo {author} {\bibfnamefont
  {R.}~\bibnamefont {Kirchschlager}}, \bibinfo {author} {\bibfnamefont
  {E.}~\bibnamefont {Lausecker}}, \bibinfo {author} {\bibfnamefont
  {A.}~\bibnamefont {Truhlar}}, \bibinfo {author} {\bibfnamefont
  {M.}~\bibnamefont {Glaser}}, \bibinfo {author} {\bibfnamefont
  {A.}~\bibnamefont {Rastelli}}, \bibinfo {author} {\bibfnamefont
  {A.}~\bibnamefont {Fuhrer}}, \bibinfo {author} {\bibfnamefont
  {D.}~\bibnamefont {Loss}}, \ and\ \bibinfo {author} {\bibfnamefont
  {G.}~\bibnamefont {Katsaros}},\ }\href {\doibase
  10.1021/acs.nanolett.6b02715} {\bibfield  {journal} {\bibinfo  {journal}
  {Nano Letters}\ }\textbf {\bibinfo {volume} {16}},\ \bibinfo {pages} {6879}
  (\bibinfo {year} {2016})}\BibitemShut {NoStop}%
\bibitem [{\citenamefont {Mansir}\ \emph {et~al.}(2018)\citenamefont {Mansir},
  \citenamefont {Conti}, \citenamefont {Zeng}, \citenamefont {Pla},
  \citenamefont {Bertet}, \citenamefont {Swift}, \citenamefont {Van~de Walle},
  \citenamefont {Thewalt}, \citenamefont {Sklenard}, \citenamefont {Niquet},\
  and\ \citenamefont {Morton}}]{Mansir18}%
  \BibitemOpen
  \bibfield  {author} {\bibinfo {author} {\bibfnamefont {J.}~\bibnamefont
  {Mansir}}, \bibinfo {author} {\bibfnamefont {P.}~\bibnamefont {Conti}},
  \bibinfo {author} {\bibfnamefont {Z.}~\bibnamefont {Zeng}}, \bibinfo {author}
  {\bibfnamefont {J.~J.}\ \bibnamefont {Pla}}, \bibinfo {author} {\bibfnamefont
  {P.}~\bibnamefont {Bertet}}, \bibinfo {author} {\bibfnamefont {M.~W.}\
  \bibnamefont {Swift}}, \bibinfo {author} {\bibfnamefont {C.~G.}\ \bibnamefont
  {Van~de Walle}}, \bibinfo {author} {\bibfnamefont {M.~L.~W.}\ \bibnamefont
  {Thewalt}}, \bibinfo {author} {\bibfnamefont {B.}~\bibnamefont {Sklenard}},
  \bibinfo {author} {\bibfnamefont {Y.~M.}\ \bibnamefont {Niquet}}, \ and\
  \bibinfo {author} {\bibfnamefont {J.~J.~L.}\ \bibnamefont {Morton}},\ }\href
  {\doibase 10.1103/PhysRevLett.120.167701} {\bibfield  {journal} {\bibinfo
  {journal} {Physical Review Letters}\ }\textbf {\bibinfo {volume} {120}},\
  \bibinfo {pages} {167701} (\bibinfo {year} {2018})}\BibitemShut {NoStop}%
\bibitem [{\citenamefont {Pla}\ \emph {et~al.}(2018)\citenamefont {Pla},
  \citenamefont {Bienfait}, \citenamefont {Pica}, \citenamefont {Mansir},
  \citenamefont {Mohiyaddin}, \citenamefont {Zeng}, \citenamefont {Niquet},
  \citenamefont {Morello}, \citenamefont {Schenkel}, \citenamefont {Morton},\
  and\ \citenamefont {Bertet}}]{Pla18}%
  \BibitemOpen
  \bibfield  {author} {\bibinfo {author} {\bibfnamefont {J.~J.}\ \bibnamefont
  {Pla}}, \bibinfo {author} {\bibfnamefont {A.}~\bibnamefont {Bienfait}},
  \bibinfo {author} {\bibfnamefont {G.}~\bibnamefont {Pica}}, \bibinfo {author}
  {\bibfnamefont {J.}~\bibnamefont {Mansir}}, \bibinfo {author} {\bibfnamefont
  {F.~A.}\ \bibnamefont {Mohiyaddin}}, \bibinfo {author} {\bibfnamefont
  {Z.}~\bibnamefont {Zeng}}, \bibinfo {author} {\bibfnamefont {Y.~M.}\
  \bibnamefont {Niquet}}, \bibinfo {author} {\bibfnamefont {A.}~\bibnamefont
  {Morello}}, \bibinfo {author} {\bibfnamefont {T.}~\bibnamefont {Schenkel}},
  \bibinfo {author} {\bibfnamefont {J.~J.~L.}\ \bibnamefont {Morton}}, \ and\
  \bibinfo {author} {\bibfnamefont {P.}~\bibnamefont {Bertet}},\ }\href
  {\doibase 10.1103/PhysRevApplied.9.044014} {\bibfield  {journal} {\bibinfo
  {journal} {Physical Review Applied}\ }\textbf {\bibinfo {volume} {9}},\
  \bibinfo {pages} {044014} (\bibinfo {year} {2018})}\BibitemShut {NoStop}%
\bibitem [{\citenamefont {Lee}\ \emph {et~al.}(2005)\citenamefont {Lee},
  \citenamefont {Fitzgerald}, \citenamefont {Bulsara}, \citenamefont {Currie},\
  and\ \citenamefont {Lochtefeld}}]{Lee05}%
  \BibitemOpen
  \bibfield  {author} {\bibinfo {author} {\bibfnamefont {M.~L.}\ \bibnamefont
  {Lee}}, \bibinfo {author} {\bibfnamefont {E.~A.}\ \bibnamefont {Fitzgerald}},
  \bibinfo {author} {\bibfnamefont {M.~T.}\ \bibnamefont {Bulsara}}, \bibinfo
  {author} {\bibfnamefont {M.~T.}\ \bibnamefont {Currie}}, \ and\ \bibinfo
  {author} {\bibfnamefont {A.}~\bibnamefont {Lochtefeld}},\ }\href {\doibase
  10.1063/1.1819976} {\bibfield  {journal} {\bibinfo  {journal} {Journal of
  Applied Physics}\ }\textbf {\bibinfo {volume} {97}},\ \bibinfo {pages}
  {011101} (\bibinfo {year} {2005})}\BibitemShut {NoStop}%
\bibitem [{\citenamefont {Sun}\ \emph {et~al.}(2007)\citenamefont {Sun},
  \citenamefont {Thompson},\ and\ \citenamefont {Nishida}}]{Sun07}%
  \BibitemOpen
  \bibfield  {author} {\bibinfo {author} {\bibfnamefont {Y.}~\bibnamefont
  {Sun}}, \bibinfo {author} {\bibfnamefont {S.~E.}\ \bibnamefont {Thompson}}, \
  and\ \bibinfo {author} {\bibfnamefont {T.}~\bibnamefont {Nishida}},\ }\href
  {\doibase 10.1063/1.2730561} {\bibfield  {journal} {\bibinfo  {journal}
  {Journal of Applied Physics}\ }\textbf {\bibinfo {volume} {101}},\ \bibinfo
  {pages} {104503} (\bibinfo {year} {2007})}\BibitemShut {NoStop}%
\bibitem [{\citenamefont {Onodera}\ and\ \citenamefont
  {Okazaki}(1966)}]{Onodera66}%
  \BibitemOpen
  \bibfield  {author} {\bibinfo {author} {\bibfnamefont {Y.}~\bibnamefont
  {Onodera}}\ and\ \bibinfo {author} {\bibfnamefont {M.}~\bibnamefont
  {Okazaki}},\ }\href {\doibase 10.1143/JPSJ.21.2400} {\bibfield  {journal}
  {\bibinfo  {journal} {Journal of the Physical Society of Japan}\ }\textbf
  {\bibinfo {volume} {21}},\ \bibinfo {pages} {2400} (\bibinfo {year}
  {1966})}\BibitemShut {NoStop}%
\bibitem [{Note8()}]{Note8}%
  \BibitemOpen
  \bibinfo {note} {The double group $D_{2h}$ has two possible irreducible
  representations for a Kramers doublet that differ by the sign of $\Gamma
  _S(\sigma _{xy})$. The conclusions are, therefore, the same whether the
  Kramers doublet belongs to one or the other irreducible
  representation.}\BibitemShut {Stop}%
\bibitem [{\citenamefont {Graf}\ and\ \citenamefont {Vogl}(1995)}]{Vogl95}%
  \BibitemOpen
  \bibfield  {author} {\bibinfo {author} {\bibfnamefont {M.}~\bibnamefont
  {Graf}}\ and\ \bibinfo {author} {\bibfnamefont {P.}~\bibnamefont {Vogl}},\
  }\href {\doibase 10.1103/PhysRevB.51.4940} {\bibfield  {journal} {\bibinfo
  {journal} {Physical Review B}\ }\textbf {\bibinfo {volume} {51}},\ \bibinfo
  {pages} {4940} (\bibinfo {year} {1995})}\BibitemShut {NoStop}%
\bibitem [{Note9()}]{Note9}%
  \BibitemOpen
  \bibinfo {note} {We assume $g_0=2$ in Eq. (\ref {eqHbloch}).}\BibitemShut
  {Stop}%
\bibitem [{\citenamefont {Sleijpen}\ and\ \citenamefont {Van~der
  Vorst}(2000)}]{Sleijpen00}%
  \BibitemOpen
  \bibfield  {author} {\bibinfo {author} {\bibfnamefont {G.}~\bibnamefont
  {Sleijpen}}\ and\ \bibinfo {author} {\bibfnamefont {H.}~\bibnamefont {Van~der
  Vorst}},\ }\href {\doibase 10.1137/S0036144599363084} {\bibfield  {journal}
  {\bibinfo  {journal} {SIAM Review}\ }\textbf {\bibinfo {volume} {42}},\
  \bibinfo {pages} {267} (\bibinfo {year} {2000})}\BibitemShut {NoStop}%
\bibitem [{\citenamefont {Bai}\ \emph {et~al.}(2000)\citenamefont {Bai},
  \citenamefont {Demmel}, \citenamefont {Dongarra}, \citenamefont {Ruhe},\ and\
  \citenamefont {van~der Vorst}}]{Templates00}%
  \BibitemOpen
  \bibinfo {editor} {\bibfnamefont {Z.}~\bibnamefont {Bai}}, \bibinfo {editor}
  {\bibfnamefont {J.}~\bibnamefont {Demmel}}, \bibinfo {editor} {\bibfnamefont
  {J.}~\bibnamefont {Dongarra}}, \bibinfo {editor} {\bibfnamefont
  {A.}~\bibnamefont {Ruhe}}, \ and\ \bibinfo {editor} {\bibfnamefont
  {H.}~\bibnamefont {van~der Vorst}},\ eds.,\ \href {\doibase
  10.1137/1.9780898719581} {\emph {\bibinfo {title} {Templates for the Solution
  of Algebraic Eigenvalue Problems: A Practical Guide}}}\ (\bibinfo
  {publisher} {SIAM},\ \bibinfo {address} {Philadelphia},\ \bibinfo {year}
  {2000})\BibitemShut {NoStop}%
\end{thebibliography}
\end{document}